\documentclass[preprint,showpacs,preprintnumbers,nofootinbib,amsmath,amssymb]{revtex4}
\usepackage{hyperref,slashed,color}
\usepackage{graphicx,subfig}

\begin{document}
\preprint{TUHEP-TH-07157}
\title{Dynamical Computation on Coefficients of\\ Electroweak Chiral Lagrangian
from One-doublet and Topcolor-assisted Technicolor Models}

\author{Hong-Hao Zhang\footnote{Email: \href{mailto:zhanghonghao@tsinghua.org.cn}{zhanghonghao@tsinghua.org.cn}.},
Shao-Zhou Jiang\footnote{Email:
\href{mailto:jsz@mails.tsinghua.edu.cn}{jsz@mails.tsinghua.edu.cn}.},
and Qing Wang\footnote{Email:
\href{mailto:wangq@mail.tsinghua.edu.cn}{wangq@mail.tsinghua.edu.cn}.}\footnote{corresponding
author}}

\affiliation{Department of Physics, Tsinghua University, Beijing 100084, China\footnote{mailing address} \\
    Center for High Energy Physics, Tsinghua University, Beijing 100084, China}

\begin{abstract}
Based on previous studies deriving the chiral Lagrangian for pseudo
scalar mesons from the first principle of QCD, we derive the
electroweak chiral Lagrangian and build up a formulation for
computing its coefficients from one-doublet technicolor model and a
schematic topcolor-assisted technicolor model. We find that the
coefficients of the electroweak chiral Lagrangian for the
topcolor-assisted technicolor model are divided into three parts:
direct TC2 interaction part, TC1 and TC2 induced effective $Z'$
particle contribution part, and ordinary quarks contribution part.
The first two parts are computed in this paper and we show that the
direct TC2 interaction part is the same as that in the one-doublet
technicolor model, while effective $Z'$ contributions are at least
proportional to the $p^2$ order parameter $\beta_1$ in the
electroweak chiral Lagrangian  and typical features of
topcolor-assisted technicolor model are that it only allows positive
$T$ and $U$ parameters and the $T$ parameter varies in the range
$0\sim 1/(25\alpha)$, the upper bound of $T$ parameter will decrease
as long as $Z'$ mass become large. The $S$ parameter can be either
positive or negative depending on whether the $Z'$ mass is large or
small. The $Z'$ mass is also bounded above and the upper bound
depend on value of $T$ parameter. We obtain the values for all the
coefficients of the electroweak chiral Lagrangian up to order of
$p^4$.
\end{abstract}
 \maketitle
\tableofcontents
\section{Introduction}

 The electroweak symmetry breaking mechanism (EWSBM) remains an intriguing puzzle for particle physics,
 although the Standard Model (SM) provides us with a version of it through introducing a higgs boson into the
 theory which suffers from triviality and unnaturalness problems. Beyond SM, numerous new physics models are
invented which exhibit many alternative EWSBMs.  With the present
situation that higgs is still not found in experiment, all new
physics models at low energy region should be described by a theory
which not only must match all present experiment data, but also have
no higgs. This theory is the well-known electroweak chiral
Lagrangian (EWCL) \cite{EWCL0,EWCL,EWCLmatter} which offers the most
general and economic description of electroweak interaction at low
energy region. With EWCL, new physics models at low energies can be
parameterized by a set of coefficients, it universally describes all
possible electroweak interactions among existing particles and
offers a model independent platform for us to investigate various
kinds EWSBMs. Starting from this platform, further phenomenological
research focus on finding effective physical processes to fix the
certain coefficients of EWCL \cite{Alam98,He97,Dobado95}, and
theoretical studies concentrate on consistency of EWCL itself such
as gauge invariance \cite{GaugeInv} and computing the values of the
coefficients for SM with heavy higgs \cite{HeavyHiggs}. Systematic
theoretical computation of the EWCL coefficients for other new
physics models have not been presented in the literature. The
possible reasons are that for weakly coupled models, since one can
perform perturbative computations, people prefer to directly discuss
physics from the model and then are reluctant to pay the extra price
to compute EWCL coefficients. While for strongly coupled models,
non-perturbative difficulties for a long time prevent people to
perform dynamical computations, only for special coefficients such
as $S$ parameter, some non-perturbative technique may be applied to
perform calculations \cite{STU} or for special QCD-like technicolor
models, in terms of their similarities with QCD, one can estimate
the coefficients of EWCL in terms of their partners fixed by
experimental data in corresponding QCD chiral Lagrangian. The
estimation of EWCL coefficients for various models is of special
importance in the sense that at present we already have some
quantitative constraints on them, such as those for the $S,T,U$
parameters and more generally for anomalous triple and quartic
couplings \cite{PDG06}, along with the experimental progress, more
constraints will be obtained. Once we know the values of the
coefficients for detailed models,  these constraints can directly be
used to judge the correctness of the model. It is the purpose of
this paper to develop a formulation to systematically compute
coefficients of EWCL for strongly coupled new physics models. For
simplicity, in this work we only discuss the bosonic part of EWCL
and leave the matter part for future investigations. The basis of
our formulation is the knowledge and experiences we obtained
previously from a series of works deriving the chiral Lagrangian for
pseudo scalar mesons from QCD first principles \cite{Wang00} and
calculate coefficients in it \cite{Wang02,Yang02,Ma03}, where we
found confidence and reliability in this work. In fact, the formal
derivation from a general underlying technicolor model to EWCL was
already achieved in Ref.\cite{WQTechnicolor01} in which except
deriving EWCL, coefficients of EWCL are formally expressed in terms
of Green's functions in underlying technicolor model. Once we know
how to compute these Green's functions, we obtain the corresponding
EWCL coefficients. The pity is that the computation is
nonperturbative, therefore not easy to achieve, it is the aim of
this paper to solve this nonperturbative dynamical computation
problems.

As the first step of performing dynamical computations, we
especially care about the reliability of the formulation we will
develop. We take one-doublet technicolor model
\cite{Weinberg7579,Susskind78,Farhi80,Hill02} as the prototype to
build up our formulation. Although this model as the earliest and
simplest dynamical electroweak symmetry breaking model was already
denied by experiment in the sense that it results in too large a
value for the $S$ parameter, but due to the following reasons, we
still start our investigations from here. First, it is similar as
QCD in the theory structures which enable us to easily generalize
the techniques developed in dealing with the QCD chiral Lagrangian
to this model and we call this generalized formulation the dynamical
computation prescription. Second due to their similarities with
conventional QCD, the coefficients of their EWCLs can be estimated
by just scaling-up corresponding coefficients in QCD
Gasser-Leutwyler chiral Lagrangian for pseudo scalar mesons
\cite{Gasser8384} and  we call this formulation the
Gasser-Leutwyler's prescription which naively is only applicable for
those QCD-like models. So for QCD-like models, we have two
prescriptions which enable us to compare them with each other to
check the correctness and increase the reliability of our
formulations. Beyond the traditional one-doublet technicolor model,
we choose the topcolor-assisted technicolor model as the first real
practice model to perform our computations. The reason to take it is
that this model is not QCD-like and is active on the market now
which does not seriously contradict with experimental data as the
case of one-doublet technicolor model and the dynamics responsible
for electroweak symmetry breaking is similar to that in the
one-doublet technicolor model. We will find
 that the coefficients for topcolor-assisted technicolor model can
 be divided into three parts: direct TC2 interaction part,TC1 and TC2
induced effective $Z'$ particle contribution part and ordinary
quarks contribution part. The first two parts are computed in this
paper and we show that direct TC2 interaction part is same as that
in one-doublet technicolor model, while TC1 and TC2 induced
effective $Z'$ particle contributions is at least proportional to
$p^2$ order parameter $\beta_1$ in EWCL and  typical features of
topcolor-assisted technicolor model are that it only allows positive
$T$ and $U$ parameters and the $T$ parameter varies in the range
$0\sim 1/(25\alpha)$,the upper bound of $T$ parameter will decrease
as long as $Z'$ mass become large. The $S$ parameter can be either
positive or negative depending on whether the $Z'$ mass is large or
small. The $Z'$ mass is also bounded above and the upper bound
depend on value of $T$ parameter. We obtain the values for all the
coefficients of the electroweak chiral Lagrangian up to order of
$p^4$.

This paper is organized as follows. Section II is the basics of the
work in which we discuss one-doublet technicolor model. We first
review the Gasser-Leutwyler's prescription,  then build up our
dynamical computation prescription, we show how to consistently set
in the dynamical computation equation (the Schwinger-Dyson equation)
into our formulation. We make comparison between two prescriptions
to check validity of the results from our dynamical computation
prescription. Section III is the main part of this work in which we
apply our formulations developed in one-doublet technicolor model to
 topcolor-assisted technicolor model. We perform dynamical calculations
 on technicolor interactions and then
integrate out colorons and $Z'$ to compute EWCL coefficients. Since
this is the first time to systematically perform dynamical
computations on the strongly coupled models, we emphasize the
technical side more than physics analysis and display the
computation procedure a little bit more in detail.
 Section IV is the conclusion. In the appendices, we list some requisite formulae.

\section{Derivation of the Electroweak Chiral Lagrangian from the One-doublet Technicolor
 Model}

 Consider the one-doublet technicolor (TC) model proposed by
Weinberg and Susskind independently
\cite{Weinberg7579,Susskind78,Farhi80,Hill02}. The techniquarks are
assigned to $(SU(N)_{TC}, SU(3)_C, SU(2)_L, U(1)_Y)$ as $\psi_L\sim
(N, 1, 2, 0)$,~$U_R=(1/2+\tau^3/2)\psi_R\sim(N, 1, 1,
1/2)$,~$D_R=(1/2-\tau^3/2)$,~$\psi_R\sim(N, 1, 1, -1/2)$. With these
assignments, the techniquarks have electric charges as defined by
$Q=T_3+Y$, of $+1/2$ for $U$ and $-1/2$ for $D$. It can be shown
below, by dynamical analysis through the Schwinger-Dyson equation,
that the $SU(N)_{TC}$ interaction induces the techniquark condensate
$\langle \bar{\psi}\psi\rangle\neq 0$, which will trigger the
electroweak symmetry breaking $SU(2)_L\times U(1)_Y\rightarrow
U(1)_{\rm EM}$. Neglecting ordinary fermions and gluons, we focus on
the action of the techniquark, technicolor-gauge-boson and
electroweak-gauge-boson sector, {\it i.e.} the electroweak symmetry
breaking sector (SBS) of this model,
\begin{eqnarray}
S_{\rm SBS}&=&\int d^4x\bigg[-\frac{1}{4}F_{\mu\nu}^\alpha
F^{\alpha,\mu\nu}-\frac{1}{4}W_{\mu\nu}^aW^{a,\mu\nu}
-\frac{1}{4}B_{\mu\nu}B^{\mu\nu}\nonumber\\
&&+\bar{\psi}\bigg(i\slashed{\partial}-g_{\rm TC}t^\alpha
\slashed{G}^\alpha-g_2\frac{\tau^a}{2}\slashed{W}^a
P_L-g_1\frac{\tau^3}{2}\slashed{B}P_R\bigg)\psi\bigg]\label{S1Ddef}\,,
\end{eqnarray}
where $g_{\rm TC}$, $g_2$ and $g_1$ ($G_\mu^\alpha$, $W_\mu^a$ and
$B_\mu$) are the coupling constants (gauge fields) of
$SU(N)_{TC}\times SU(2)_L\times U(1)_Y$ with technicolor index
$\alpha$ ($\alpha=1,2,\ldots,N^2-1$) and weak index $a$ ($a=1,2,3$)
respectively; and $F_{\mu\nu}^\alpha$, $W_{\mu\nu}^a$ and
$B_{\mu\nu}$ are the corresponding field strength tensors;
$t^\alpha$ ($\alpha=1,2,\ldots,N^2-1$) are the generators for the
fundamental representation of $SU(N)_{TC}$, while $\tau^a$
($a=1,2,3$) are Pauli matrices; and the left and right chirality
projection operators $P_{^L_R}=(1\mp\gamma_5)/2$.

To derive low energy effective electroweak chiral Lagrangian from
the one-doublet TC model, we need to integrate out the technigluons
and techniquarks above the electroweak scale which can be formulated
as
\begin{eqnarray}
\int{\cal D}G_\mu^\alpha{\cal D}\bar{\psi}{\cal
D}\psi\exp\bigg(\,iS_{\rm
SBS}\big[G_\mu^\alpha,W_\mu^a,B_\mu,\bar{\psi},\psi\big]\bigg)=\int{\cal
D}\mu(U)\exp\bigg(iS_{\rm
eff}[U,W_\mu^a,B_\mu]\bigg)\;,\label{strategy}
\end{eqnarray}
where $U(x)$ is a dimensionless unitary unimodular matrix field in
the electroweak chiral Lagrangian, and ${\cal D}\mu(U)$ denotes the
corresponding functional integration measure.

As mentioned in previous section, there are two different
approaches, one is the Gasser-Leutwyler's prescription, the other is
the dynamical computation prescription. The second approach we
developed in this paper is relatively easy to be generalized to more
complicated theories. We will compare the results obtained in both
approaches.

\subsection{The Gasser-Leutwyler's Prescription}

As we will see, it is easy to relate QCD-like models to chiral
Lagrangian using the Gasser-Leutwyler's prescription. To begin, we
substitute (\ref{S1Ddef}) into the left-hand side of
Eq.(\ref{strategy}) and
 the result path integral involved technicolor interaction
is analogous to QCD and then we can use technique developed by
Gasser and Leutwyler relating it with the path integral of chiral
Lagrangian for goldstone bosons induced from SBS \cite{Gasser8384},
\begin{eqnarray}
&&\frac{\int{\cal D}G_\mu^\alpha{\cal D}\bar{\psi}{\cal
D}\psi\exp\bigg\{i\int d^4x\bigg[-\frac{1}{4}F_{\mu\nu}^\alpha
F^{\alpha,\mu\nu} +\bar{\psi}\bigg(i\slashed{\partial}-g_{\rm
TC}t^\alpha \slashed{G}^\alpha-g_2\frac{\tau^a}{2}\slashed{W}^a
P_L-g_1\frac{\tau^3}{2}\slashed{B}P_R\bigg)\psi\bigg]\bigg\}}{\int{\cal
D}\bar{\psi}{\cal D}\psi\exp\{i\int
d^4x\bar{\psi}[i\slashed{\partial}-g_2\frac{\tau^a}{2}\slashed{W}^a
P_L-g_1\frac{\tau^3}{2}\slashed{B}P_R]\psi\}}\nonumber\\
&&\qquad=\int{\cal D}\mu(\tilde{U})\exp\{iS_{\mbox{\scriptsize
TC-induced eff}}[\tilde{U},W,B]\}\;,\label{TCinducedef}
\end{eqnarray}
in which the denominator of the left hand side of above equation is
introduced to insure the technicolor-induced chiral effective action
$S_{\mbox{\scriptsize TC-induced eff}}[\tilde{U},W,B]$ normalized as
zero when we switch off technicolor interactions by setting
$g_\mathrm{TC}=0$ and $S_{\mbox{\scriptsize TC-induced
eff}}[\tilde{U},W,B]$ can be written as
\begin{eqnarray}
S_{\mbox{\scriptsize TC-induced eff}}[\tilde{U},W,B]&=&\int d^4x
\bigg[\frac{(F_0^{\rm 1D})^2}{4} {\rm
tr}[(\nabla^\mu\tilde{U}^\dag)(\nabla_\mu\tilde{U})]+ L_1^{\rm
1D}[{\rm
tr}(\nabla^\mu\tilde{U}^\dag\nabla_\mu\tilde{U})]^2\nonumber\\&&+L_2^{\rm
1D}{\rm tr}[\nabla_\mu\tilde{U}^{\dag}\nabla_\nu\tilde{U}] {\rm
tr}[\nabla^\mu\tilde{U}^{\dag}\nabla^\nu\tilde{U}]+L_3^{\rm 1D}{\rm
tr}[(\nabla^\mu\tilde{U}^\dag\nabla_\mu\tilde{U})^2]\nonumber\\
&&-iL_9^{\rm 1D}{\rm
tr}[F_{\mu\nu}^R\nabla^\mu\tilde{U}\nabla^\nu\tilde{U}^{\dag}
+F_{\mu\nu}^L\nabla^\mu\tilde{U}^\dag\nabla^\nu\tilde{U}]+L_{10}^{\rm
1D}{\rm
tr}[\tilde{U}^{\dag}F_{\mu\nu}^R\tilde{U}F^{L,\mu\nu}]\nonumber\\
&& +H_1^{\rm 1D}{\rm
tr}[F_{\mu\nu}^RF^{R,\mu\nu}+F_{\mu\nu}^LF^{L,\mu\nu}]\bigg]+O(p^6)\;,\label{action-QCD-like}
\end{eqnarray}
where the coefficients $F_0^{\rm 1D}$, $L_1^{\rm 1D}$, $L_2^{\rm
1D}$, $L_3^{\rm 1D}$, $L_{10}^{\rm 1D}$, $H_1^{\rm 1D}$ arise from
$SU(N)_{\mathrm{TC}}$ dynamics at the scale of 250 GeV, and where
\begin{eqnarray}
&&\hspace{-0.5cm}\nabla_\mu\tilde{U}\equiv\partial_\mu\tilde{U}-ir_\mu\tilde{U}+i\tilde{U}l_\mu\;,
\qquad\nabla_{\mu}\tilde{U}^\dag=-\tilde{U}^\dag(\nabla_{\mu}\tilde{U})\tilde{U}^\dag
=\partial_\mu\tilde{U}^\dag-il_\mu\tilde{U}^\dag+i\tilde{U}^\dag
r_\mu\;,\nonumber\\
&&\hspace{-0.5cm}F_{\mu\nu}^R\equiv i[\partial_\mu-ir_\mu,
\partial_\nu-ir_\nu]\hspace{0.5cm} F_{\mu\nu}^L\equiv
i[\partial_\mu-il_\mu,
\partial_\nu-il_\nu]\hspace{0.5cm}r_\mu\equiv-g_1\frac{\tau^3}{2}B_\mu\hspace{0.5cm}
l_\mu\equiv-g_2\frac{\tau^a}{2}W_\mu^a\;.~~~~\label{eq-U-tilde-definition}
\end{eqnarray}
Note that conventional $\tilde{U}$ field in
Eq.(\ref{action-QCD-like})  given in second paper of
\cite{Gasser8384} is $3\times 3$ unitary matrix. However, for the
$SU(2)_L\times SU(2)_R$ electroweak chiral Lagrangian we considered
in this paper, $\tilde{U}$ is a $2\times 2$ unitary matrix, and thus
the $L_1^{\rm 1D}$ term and the $L_3^{\rm 1D}$ term in present
situation are linearly related,
\begin{eqnarray}
L_3^{\rm 1D}{\rm
tr}[(\nabla^\mu\tilde{U}^\dag\nabla_\mu\tilde{U})^2]=L_3^{\rm
1D}{\rm tr}\bigg\{\big[\tilde{U}^\dag(\nabla^\mu\tilde{U})
\tilde{U}^\dag(\nabla^\mu\tilde{U})\big]^2\bigg\} =\frac{L_3^{\rm
1D}}{2}\bigg\{{\rm
tr}\big[\tilde{U}^\dag(\nabla^\mu\tilde{U})\tilde{U}^\dag(\nabla^\mu\tilde{U})\big]\bigg\}^2
\end{eqnarray}
 Comparing covariant derivative for $\tilde{U}$ given in
 (\ref{eq-U-tilde-definition}) and covariant derivative given in
 Ref.\cite{EWCL}, we find we must recognize $\tilde{U}^\dag=U$, $\nabla_\mu\tilde{U}^\dag=D_\mu
 U$, $F_{\mu\nu}^R=-g_1\frac{\tau^3}{2}B_{\mu\nu}$ and
$F_{\mu\nu}^L=-g_2\frac{\tau^a}{2}W_{\mu\nu}^a$. Substitute them
back into Eq.(\ref{action-QCD-like}), we obtain
\begin{eqnarray}
S_{\mbox{\scriptsize TC-induced eff}}[U,W,B]&=&\int d^4x
\bigg[-\frac{(F_0^{\rm 1D})^2}{4} {\rm tr}(X_\mu X^\mu)+(L_1^{\rm
1D}+\frac{L_3^{\rm 1D}}{2})[{\rm tr}(X_\mu
X^\mu)]^2\nonumber\\
&&+L_2^{\rm 1D}[{\rm tr}(X_\mu X_\nu)]^2-i\frac{L_9^{\rm
1D}}{2}g_1B_{\mu\nu}{\rm tr}(\tau^3X^\mu X^\nu)-iL_9^{\rm 1D}{\rm
tr}(\overline{W}_{\mu\nu}X^\mu X^\nu)\nonumber\\
&&+\frac{L_{10}^{\rm 1D}}{2}g_1B_{\mu\nu}{\rm
tr}(\tau^3\overline{W}^{\mu\nu})+\frac{H_1^{\rm
1D}}{2}g_1^2B_{\mu\nu}B^{\mu\nu}+H_1^\mathrm{1D}{\rm
tr}(\overline{W}_{\mu\nu}\overline{W}^{\mu\nu})\bigg]\;.\label{action-QCD-like2}
\end{eqnarray}
where
\begin{eqnarray}
X_\mu\equiv U^\dag(D_\mu U)\hspace{2cm}\overline{W}_{\mu\nu}\equiv
U^\dag g_2W_{\mu\nu}U\;,
\end{eqnarray}
We have reformulated the EWCL in terms of $X_\mu$ and $\tau^3$
instead of $V_\mu$ and $T$ in Ref.\cite{EWCL}, the corresponding
relations are given in Appendix A. Comparing
(\ref{action-QCD-like2}) with the standard electroweak chiral
Lagrangian given in Ref.\cite{EWCL},  we find
\begin{eqnarray}
&&f^2=(F_0^{\rm 1D})^2\,,\qquad \beta_1=0\,,\qquad
\alpha_1=L_{10}^{\rm 1D}\,,\qquad
\alpha_2=\alpha_3=-\frac{L_9^{\rm 1D}}{2}\,,\qquad \alpha_4=L_2^{\rm 1D}\,,\nonumber\\
&&\alpha_5=L_1^{\rm 1D}+\frac{L_3^{\rm 1D}}{2}\,,\qquad
\alpha_i=0~~(i=6,7,\ldots,14)\,.\label{Gprescript}
\end{eqnarray}
Note that in (\ref{action-QCD-like2}), the term with coefficient
$H_1^\mathrm{1D}$ do not affect the result $\alpha_i$ coefficients.
Not like the original case of QCD, $H_1^\mathrm{1D}$ now is a finite
constant. The divergences are from the term
$\mathrm{Tr}\log(i\slashed{\partial}-g_2\frac{\tau^a}{2}\slashed{W}^a
P_L-g_1\frac{\tau^3}{2}\slashed{B}P_R)$ in (\ref{TCinducedef}) which
will only contribute $\mathrm{tr}(W_{\mu\nu}W^{\mu\nu})$ and
$B_{\mu}B^{\mu}$ terms with divergent coefficients due to gauge
invariance. These divergent coefficients in combined with
$H_1^\mathrm{1D}$ will cause wave function renormalization
corrections for $W^a_\mu$ and $B_\mu$ fields which further lead
redefinitions of $W^a_\mu$ and $B_\mu$ fields and their gauge
couplings $g_2$ and $g_1$. This renormalization procedure will have
no effects on our EWCL, since all electroweak gauge fields appeared
in EWCL are as form of $g_2W^a_\mu$ and $g_1B_\mu$ which are
renormalization invariant quantities. Due to this consideration, in
rest of this paper, we just left the wave function corrections to
electroweak gauge fields $W^a_\mu$ and $B_\mu$ in the theory and
skip the corresponding renormalization procedure.

\subsection{The Dynamical Computation Prescription}

Now we develop a dynamical computation program and we will apply it
to more complicated model in next section.

We first review the derivation process given in
Ref.\cite{WQTechnicolor01} and start with introducing a local
$2\times2$ operator $O(x)$ as $O(x)\equiv{\rm
tr}_{lc}[\psi_L(x)\bar{\psi}_R(x)]$ with ${\rm tr}_{lc}$ is the
trace with respect to Lorentz and technicolor indices. The
transformation of $O(x)$ under $SU(2)_L\times U(1)_Y$ is
$O(x)\rightarrow V_L(x)O(x)V_R^\dag(x)$. Then we decompose $O(x)$ as
$O(x)=\xi_L^\dag(x)\sigma(x)\xi_R(x)$ with the $\sigma(x)$
represented by a hermitian matrix describes the modular degree of
freedom; while $\xi_L(x)$ and $\xi_R(x)$ represented by unitary
matrices describe the phase degree of freedom of $SU(2)_L$ and
$U(1)_Y$ respectively. Their transformation under $SU(2)_L\times
U(1)_Y$ are $\sigma(x)\rightarrow
h(x)\sigma(x)h^\dag(x)$,~$\xi_L(x)\rightarrow
h(x)\xi_L(x)V_L^\dag(x)$ and $\xi_R(x)\rightarrow
h(x)\xi_R(x)V_R^\dag(x)$ with $h(x)=e^{i\theta_h(x)\tau^3/2}$
belongs to an induced hidden local $U(1)$ symmetry group. Now we
define a new field $U(x)$ as $U(x)\equiv\xi_L^\dag(x)\xi_R(x)$ which
 is the nonlinear realization of the goldstone boson fields in the
electroweak chiral Lagrangian. Subtracting the $\sigma(x)$ field, we
find that the present decomposition results in a constraint
$\xi_L(x)O(x)\xi_R^\dag(x)-\xi_R(x)O^\dag(x)\xi_L^\dag(x)=0$, the
functional expression of it is
\begin{eqnarray}
\int\mathcal{D}\mu(U)\mathcal{F}[O]\delta(\xi_LO\xi_R^\dag-\xi_RO^\dag\xi_L^\dag)
=\mathrm{const.}\;,\label{eq-functional-identity}
\end{eqnarray}
where $\mathcal{D}\mu(U)$ is an effective invariant integration
measure; $\mathcal{F}[O]$ only depends on $O$, and it compensates
the integration to make it to be a constant. It is easy to show that
$\mathcal{F}[O]$ is invariant under $SU(2)_L\times U(1)_Y$
transformations. Substituting Eq.\eqref{eq-functional-identity} into
the left-hand side of Eq.(\ref{strategy}), we have
\begin{eqnarray}
\int{\cal D}G_\mu^\alpha{\cal D}\bar{\psi}{\cal
D}\psi\exp\bigg(\,iS_{\rm
SBS}\big[G_\mu^\alpha,W_\mu^a,B_\mu,\bar{\psi},\psi\big]\bigg)
=\int\mathcal{D}\mu(U)\exp\bigg(iS_{\rm
eff}[U,W_\mu^a,B_\mu]\bigg)\;,\label{SBSdef}
\end{eqnarray}
where
\begin{eqnarray}
&&\hspace{-0.5cm}S_{\rm eff}[U,W_\mu^a,B_\mu]\label{action-eff1a}\\
&=&\int
d^4x(-\frac{1}{4}W_{\mu\nu}^aW^{a,\mu\nu}-\frac{1}{4}B_{\mu\nu}B^{\mu\nu})
-i\log\int{\cal D}G_\mu^\alpha{\cal D}\bar{\psi}{\cal
D}\psi\,\mathcal{F}[O]\delta(\xi_LO\xi_R^\dag-\xi_RO^\dag\xi_L^\dag)\nonumber\\
&&\times \exp\bigg\{i\int d^4x\bigg[-\frac{1}{4}F_{\mu\nu}^\alpha
F^{\alpha,\mu\nu} +\bar{\psi}\bigg(i\slashed{\partial}-g_{\rm
TC}t^\alpha \slashed{G}^\alpha-g_2\frac{\tau^a}{2}\slashed{W}^a
P_L-g_1\frac{\tau^3}{2}\slashed{B}P_R\bigg)\psi\bigg]\bigg\}\nonumber
\end{eqnarray}
To match the correct normalization, we introduce in the argument of
logarithm function the normalization factor $\int{\cal
D}\bar{\psi}{\cal D}\psi e^{i\int
d^4x\bar{\psi}(i\slashed{\partial}-g_2\frac{\tau^a}{2}\slashed{W}^a
P_L-g_1\frac{\tau^3}{2}\slashed{B}P_R)\psi}=
\exp\mathrm{Tr}\log(i\slashed{\partial}-g_2\frac{\tau^a}{2}\slashed{W}^a
P_L-g_1\frac{\tau^3}{2}\slashed{B}P_R)$ and then take a special
$SU(2)_L\times U(1)_Y$ rotation, as $V_L(x)=\xi_L(x)$ and
$V_R(x)=\xi_R(x)$, on both numerator and denominator of the
normalization factor
\begin{eqnarray}
&&\hspace{-0.5cm}S_{\rm eff}[U,W_\mu^a,B_\mu]\label{action-eff1}\\
&=&\int
d^4x(-\frac{1}{4}W_{\mu\nu}^aW^{a,\mu\nu}-\frac{1}{4}B_{\mu\nu}B^{\mu\nu})
+\mathrm{Tr}\log(i\slashed{\partial}-g_2\frac{\tau^a}{2}\slashed{W}^a
P_L-g_1\frac{\tau^3}{2}\slashed{B}P_R)\nonumber\\
&&\hspace{-0.5cm}-i\log\frac{\int{\cal D}G_\mu^\alpha{\cal
D}\bar{\psi}_\xi{\cal
D}\psi_\xi\,\mathcal{F}[O_\xi]\delta(O_\xi-O_\xi^\dag)e^{i\int
d^4x[-\frac{1}{4}F_{\mu\nu}^\alpha F^{\alpha,\mu\nu}
+\bar{\psi}_\xi(i\slashed{\partial}-g_{\rm TC}t^\alpha
\slashed{G}^\alpha-g_2\frac{\tau^a}{2}\slashed{W}_\xi^a
P_L-g_1\frac{\tau^3}{2}\slashed{B}_\xi P_R)\psi_\xi]}}{\int{\cal
D}\bar{\psi}_\xi{\cal D}\psi_\xi e^{i\int
d^4x\bar{\psi}_\xi(i\slashed{\partial}-g_2\frac{\tau^a}{2}\slashed{W}_\xi^a
P_L-g_1\frac{\tau^3}{2}\slashed{B}_\xi P_R)\psi_\xi}}\;,\nonumber
\end{eqnarray}
where rotated fields are denoted as follows
\begin{eqnarray}
&&\hspace{-0.5cm}\psi_\xi=P_L\xi_L(x)\psi_L(x)+P_R\xi_R(x)\psi_R(x)\,,
\qquad O_\xi(x)\equiv\xi_L(x)O(x)\xi_R^\dag(x)\,,\\
&&\hspace{-0.5cm}g_2\frac{\tau^a}{2}W_{\xi,\mu}^a(x)\equiv
\xi_L(x)[g_2\frac{\tau^a}{2}W_{\mu}^a(x)-i\partial_\mu]\xi_L^\dag(x)\hspace{0.5cm}
g_1\frac{\tau^3}{2}B_{\xi,\mu}(x)\equiv
\xi_R(x)[g_1\frac{\tau^3}{2}B_{\mu}(x)-i\partial_\mu]\xi_R^\dag(x)\,.\nonumber
\end{eqnarray}
In (\ref{action-eff1}), the possible anomalies caused by this
special chiral rotation are canceled between the numerator and the
denominator. Thus Eq.\eqref{action-eff1} can be written as
\begin{subequations}
\label{action-eff2}
\begin{eqnarray}
S_{\rm eff}[U,W_\mu^a,B_\mu]=\int
d^4x(-\frac{1}{4}W_{\mu\nu}^aW^{a,\mu\nu}-\frac{1}{4}B_{\mu\nu}B^{\mu\nu})
+S_{\rm anom}[U,W_\mu^a,B_\mu]+S_{\rm
norm}[U,W_\mu^a,B_\mu],~~~\label{action-eff2}
\end{eqnarray}
where
\begin{eqnarray} S_{\rm norm}[U,W_\mu^a,B_\mu]&=&-i\log\int{\cal
D}G_\mu^\alpha{\cal D}\bar{\psi}_\xi{\cal
D}\psi_\xi\,\mathcal{F}[O_\xi]\delta(O_\xi-O_\xi^\dag)\exp\bigg\{i\int
d^4x\bigg[-\frac{1}{4}F_{\mu\nu}^\alpha F^{\alpha,\mu\nu}\nonumber\\
&&+\bar{\psi}_\xi\bigg(i\slashed{\partial}-g_{\rm TC}t^\alpha
\slashed{G}^\alpha-g_2\frac{\tau^a}{2}\slashed{W}_\xi^a
P_L-g_1\frac{\tau^3}{2}\slashed{B}_\xi
P_R\bigg)\psi_\xi\bigg]\bigg\}\;.\label{action-eff2-norm}
\end{eqnarray}
and
\begin{eqnarray}
iS_{\rm anom}[U,W_\mu^a,B_\mu]&=&
\mathrm{Tr}\log(i\slashed{\partial}-g_2\frac{\tau^a}{2}\slashed{W}^a
P_L-g_1\frac{\tau^3}{2}\slashed{B}P_R)\nonumber\\
&&-\mathrm{Tr}\log(i\slashed{\partial}-g_2\frac{\tau^a}{2}\slashed{W}_\xi^a
P_L-g_1\frac{\tau^3}{2}\slashed{B}_\xi
P_R)\,,\label{action-eff2-anom}
\end{eqnarray}
\end{subequations}
 It is worthwhile to show the transformations of the rotated
fields under $SU(2)_L\times U(1)_Y$ are $\psi_\xi(x)\rightarrow
h(x)\psi_\xi(x)$,~$O_\xi(x)\rightarrow h(x)O_\xi(x)h^\dag(x)$ with
$h(x)$ defined previously describes an invariant hidden local $U(1)$
symmetry. Thus, the chiral symmetry $SU(2)_L\times U(1)_Y$
covariance of the unrotated fields has been transferred totally to
the hidden symmetry $U(1)$ covariance of the rotated fields. We can
further find combination of electroweak gauge fields
$g_2\frac{\tau^a}{2}W_{\xi,\mu}^a(x)-g_1\frac{\tau^3}{2}B_{\xi,\mu}(x)
\rightarrow h(x)
[g_2\frac{\tau^a}{2}W_{\xi,\mu}^a(x)-g_1\frac{\tau^3}{2}B_{\xi,\mu}(x)]h^\dag(x)$
transforms covariantly, while alternative combination
$g_2\frac{\tau^a}{2}W_{\xi,\mu}^a(x)+g_1\frac{\tau^3}{2}B_{\xi,\mu}(x)
\rightarrow h(x)
[g_2\frac{\tau^a}{2}W_{\xi,\mu}^a(x)+g_1\frac{\tau^3}{2}B_{\xi,\mu}(x)
-2i\partial_\mu]h^\dag(x)$ transforms as the "gauge field" of the
hidden local $U(1)$ symmetry.

With technique used in Ref.\cite{Wang00}, the integration over
technigulon fields in Eq.\eqref{action-eff2-norm} can be formally
integrated out with help of full $n$-point Green's function of the
$G_\mu^\alpha$-field
$G_{\mu_1\ldots\mu_n}^{\alpha_1\ldots\alpha_n}$, thus
Eq.\eqref{action-eff2-norm} after integration becomes
\begin{eqnarray}
e^{iS_{\rm norm}[U,W_\mu^a,B_\mu]}&=&\int{\cal D}\bar{\psi}_\xi{\cal
D}\psi_\xi\,\mathcal{F}[O_\xi]\delta(O_\xi-O_\xi^\dag)\exp\bigg[i\int
d^4x\bar{\psi}_\xi\bigg(i\slashed{\partial}-g_2\frac{\tau^a}{2}\slashed{W}_\xi^a
P_L-g_1\frac{\tau^3}{2}\slashed{B}_\xi
P_R\bigg)\psi_\xi\nonumber\\
&&+\sum_{n=2}^\infty\int d^4x_1\ldots d^4x_n\frac{(-ig_{\rm
TC})^n}{n!}G_{\mu_1\ldots\mu_n}^{\alpha_1\ldots\alpha_n}(x_1,\ldots,x_n)
J_{\xi,\alpha_1}^{\mu_1}(x_1)\ldots
J_{\xi,\alpha_n}^{\mu_n}(x_n)\bigg]\;.\label{action-eff2-norm2}
\end{eqnarray}
where effective sources $J_{\xi,\alpha}^\mu(x)$ are identified as
$J_{\xi,\alpha}^\mu(x)\equiv \bar{\psi}_\xi(x)
t^\alpha\gamma^\mu\psi_\xi(x)$.

\subsubsection{Schwinger-Dyson Equation for Techniquark Propagator}

To show that the technicolor interaction indeed induces the
condensate $\langle\bar{\psi}\psi\rangle\neq0$ which triggers the
electroweak symmetry breaking, we investigate the behavior of the
techniquark propagator $S^{\sigma\rho}(x,x^\prime)\equiv
\langle\psi_\xi^\sigma(x)\bar{\psi}_\xi^\rho(x^\prime)\rangle$ in
the following. {\it Neglecting the factor
$\mathcal{F}[O_\xi]\delta(O_\xi-O_\xi^\dag)$ in
Eq.\eqref{action-eff2-norm2}}, the total functional derivative of
the integrand with respect to $\bar{\psi}_\xi^\sigma(x)$ is zero,
(here and henceforth the suffixes $\sigma$ and $\rho$  are short
notations for Lorentz spinor, techniflavor and technicolor indices,)
{\it i.e.},
\begin{eqnarray}
0&=&\int{\cal D}\bar{\psi}_\xi{\cal
D}\psi_\xi\frac{\delta}{\delta\bar{\psi}_\xi^\sigma(x)}\exp\bigg[\int
d^4x(\bar{\psi}_\xi I+\bar{I}\psi_\xi)+i\int
d^4x\bar{\psi}_\xi\bigg(i\slashed{\partial}-g_2\frac{\tau^a}{2}\slashed{W}_\xi^a
P_L-g_1\frac{\tau^3}{2}\slashed{B}_\xi P_R\bigg)\psi_\xi\nonumber\\
&&+\sum_{n=2}^\infty\int d^4x_1\ldots d^4x_n\frac{(-ig_{\rm
TC})^n}{n!}G_{\mu_1\ldots\mu_n}^{\alpha_1\ldots\alpha_n}(x_1,\ldots,x_n)
J_{\xi,\alpha_1}^{\mu_1}(x_1)\ldots
J_{\xi,\alpha_n}^{\mu_n}(x_n)\bigg]\;,
\end{eqnarray}
where $I(x)$ and $\bar{I}(x)$ are the external sources for,
respectively, $\bar{\psi}_\xi(x)$ and $\psi_\xi(x)$; and which leads
to
\begin{eqnarray}
0&=&\bigg\langle\!\!\bigg\langle I_\sigma(x)
+i\bigg[i\slashed{\partial}_x-g_2\frac{\tau^a}{2}\slashed{W}_\xi^a(x)
P_L-g_1\frac{\tau^3}{2}\slashed{B}_\xi(x) P_R\bigg]_{\sigma\tau}\psi_\xi^\tau(x)\label{eq-to-SDE1}\\
&&+\sum_{n=2}^\infty\int d^4x_2\ldots d^4x_n\frac{(-ig_{\rm
TC})^n}{(n-1)!}G_{\mu_1\ldots\mu_n}^{\alpha_1\ldots\alpha_n}(x,x_2,\ldots,x_n)
(t^{\alpha_1}\gamma^{\mu_1})_{\sigma\tau}\psi_\xi^\tau(x)
J_{\xi,\alpha_2}^{\mu_2}(x_2)\ldots
J_{\xi,\alpha_n}^{\mu_n}(x_n)\bigg\rangle\!\!\bigg\rangle_I\;,\nonumber
\end{eqnarray}
where we have defined the notation
$\langle\langle\,\cdots\rangle\rangle_I$ in this section by
\begin{eqnarray}
\big\langle\!\big\langle\mathcal{O}(x)\big\rangle\!\big\rangle_I&\equiv&\int{\cal
D}\bar{\psi}_\xi{\cal D}\psi_\xi\,\mathcal{O}(x)\exp\bigg[\int
d^4x(\bar{\psi}_\xi I+\bar{I}\psi_\xi)\nonumber\\
&&+i\int
d^4x\bar{\psi}_\xi\bigg(i\slashed{\partial}-g_2\frac{\tau^a}{2}\slashed{W}_\xi^a
P_L-g_1\frac{\tau^3}{2}\slashed{B}_\xi P_R\bigg)\psi_\xi\nonumber\\
&&+\sum_{n=2}^\infty\int d^4x_1\ldots d^4x_n\frac{(-ig_{\rm
TC})^n}{n!}G_{\mu_1\ldots\mu_n}^{\alpha_1\ldots\alpha_n}(x_1,\ldots,x_n)
J_{\xi,\alpha_1}^{\mu_1}(x_1)\ldots
J_{\xi,\alpha_n}^{\mu_n}(x_n)\bigg]\;.
\end{eqnarray}
Taking functional derivative of Eq.\eqref{eq-to-SDE1} with respect
to $I_\rho(y)$, and subsequently setting $I=\bar{I}=0$, we obtain
\begin{eqnarray}
0&=&\delta_{\sigma\rho}\delta(x-y)
+i\bigg[i\slashed{\partial}_x-g_2\frac{\tau^a}{2}\slashed{W}_\xi^a(x)
P_L-g_1\frac{\tau^3}{2}\slashed{B}_\xi(x)
P_R\bigg]_{\sigma\tau}\langle\psi_\xi^\tau(x)
\bar{\psi}_\xi^\rho(y)\rangle\nonumber\\
&&-\sum_{n=2}^\infty\int d^4x_2\ldots d^4x_n\frac{(-ig_{\rm
TC})^n}{(n-1)!}G_{\mu_1\ldots\mu_n}^{\alpha_1\ldots\alpha_n}(x,x_2,\ldots,x_n)\nonumber\\
&&\times\big\langle\bar{\psi}_\xi^\rho(y)
(t^{\alpha_1}\gamma^{\mu_1})_{\sigma\tau}\psi_\xi^\tau(x)
J_{\xi,\alpha_2}^{\mu_2}(x_2)\ldots
J_{\xi,\alpha_n}^{\mu_n}(x_n)\big\rangle\;,\label{eq-to-SDE2}
\end{eqnarray}
where we have defined vacuum expectation value (VEV)
$\langle\,\cdots\rangle$ by
$\big\langle\mathcal{O}(x)\big\rangle\equiv
\big\langle\!\big\langle\mathcal{O}(x)\big\rangle\big\rangle_I/
\langle\langle\,1\,\rangle\rangle_I|_{I=\bar{I}=0}$. If we {\it
neglect higher-point Green's functions},and further {\it taking
factorization approximation}, {\it i.e.},
$\big\langle\bar{\psi}_\xi^\rho(y)\psi_\xi^\tau(x)
\bar{\psi}_\xi^\gamma(x_2)\psi_\xi^\delta(x_2)\big\rangle\approx
\big\langle\bar{\psi}_\xi^\rho(y)\psi_\xi^\tau(x)\big\rangle
\big\langle\bar{\psi}_\xi^\gamma(x_2)\psi_\xi^\delta(x_2)\big\rangle
-\big\langle\bar{\psi}_\xi^\rho(y)\psi_\xi^\delta(x_2)\big\rangle
\big\langle\bar{\psi}_\xi^\gamma(x_2)\psi_\xi^\tau(x)\big\rangle$,
we obtain
\begin{eqnarray}
0&=&\delta_{\sigma\rho}\delta(x-y)
+i\bigg[i\slashed{\partial}_x-g_2\frac{\tau^a}{2}\slashed{W}_\xi^a(x)
P_L-g_1\frac{\tau^3}{2}\slashed{B}_\xi(x)
P_R\bigg]_{\sigma\tau}\langle\psi_\xi^\tau(x)
\bar{\psi}_\xi^\rho(y)\rangle\nonumber\\
&&-g_{\rm TC}^2\int
d^4x_2\,G_{\mu_1\mu_2}^{\alpha_1\alpha_2}(x,x_2)(t^{\alpha_1}\gamma^{\mu_1})_{\sigma\tau}
(t^{\alpha_2}\gamma^{\mu_2})_{\gamma\delta}
\big\langle\bar{\psi}_\xi^\rho(y)\psi_\xi^\delta(x_2)\big\rangle
\big\langle\bar{\psi}_\xi^\gamma(x_2)\psi_\xi^\tau(x)\big\rangle\;,\label{eq-to-SDE4}
\end{eqnarray}
where we have used
$\big\langle\bar{\psi}_\xi(x_2)t^{\alpha_2}\gamma^{\mu_2}\psi_\xi(x_2)\big\rangle=0$,
which comes from the Lorentz and gauge invariance of vacuum. We
denote the technifermion propagator
$S^{\sigma\rho}(x,x^\prime)\equiv
\langle\psi_\xi^\sigma(x)\bar{\psi}_\xi^\rho(x^\prime)\rangle$,
multiplying inverse of technifermion propagator in both sides of
Eq.\eqref{eq-to-SDE4}, it then is written as
 the Schwinger-Dyson equation (SDE) for techniquark
propagator,
\begin{eqnarray}
0&=&S^{-1}_{\sigma\rho}(x,y)
+i\bigg[i\slashed{\partial}_x-g_2\frac{\tau^a}{2}\slashed{W}_\xi^a(x)
P_L-g_1\frac{\tau^3}{2}\slashed{B}_\xi(x)
P_R\bigg]_{\sigma\rho}\delta(x-y)\nonumber\\
&&-g_{\rm TC}^2G_{\mu_1\mu_2}^{\alpha_1\alpha_2}(x,y)
\bigg[t^{\alpha_1}\gamma^{\mu_1}S(x,y)
t^{\alpha_2}\gamma^{\mu_2}\bigg]_{\sigma\rho}\;.\label{eq-SDE1}
\end{eqnarray}
By defining techniquark self energy $i\Sigma$ as
\begin{eqnarray}
i\Sigma_{\sigma\rho}(x,y)\equiv S^{-1}_{\sigma\rho}(x,y)
+i\bigg[i\slashed{\partial}_x-g_2\frac{\tau^a}{2}\slashed{W}_\xi^a(x)
P_L-g_1\frac{\tau^3}{2}\slashed{B}_\xi(x)
P_R\bigg]_{\sigma\rho}\delta(x-y)\;,
\end{eqnarray}
the SDE \eqref{eq-SDE1} can also be written as
\begin{eqnarray}
i\Sigma_{\sigma\rho}(x,y)=g_{\rm
TC}^2G_{\mu_1\mu_2}^{\alpha_1\alpha_2}(x,y)
\bigg[t^{\alpha_1}\gamma^{\mu_1}S(x,y)
t^{\alpha_2}\gamma^{\mu_2}\bigg]_{\sigma\rho}\;.\label{eq-SDE2}
\end{eqnarray}
Moreover, from the fact that technigluon propagator is diagonal in
the adjoint representation space of technicolor group, {\it i.e.},
$G_{\mu\nu}^{\alpha\beta}(x,y)=\delta^{\alpha\beta}G_{\mu\nu}(x,y)$,
and techniquark propagator is diagonal in the techniquark
representation space of technicolor group, and also $(t^\alpha
t^\alpha)_{ab}=C_2(N)\delta_{ab}$ for the fundamental representation
of $SU(N)$, Eq.\eqref{eq-SDE2} is diagonal in technicolor indices
$a,b$ and diagonal part satisfy
\begin{eqnarray} i\Sigma_{\eta\zeta}^{ij}(x,y)=C_2(N)g_{\rm
TC}^2G_{\mu_1\mu_2}(x,y) [\gamma^{\mu_1}S(x,y)
\gamma^{\mu_2}]_{\eta\zeta}^{ij}\;,\label{eq-SDE3}
\end{eqnarray}
where $\{i,~j\}$, and $\{\eta,~\zeta\}$ are, respectively,
techniflavor and Lorentz spinor indices; and the Casimir operator
$C_2(N)=(N^2-1)/(2N)$.

\vspace*{0.5cm}\noindent$\clubsuit$~~$B_{\xi,\mu}=W_{\xi,\mu}^a=0$
Case: the Gap Equation

\vspace*{0.3cm}The technigluon propagator in Landau gauge is
$G_{\mu\nu}^{\alpha\beta}(x,y)=\delta^{\alpha\beta}\int\frac{d^4p}{(2\pi)^4}e^{-ip(x-y)}G_{\mu\nu}(p^2)$
 with $G_{\mu\nu}(p^2)=\frac{i}{-p^2[1+\Pi(-p^2)]}(g_{\mu\nu}-p_\mu
p_\nu/p^2)$. And the techniquark self energy and propagator are
respectively
\begin{eqnarray}
\Sigma_{\eta\zeta}^{ij}(x,y)=\int\frac{d^4p}{(2\pi)^4}e^{-ip(x-y)}\,\Sigma_{\eta\zeta}^{ij}(-p^2)\hspace{1cm}
S_{\eta\zeta}^{ij}(x,y)=\int\frac{d^4p}{(2\pi)^4}e^{-ip(x-y)}S_{\eta\zeta}^{ij}(p)\;,
\end{eqnarray}
with
$S_{\eta\zeta}^{ij}(p)=i[1/(\slashed{p}-\Sigma(-p^2)]_{\eta\zeta}^{ij}$.
Substituting above results into the SDE \eqref{eq-SDE3}, we have
\begin{eqnarray}
\Sigma_{\eta\zeta}^{ij}(-p^2)=\int\frac{d^4q}{(2\pi)^4}\frac{-C_2(N)g_{\rm
TC}^2}{(p-q)^2[1+\Pi(-(p-q)^2)]} \bigg[g_{\mu\nu}-\frac{(p-q)_\mu
(p-q)_\nu}{(p-q)^2}\bigg]\bigg[\gamma^\mu\frac{i}{\slashed{q}-\Sigma(-q^2)}\gamma^\nu\bigg]_{\eta\zeta}^{ij}
~~~\label{eq-SDE4}
\end{eqnarray}
from which we can see that the solution of the techniquark self
energy must be diagonal in techniflavor space, since the integration
kernel is independent of  techniflavor indices, {\it i.e.},
$\Sigma_{\eta\zeta}^{ij}(-p^2)=\delta^{ij}\Sigma_{\eta\zeta}(-p^2)$.
With the assumption that the techniquark self energy is diagonal and
equal in Lorentz spinor space, leads to the following two equations
\begin{subequations}
\begin{eqnarray}
&&i\,\Sigma(-p^2)=3C_2(N)\int\frac{d^4q}{4\pi^3}
\frac{\alpha_{\mathrm{TC}}[-(p-q)^2]}{(p-q)^2}
\frac{\Sigma(-q^2)}{q^2-\Sigma^2(-q^2)}\;,\label{eq-SDE6a}\\
&&0=\int d^4q\,\frac{\alpha_{\mathrm{TC}}[-(p-q)^2]}{(p-q)^2}
\bigg[g_{\mu\nu}-\frac{(p-q)_\mu (p-q)_\nu}{(p-q)^2}\bigg]
\gamma^\mu\frac{\slashed{q}}{q^2-\Sigma^2(-q^2)}\gamma^\nu\;.\label{eq-SDE6b}
\end{eqnarray}
\end{subequations}
In which we have labeled the integration kernel with running
coupling constant $\alpha_\mathrm{TC}(-p^2)\equiv
g_\mathrm{TC}^2(-p^2)/(4\pi) =g_\mathrm{TC}^2/(4\pi[1+\Pi(-p^2)])$.
Eq.\eqref{eq-SDE6b} is automatically satisfied when taken {\it the
approximation},
$\alpha_\mathrm{TC}[(p_E-q_E)^2]=\alpha_\mathrm{TC}(p_E^2)\theta(p_E^2-q_E^2)
+\alpha_\mathrm{TC}(q_E^2)\theta(q_E^2-p_E^2)$. The other equation
\eqref{eq-SDE6a} can be written in Euclidean space as
\begin{eqnarray}
\Sigma(p_E^2)=3C_2(N)\int\frac{d^4q_E}{4\pi^3}
\frac{\alpha_\mathrm{TC}[(p_E-q_E)^2]}{(p_E-q_E)^2}
\frac{\Sigma_(q_E^2)}{q_E^2+\Sigma^2(q_E^2)}\;,\label{eq-SDE8a}
\end{eqnarray}
If there is nonzero solution for above equation, we will obtain
nonzero techniquark condensate
$\langle\bar{\psi}_\xi^k\psi_\xi^j\rangle$ with $k$ and $j$
techniflavor indices,
\begin{eqnarray}
\langle\bar{\psi}_\xi^k(x)\psi_\xi^j(x)\rangle
&=&-\mathrm{tr}_{lc}[S^{jk}(x,x)]
=-4N\delta^{jk}\int\frac{d^4p_E}{(2\pi)^4}\frac{\Sigma(p_E^2)}{p_E^2
+\Sigma^2(p_E^2)}\;,
\end{eqnarray}
where ${\rm tr}_{lc}$ is the trace with respect to Lorentz,
technicolor indices, and where we have used that the techniquark
self energy, the solution of the SDE, must be diagonal in
techniflavor space. Thus, nonzero techniquark self energy can give a
nontrivial diagonal condensate $\langle\bar{\psi}\psi\rangle\neq0$,
which spontaneously breaks $SU(2)_L\times U(1)_Y$ to
$U(1)_{\mathrm{em}}$.

\vspace*{0.5cm}\noindent$\clubsuit$~~$B_{\xi,\mu}\neq0$ and
$W_{\xi,\mu}^a\neq0$ Case: the Lowest-order Approximation

\vspace*{0.3cm}In the following we consider the effects of the
nonzero electroweak gauge fields $B_{\xi,\mu}$ and $W_{\xi,\mu}^a$.
The SDE \eqref{eq-SDE3} is explicitly
\begin{eqnarray}
\Sigma(x,y)&=&C_2(N)g_{\mathrm{TC}}^2G_{\mu\nu}(x,y)\nonumber\\
&&\times\gamma^\mu
\bigg[\bigg(i\slashed{\partial}_x-g_2\frac{\tau^a}{2}\slashed{W}_\xi^a(x)
P_L-g_1\frac{\tau^3}{2}\slashed{B}_\xi(x)
P_R\bigg)\delta(x-y)-\Sigma(x,y)\bigg]^{-1}\gamma^\nu\;,\label{eq-SDE-sources}
\end{eqnarray}
where the techniflavor and Lorentz spinor indices of the techniquark
self energy are implicitly contained. In this case, the self energy
can no longer be written as the function on the derivatives with
respect to spacetime, {\it i.e.},
$\Sigma(x,y)\neq\Sigma(\partial_x^2)\delta(x-y)$.

Suppose the function $\Sigma(-p^2)$ is a solution of the SDE in the
case $B_{\xi,\mu}=W_{\xi,\mu}^a=0$, that is, it satisfies the
equation
\begin{eqnarray}
\Sigma(-p^2)&=&C_2(N)g_{\mathrm{TC}}^2\int\frac{d^4q}{(2\pi)^4}
G_{\mu\nu}(q^2)\gamma^\mu
\frac{1}{\slashed{q}+\slashed{p}-\Sigma[-(q+p)^2]}\gamma^\nu\;,\label{eq-sol-SDE1}
\end{eqnarray}
where in the second equality the legality of the integration measure
translation comes from the logarithmical divergence of the fermion
self energy. Replacing the variable $p$ by $p+\Delta$ in
Eq.\eqref{eq-sol-SDE1} and subsequently integrating over $p$ with
the weight $e^{-ip(x-y)}$, we obtain, as long as $\Delta$ is
commutative with $\partial_x$ and Dirac matrices,
Eq.\eqref{eq-sol-SDE1} imply
\begin{eqnarray}
\Sigma[(\partial_x-i\Delta)^2]\delta(x-y)
=C_2(N)g_{\mathrm{TC}}^2G_{\mu\nu}(x,y)\gamma^\mu\frac{1}{i\slashed{\partial}_x
+\slashed{\Delta}-\Sigma[-(i\partial_x+\Delta)^2]}\delta(x-y)\gamma^\nu\;.~~~\label{eq-sol-SDE2}
\end{eqnarray}
Even if $\Delta$ is noncommutative with $\partial_x$ and Dirac
matrices, the above equation holds as the lowest order
approximation, for the commutator $[\slashed{\partial}, \Delta]$ is
higher order of momentum than $\Delta$ itself. Now if we take
$\Delta$ to be
$-g_2\frac{\tau^a}{2}W_{\xi}^aP_L-g_1\frac{\tau^3}{2}B_{\xi}P_R$,
{\it ignoring its noncommutative property with $\partial_x$ and
Dirac matrices}, Eq.\eqref{eq-sol-SDE2} is just the SDE
\eqref{eq-SDE-sources} in the case $B_{\xi,\mu}\neq0$ and
$W_{\xi,\mu}^a\neq0$. Thus,
$\Sigma[(\partial_\mu^x+ig_2\frac{\tau^a}{2}W_{\xi,\mu}^aP_L
+ig_1\frac{\tau^3}{2}B_{\xi,\mu}P_R)^2]\delta(x-y)$, which is the
hidden symmetry $U(1)$ covariant, can be regarded as the
lowest-order solution of Eq.\eqref{eq-SDE-sources}. To further
simplify the calculations and still keep this hidden-symmetry
covariance of the self energy, we can {\it reduce the covariant
derivative inside the self energy}
$\nabla_\mu\equiv\partial_\mu+ig_2\frac{\tau^a}{2}W_{\xi,\mu}^aP_L
+ig_1\frac{\tau^3}{2}B_{\xi,\mu}P_R$ {\it to its minimal-coupling
form}
\begin{eqnarray}
\overline{\nabla}_\mu\equiv\partial_\mu
+\frac{i}{2}[g_2\frac{\tau^a}{2}W_{\xi,\mu}^a(x)+g_1\frac{\tau^3}{2}B_{\xi,\mu}(x)]\;.
\end{eqnarray}
In which as we mentioned before
$[g_2\frac{\tau^a}{2}W_{\xi,\mu}^a(x)+g_1\frac{\tau^3}{2}B_{\xi,\mu}(x)]/2$
transforms as a gauge field under hidden $U(1)$ symmetry
transformations. Thus, if the function
$\Sigma(\partial_x^2)\delta(x-y)$ is the self-energy solution of the
SDE in the case $B_{\xi,\mu}=W_{\xi,\mu}^a=0$, we can replace its
argument $\partial_x$ by the minimal-coupling covariant derivative
$\overline{\nabla}_x$, {\it i.e.},
$\Sigma(\overline{\nabla}_x^2)\delta(x-y)$, as an approximate
solution of the SDE in the case $B_{\xi,\mu}\neq0$ and
$W_{\xi,\mu}^a\neq0$.

\subsubsection{Effective Action}

The exponential terms on the right-hand side of
Eq.\eqref{action-eff2-norm2} can be written explicitly as
\begin{eqnarray}
&&\hspace{-0.5cm}\sum_{n=2}^\infty\int d^4x_1\ldots
d^4x_n\frac{(-ig_{\rm
TC})^n}{n!}G_{\mu_1\ldots\mu_n}^{\alpha_1\ldots\alpha_n}(x_1,\ldots,x_n)
J_{\xi,\alpha_1}^{\mu_1}(x_1)\ldots J_{\xi,\alpha_n}^{\mu_n}(x_n)\nonumber\\
&\approx&\int d^4xd^4x^\prime
\bar{\psi}_\xi^\sigma(x)\Pi_{\sigma\rho}(x,x^\prime)\psi_\xi^\rho(x^\prime)\;,
\end{eqnarray}
where we have taken the approximation of {\it replacing the
summation over $2n$-fermion interactions with parts of them by their
vacuum expectation values}, that is,
\begin{eqnarray}
\Pi_{\sigma\rho}(x,x^\prime)&=&\sum_{n=2}^\infty\Pi_{\sigma\rho}^{(n)}(x,x^\prime)\;,\\
\Pi_{\sigma\rho}^{(n)}(x,x^\prime)&=&n\times\int d^4x_2\ldots
d^4x_{n-1}\frac{(-ig_{\rm
TC})^n}{n!}G_{\mu_1\ldots\mu_n}^{\alpha_1\ldots\alpha_n}(x,x_2\ldots,x_{n-1},x^\prime)
\bigg\langle(t^{\alpha_1}\gamma^{\mu_1})_{\sigma\sigma_1}\psi_\xi^{\sigma_1}(x)\nonumber\\
&&\times\bar{\psi}_\xi(x_2)
t^{\alpha_2}\gamma^{\mu_2}\psi_\xi(x_2)\ldots
\bar{\psi}_\xi(x_{n-1})
t^{\alpha_{n-1}}\gamma^{\mu_{n-1}}\psi_\xi(x_{n-1})\bar{\psi}_\xi^{\rho_n}(x^\prime)
(t^{\alpha_n}\gamma^{\mu_n})_{\rho_n\rho}\bigg\rangle
\end{eqnarray}
where the factor $n$ comes from $n$ different choices of unaveraged
$\bar{\psi}_\xi\psi_\xi$, and the lowest term of which is
\begin{eqnarray}
\Pi_{\sigma\rho}^{(2)}(x,x^\prime)&=&-g_{\mathrm{TC}}^2G_{\mu_1\mu_2}^{\alpha_1\alpha_2}(x,x^\prime)
\bigg[t^{\alpha_1}\gamma^{\mu_1}S(x,y)
t^{\alpha_2}\gamma^{\mu_2}\bigg]_{\sigma\rho}\;.\label{eq-Pi-2}
\end{eqnarray}
Comparing Eq.\eqref{eq-Pi-2} with Eq.\eqref{eq-SDE2}, we have
\begin{eqnarray}
i\Pi_{\sigma\rho}^{(2)}(x,x^\prime)=\Sigma_{\sigma\rho}(x,x^\prime)
\approx\Sigma_{\sigma\rho}(\overline{\nabla}_x^2)\delta(x-y)\;.
\label{eq-Pi-Sigma}
\end{eqnarray}

{\it By neglecting the factor
$\mathcal{F}[O_\xi]\delta(O_\xi-O_\xi^\dag)$ in
Eq.\eqref{action-eff2-norm2}}, we have
\begin{eqnarray}
e^{iS_{\rm norm}[U,W_\mu^a,B_\mu]}&\approx&\int{\cal
D}\bar{\psi}_\xi{\cal D}\psi_\xi\exp\bigg[i\int
d^4x\bar{\psi}_\xi\bigg(i\slashed{\partial}-g_2\frac{\tau^a}{2}\slashed{W}_\xi^a
P_L-g_1\frac{\tau^3}{2}\slashed{B}_\xi P_R\bigg)\psi_\xi\nonumber\\
&&+\int d^4xd^4x^\prime
\bar{\psi}_\xi^\sigma(x)\Pi_{\sigma\rho}(x,x^\prime)\psi_\xi^\rho(x^\prime)\bigg]\nonumber\\
&\approx&\mathrm{Det}[i\slashed{\partial}-g_2\frac{\tau^a}{2}\slashed{W}_\xi^a
P_L-g_1\frac{\tau^3}{2}\slashed{B}_\xi
P_R-\Sigma(\overline{\nabla}^2)]
\end{eqnarray}
where in the second equality we have taken further approximation of
{\it keeping only the lowest order, {\it i.e.}
$\Pi_{\sigma\rho}^{(2)}(x,x^\prime)$, of
$\Pi_{\sigma\rho}(x,x^\prime)$}. With these three approximations, we
have
\begin{eqnarray}
S_{\rm
norm}[U,W_\mu^a,B_\mu]=-i\mathrm{Tr}\log[i\slashed{\partial}-g_2\frac{\tau^a}{2}\slashed{W}_\xi^a
P_L-g_1\frac{\tau^3}{2}\slashed{B}_\xi
P_R-\Sigma(\overline{\nabla}^2)]\;,\label{normdef}
\end{eqnarray}

As done in \cite{Wang02}, we can parameterize the normal part of the
effective action as follows
\begin{eqnarray}
&&\hspace{-0.5cm}S_{\rm
norm}[U,W_\mu^a,B_\mu]=-i\mathrm{Tr}\log[i\slashed{\partial}+\slashed{v}
+\slashed{a}\gamma_5-\Sigma(\overline{\nabla}^2)]\nonumber\\
&=&\int d^4x\,\mathrm{tr}_f\bigg[(F_0^{\rm
1D})^2a^2-\mathcal{K}_1^{\rm 1D}(d_\mu a^\mu)^2-\mathcal{K}_2^{\rm
1D}(d_\mu a_\nu-d_\nu a_\mu)^2+\mathcal{K}_3^{\rm
1D}(a^2)^2+\mathcal{K}_4^{\rm
1D}(a_\mu a_\nu)^2\nonumber\\
&&-\mathcal{K}_{13}^{\rm
1D}V_{\mu\nu}V^{\mu\nu}+i\mathcal{K}_{14}^{\rm 1D}a_\mu a_\nu
V^{\mu\nu}\bigg]+\mathcal{O}(p^6)\;,\label{action-GND-1doublet-norm}
\end{eqnarray}
where the fields $v_\mu$, $a_\mu$ are identified with
$v_\mu\equiv-\frac{1}{2}(g_2\frac{\tau^a}{2}W_{\xi,\mu}^a
+g_1\frac{\tau^3}{2}B_{\xi,\mu})$ and
$a_\mu\equiv\frac{1}{2}(g_2\frac{\tau^a}{2}W_{\xi,\mu}^a
-g_1\frac{\tau^3}{2}B_{\xi,\mu})$ and $d_\mu a_\nu\equiv\partial_\mu
a_\nu-i[v_\mu, a_\nu]$,~$V_{\mu\nu}\equiv i[\partial_\mu-iv_\mu,
\partial_\nu-iv_\nu]$. $\mathcal{K}_i^{\rm 1D}$ coefficients with superscript
1D to denote present one doublet model are functions of techniquark
self energy $\Sigma(p^2)$ and detail expressions are already written
down in (36) of Ref.\cite{Wang02} with the replacement of
$N_c\rightarrow N$.

For anomaly part, compare (\ref{action-eff2-anom}) and
(\ref{normdef}), we find its $U$ field dependent part can be
produced from normal part by vanishing techniquark self energy
$\Sigma$, i.e.
\begin{eqnarray}
iS_{\rm anom}[U,W_\mu^a,B_\mu]=
\mathrm{Tr}\log(i\slashed{\partial}-g_2\frac{\tau^a}{2}\slashed{W}^a
P_L-g_1\frac{\tau^3}{2}\slashed{B}P_R) -iS_{\rm
norm}[U,W_\mu^a,B_\mu]|_{\Sigma=0}
\end{eqnarray}
Notice that $U$ field independent part of pure gauge field part is
irrelevant to EWCL. Combine with (\ref{action-GND-1doublet-norm}),
above relation imply
\begin{eqnarray}
iS_{\rm
anom}[U,W_\mu^a,B_\mu]&=&\mathrm{Tr}\log(i\slashed{\partial}-g_2\frac{\tau^a}{2}\slashed{W}^a
P_L-g_1\frac{\tau^3}{2}\slashed{B}P_R)+i\int
d^4x\,\mathrm{tr}_f\bigg[-\mathcal{K}_1^{\rm 1D,(anom)}(d_\mu
a^\mu)^2\nonumber\\
&&-\mathcal{K}_2^{\rm 1D,(anom)}(d_\mu a_\nu-d_\nu
a_\mu)^2+\mathcal{K}_3^{\rm 1D,(anom)}(a^2)^2+\mathcal{K}_4^{\rm
1D,(anom)}(a_\mu a_\nu)^2\nonumber\\
&&-\mathcal{K}_{13}^{\rm
1D,(anom)}V_{\mu\nu}V^{\mu\nu}+i\mathcal{K}_{14}^{\rm
1D,(anom)}a_\mu a_\nu V^{\mu\nu}\bigg]+\mathcal{O}(p^6)\;,
\end{eqnarray}
with
\begin{eqnarray}
{\cal K}_i^{\rm 1D,(anom)}=-{\cal K}_i^{\rm
1D}|_{\Sigma=0}\hspace{2cm}i=1,2,3,4,13,14
\end{eqnarray}
where we have used result that $F_0^{\rm 1D}|_{\Sigma=0}=0$. Combine
normal and anomaly part contribution together, with help of
(\ref{action-eff2}), we finally find
\begin{eqnarray}
S_{\rm eff}[U,W_\mu^a,B_\mu]&=&-\frac{1}{4}\int
d^4x(W_{\mu\nu}^aW^{a,\mu\nu}+B_{\mu\nu}B^{\mu\nu})+
\mathrm{Tr}\log(i\slashed{\partial}-g_2\frac{\tau^a}{2}\slashed{W}^a
P_L-g_1\frac{\tau^3}{2}\slashed{B}P_R)\nonumber\\
&&+i\int d^4x\,\mathrm{tr}_f\bigg[(F_0^{\rm
1D})^2a^2-\mathcal{K}_1^{{\rm 1D},\Sigma\neq0}(d_\mu
a^\mu)^2-\mathcal{K}_2^{{\rm 1D},\Sigma\neq0}(d_\mu a_\nu-d_\nu
a_\mu)^2+\mathcal{K}_3^{{\rm
1D},\Sigma\neq0}(a^2)^2\nonumber\\
&&+\mathcal{K}_4^{{\rm 1D},\Sigma\neq0}(a_\mu
a_\nu)^2-\mathcal{K}_{13}^{{\rm
1D},\Sigma\neq0}V_{\mu\nu}V^{\mu\nu}+i\mathcal{K}_{14}^{{\rm
1D},\Sigma\neq0}a_\mu a_\nu
V^{\mu\nu}\bigg]+\mathcal{O}(p^6)\;,\label{action-GND-1doublet-full}
\end{eqnarray}
with $\mathcal{K}_i^{{\rm 1D},\Sigma\neq0}$ be $\Sigma$ dependent
part of $\mathcal{K}_i$
\begin{eqnarray}
\mathcal{K}_i^{{\rm 1D},\Sigma\neq0}\equiv\mathcal{K}_i^{\rm
1D}-\mathcal{K}_i^{\rm 1D}|_{\Sigma=0}\hspace{2cm}i=1,2,3,4,13,14
\end{eqnarray}
After some algebras, terms in Eq.\eqref{action-GND-1doublet-full}
can be reexpressed in terms of $X_\mu$ and $\overline{W}_{\mu\nu}$
which are just standard EWCL given in Ref.\cite{EWCL} with
coefficients
\begin{eqnarray}
&&\hspace{-0.5cm}f^2=(F_0^{\rm 1D})^2\;,\qquad \beta_1=0\;,\qquad
\alpha_1=\frac{\mathcal{K}_{2}^{{\rm
1D},\Sigma\neq0}-\mathcal{K}_{13}^{{\rm
1D},\Sigma\neq0}}{2}\;,\qquad
\alpha_2=\alpha_3=-\frac{\mathcal{K}_{13}^{{\rm
1D},\Sigma\neq0}}{4}+\frac{\mathcal{K}_{14}^{{\rm
1D},\Sigma\neq0}}{16}\;,\nonumber\\
&&\hspace{-0.5cm}\alpha_4=\frac{\mathcal{K}_{4}^{{\rm
1D},\Sigma\neq0}+2\mathcal{K}_{13}^{{\rm
1D},\Sigma\neq0}-\mathcal{K}_{14}^{{\rm 1D},\Sigma\neq0}}{16}\;,
\qquad \alpha_5=\frac{\mathcal{K}_{3}^{{\rm
1D},\Sigma\neq0}-\mathcal{K}_{4}^{{\rm
1D},\Sigma\neq0}-4\mathcal{K}_{13}^{{\rm 1D},\Sigma\neq0}
+2\mathcal{K}_{14}^{\Sigma\neq0}}{32}\;.\label{Dprescript}
\end{eqnarray}
With formulae of $\mathcal{K}_i$ coefficients given in
Ref.\cite{Wang02}, we can substitute the solution of SD equation
 (\ref{eq-SDE8a}) into them and then obtain numerical results for
 those nonzero $\alpha_i$ coefficients. In TABLE I, we list down the numerical calculation results for
different kind of dynamics.

\begin{table}[h]
{\small{\bf TABLE I}. The obtained nonzero values of the $O(p^4)$
coefficients $\alpha_1,\alpha_2=\alpha_3,\alpha_4,\alpha_5$  for one
doublet technicolor model with the conventional strong interaction
QCD theory values given in Ref.\cite{Wang02} for model A and
experimental values for comparison. $\Lambda_{\rm TC}$ and
$\Lambda_{\rm QCD}$ are in TeV, they are determined by $f=250$GeV
and $f_\pi=93$MeV respectively. The coefficients are in units of
$10^{-3}$. QCD values are taken by using relation
(\ref{Gprescript}). }

\vspace*{0.3cm}\begin{tabular}{c c|c c c c}\hline\hline
 $N$&$\Lambda_{\rm TC}$&$\alpha_1$ &$\alpha_2=\alpha_3$ &$\alpha_4$
 &$\alpha_5$\\ \hline
3&1.34&-6.90&-2.43&2.02&-2.69\\
4&1.15&-9.26&-3.28&2.87&-3.69\\
5&1.03&-11.6&-4.11&3.60&-4.62\\
6&0.94&-13.9&-4.93&4.32&-5.54\\ \hline QCD
Theor&~~$\Lambda_{\rm QCD}$=0.484$*10^{-3}$&-7.06&-2.54&2.20&-2.81\\
QCD Expt&&$-6.0\pm0.7$&$-2.7\pm0.4$&$1.7\pm0.7$&$-1.3\pm1.5$\\
 \hline\hline
\end{tabular}
\end{table}
To obtain above numerical result, we have solved Schwinger-Dyson
equation (\ref{eq-SDE8a}) with following  running coupling which was
used as model A in Ref.\cite{Wang02}
\begin{eqnarray}
\alpha_{\rm
TC}(p^2)=\left\{\begin{array}{lcl}7\frac{12\pi}{(11N-2N_f)},&
\hspace{0.5cm}& \mbox{for}~\ln(p^2/\Lambda_{\rm TC}^2)\leq -2;\\
\{7-\frac{4}{5}[2+\ln(p^2/\Lambda_{\rm
TC}^2)]^2\}\frac{12\pi}{(11N-2N_f)},&&
\mbox{for}~-2\leq\ln(p^2/\Lambda_{\rm TC}^2)\leq 0.5;\\
\displaystyle\frac{1}{\ln(p^2/\Lambda^2_{\rm
TC})}\frac{12\pi}{(11N-2N_f)}, &&
\mbox{for}~0.5\leq\ln(p^2/\Lambda_{\rm
TC}^2)\end{array}\right.~~\label{runningcoupling}
\end{eqnarray}
In which the fermion number is taken to be $N_f=2$ corresponding to
present one doublet techniquark. Although there is a dimensional
parameter $\Lambda_{\rm TC}$ appear in $\alpha_{\rm TC}(p^2)$,
except dimensional coefficient $F_0^{\rm 1D}$, all dimensionless
result coefficients $\alpha_i,~~i=1,2,3,4,5$ are independent of this
parameter. This can be seen as follows, if we scale up $\Lambda_{\rm
TC}$ as $\lambda\Lambda_{\rm TC}$, $\alpha_{\rm TC}(p)$ defined
above satisfy $\alpha_{\rm TC}(p^2)|_{\lambda\Lambda_{\rm
TC}}=\alpha_{\rm TC}(\lambda^{-2} p^2)|_{\Lambda_{\rm TC}}$ which,
by (\ref{eq-SDE8a}), result a scaling-up techniquark self energy
$\Sigma(p^2)|_{\lambda\Lambda_{\rm
TC}}=\lambda\Sigma(\lambda^{-2}p^2)|_{\Lambda_{\rm TC}}$, since an
alternative expression of (\ref{eq-SDE8a}) is
\begin{eqnarray}
\lambda\Sigma(\lambda^{-2}p_E^2)=3C_2(N)\int\frac{d^4q_E}{4\pi^3}
\frac{\alpha_{\mathrm{TC}}[\lambda^{-2}(p_E-q_E)^2]}{(p_E-q_E)^2}
\frac{\lambda\Sigma(\lambda^{-2}q_E^2)}{q_E^2+\lambda^2\Sigma^2(\lambda^{-2}q_E^2)}\;.\label{eq-SDE8a1}
\end{eqnarray}
Further from (36) of Ref.\cite{Wang02}, we find coefficients
$\mathcal{K}_i^{\Sigma\neq 0},~~i=1,2,3,4,13,14$ are invariant and
$F_0$ is changed to $\lambda F_0$ under exchanging
$\Sigma(p^2)\rightarrow\lambda\Sigma(\lambda^{-2}p^2)$ if we take
cutoff in the formulae $\Lambda\rightarrow\infty$. Due to this
invariance for $\mathcal{K}_i^{\Sigma\neq 0},~~i=1,2,3,4,13,14$,
from (\ref{1Dresult0}), we can see then $\alpha_i,~~i=1,2,3,4,5$ are
independent of $\Lambda_{\rm TC}$ and $F_0^{\rm 1D}$ scales same as
$\Lambda_{\rm TC}$. It is this scale dependence for $F_0$ which
makes ordinary QCD contribution small to electroweak symmetry
breaking and leads necessity for new interactions at higher energy
scale. The scale relations above are result of present rough
approximations, it will simplify our future computations very much
in next section.

From TABLE I, we see that within error of our approximation, the
numerical result exhibit the scaling-up behavior among different
$N$.

\subsection{Comparison and Discussion on Two Prescriptions}

Compare results from Gasser-Leutwyler's prescription and dynamical
computation prescription, (\ref{Gprescript}) and (\ref{Dprescript}),
we find results are same as long as we identify
\begin{eqnarray}
&&H_1^{\rm 1D}=-\frac{\mathcal{K}_{2}^{{\rm
1D},\Sigma\neq0}+\mathcal{K}_{13}^{{\rm 1D},\Sigma\neq0}}{4}\qquad
L_{10}^{\rm 1D}=\frac{\mathcal{K}_{2}^{{\rm
1D},\Sigma\neq0}-\mathcal{K}_{13}^{{\rm 1D},\Sigma\neq0}}{2}\nonumber\\
&& L_9^{\rm 1D}=\frac{\mathcal{K}_{13}^{{\rm
1D},\Sigma\neq0}}{2}-\frac{\mathcal{K}_{14}^{{\rm
1D},\Sigma\neq0}}{8}\qquad L_2^{\rm 1D}=\frac{\mathcal{K}_{4}^{{\rm
1D},\Sigma\neq0}+2\mathcal{K}_{13}^{{\rm
1D},\Sigma\neq0}-\mathcal{K}_{14}^{{\rm
1D},\Sigma\neq0}}{16}\label{1Dresult0}\\
&&L_1^{\rm 1D}+\frac{L_3^{\rm 1D}}{2}=\frac{\mathcal{K}_{3}^{{\rm
1D},\Sigma\neq0}-\mathcal{K}_{4}^{{\rm
1D},\Sigma\neq0}-4\mathcal{K}_{13}^{{\rm 1D},\Sigma\neq0}
+2\mathcal{K}_{14}^{{\rm 1D},\Sigma\neq0}}{32}\nonumber
\end{eqnarray}
which is just the result (25) obtained in Ref.\cite{Wang02}. This
shows that two prescriptions are equivalent in results. The merit of
Gasser-Leutwyler's prescription is its simplicity and express result
coefficients of electroweak chiral lagrangian in terms those in
Gasser-Leutwyler chiral lagrangian for pseudo scalar mesons in
strong interaction, but we can only apply this prescription to so
called QCD-like theories for which the technicolor interaction must
be vector-like. On the other hand, dynamical computation
prescription, although much more complex but touch the dynamics
details, do not limit us in the type of detail interactions. this
has very strong potential to be applied to more complicated
theories, such as chiral-like technicolor models. Since we involve
detail dynamical computation in this prescription, not like
Gasser-Leutwyler's prescription the coefficients are expressed in
terms of strong interaction experiment
 fixed values, we can give detail theoretical computation result for all coefficients and it further
allow us to test possible effects on the coefficients from variation
of the dynamics.

The first  property qualitatively drew out from (\ref{normdef}) and
(\ref{action-eff2-anom}) for their trace operation is that all
coefficients are proportional to $N$. This is the well known
scaling-up result for one doublet technicolor model, i.e., present
coefficients can be got by (\ref{Gprescript}) but identify the $L_i$
with the corresponding Gasser-Leutwyler chiral lagrangian for pseudo
scalar mesons by an scaling-up factor $N/N_c$ with $N_c=3$ be number
of color for strong interaction. In fact, it was this direct
correspondence lead to the death of one doublet technicolor model,
since negative experiment value for $L_{10}^{\rm 1D}$ result large
positive $S=-16\pi\alpha_1=-16\pi L_{10}^{\rm 1D}$ parameter which
contradict with present electroweak precision measurement data.

The second property quantitatively drew out from (\ref{normdef}) and
(\ref{action-eff2-anom}) is that except the overall $N$ factor in
front of all coefficients,  remaining part of coefficients depends
on dynamics, so exactly speaking, they are not precisely equivalent
to their strong interaction partners. For one doublet technicolor
model, when $N\neq N_c$, not only we will have overall scaling-up
factor $N/N_c$, but also we will have different techniquark self
energy $\Sigma$ due to the difference in running coupling constant
(\ref{runningcoupling}) appeared in SDE (\ref{eq-SDE8a}). $N_f=2$
also cause similar differences. But, since the estimations over the
values in original strong interaction suffers large errors either in
experiment or theories, this difference caused by dynamics hides in
these uncertainties.

\section{Derivation of the Electroweak Chiral Lagrangian from
 a Topcolor-assisted Technicolor Model}

 There are several options in topcolor-assisted technicolor model building: (1) TC breaks both
the EW interactions and the TopC interactions; (2) TC breaks EW, and
something else breaks TopC; (3) TC breaks only TopC and something
else drives EWSB ({\it e.g.}, a fourth generation condensate driven
by TopC). For definiteness, we will focus on a skeletal model in
category (1) in the following.

Consider a schematic TC2 model proposed by C.~T. Hill \cite{Hill94}.
The technicolor group is chosen to be $G_{\rm TC}=SU(3)_{\rm
TC1}\times SU(3)_{\rm TC2}$. The gauge charge assignments of
techniquarks in $G_{\rm TC}\times SU(3)_1\times SU(3)_2\times
SU(2)_L\times U(1)_{Y_1}\times U(1)_{Y_2}$
 are shown as Table II.

\begin{table}[h]
\small{{\bf TABLE II}.~Gauge charge assignments of techniquarks  a
schematic Topcolor-assisted Technicolor model. Ordinary quarks and
additional fields (such as leptons) required for anomaly
cancellation are not shown. The techniquark condensate $\langle
\bar{Q}Q\rangle$ breaks $SU(3)_1\times SU(3)_2\times
U(1)_{Y_1}\times U(1)_{Y_2}\rightarrow SU(3)\times U(1)_Y$, while
$\langle \bar{T}T\rangle$ breaks $SU(2)_L\times U(1)_{Y}\rightarrow
U(1)_{\rm EM}$.}
\renewcommand{\arraystretch}{1.2}
\begin{tabular}{*{8}{l}}
\hline field & $SU(3)_{\rm TC1}$ & $SU(3)_{\rm TC2}$ & $SU(3)_1$ &
$SU(3)_2$ &
$SU(2)_L$ & $U(1)_{Y_1}$ & $U(1)_{Y_2}$\\
\hline \hline $Q_L$ & 3& 1& 3& 1& 1& $\frac{1}{2}$& 0\\
$Q_R$ & 3& 1& 1& 3& 1& 0& $\frac{1}{2}$\\
$T_L=(T,~B)_L$ & 1 & 3 & 1 & 1 & 2 & 0 & $\frac{1}{6}$\\
$T_R=(T,~B)_R$ & 1 & 3 & 1 & 1 & 1 & 0 & $(\frac{2}{3},~
-\frac{1}{3})$\\ \hline
\end{tabular}
\end{table}

The action of the symmetry breaking sector (SBS) then is
\begin{eqnarray}
S_{\rm SBS}&=&\int d^4x({\cal L}_{\rm gauge}+{\cal L}_{\rm
techniquark})\;,
\end{eqnarray}
with different part of Lagrangian given by
\begin{eqnarray}
 {\cal L}_{\rm gauge}&=&-\frac{1}{4}F_{1\mu\nu}^\alpha
F_1^{\alpha\mu\nu}-\frac{1}{4}F_{2\mu\nu}^\alpha
F_2^{\alpha\mu\nu}-\frac{1}{4}A_{1\mu\nu}^A
A_1^{A\mu\nu}-\frac{1}{4}A_{2\mu\nu}^A
A_2^{A\mu\nu}-\frac{1}{4}W_{\mu\nu}^a W^{a\mu\nu}\nonumber\\
&&-\frac{1}{4}B_{1\mu\nu} B_1^{\mu\nu}-\frac{1}{4}B_{2\mu\nu}
B_2^{\mu\nu}\;,\\
{\cal L}_{\rm
techniquark}&=&\bar{Q}\big(i\slashed{\partial}-g_{31}r_1^\alpha\slashed{G}_1^\alpha
-h_1\frac{\lambda^A}{2}\slashed{A}_1^AP_L-h_2\frac{\lambda^A}{2}\slashed{A}_2^AP_R
-q_1\frac{1}{2}\slashed{B}_1P_L-q_2\frac{1}{2}\slashed{B}_2P_R\big)Q\nonumber\\
&&+\bar{T}\big[i\slashed{\partial}-g_{32}r_2^\alpha\slashed{G}_2^\alpha
-g_2\frac{\tau^a}{2}\slashed{W}P_L-q_2\frac{1}{6}\slashed{B}_2P_L
-q_2(\frac{1}{6}+\frac{\tau^3}{2})\slashed{B}_2P_R\big]T\;,
\end{eqnarray}
where $g_{31}$, $g_{32}$, $h_1$, $h_2$, $g_2$, $q_1$ and $q_2$ are
the coupling constants of, respectively, $SU(3)_{\mathrm{TC1}}$,
$SU(3)_{\mathrm{TC2}}$, $SU(3)_1$, $SU(3)_2$, $SU(2)_L$,
$U(1)_{Y_1}$ and $U(1)_{Y_2}$; and the corresponding gauge fields
(field strength tensors) are denoted by $G_{1\mu}^\alpha$,
$G_{2\mu}^\alpha$, $A_{1\mu}^A$, $A_{2\mu}^A$, $W_\mu^a$, $B_{1\mu}$
and $B_{2\mu}$ ($F_{1\mu\nu}^\alpha$, $F_{2\mu\nu}^\alpha$,
$A_{1\mu\nu}^A$, $A_{2\mu\nu}^A$, $W_{\mu\nu}^a$, $B_{1\mu\nu}$ and
$B_{2\mu\nu}$) with the superscripts $\alpha$ and $A$ running from 1
to 8 and $a$ from 1 to 3; $r_1^\alpha$ and $r_2^\alpha$
($\alpha=1,\ldots,8$) are the generators of, respectively,
$SU(3)_{\mathrm{TC1}}$ and $SU(3)_{\mathrm{TC2}}$, while $\lambda^A$
($A=1,\ldots,8$) and $\tau^a$ ($a=1,2,3$) are, respectively,
Gell-Mann and Pauli matrices.  We do not consider the ordinary
quarks in this work for following considerations, as we mentioned in
the Introduction that this paper only discuss bosonic part of EWCL
 and matter part of EWCL will discussed in future.
The matter part of EWCL mainly deals with effective interactions
among ordinary fermions which certainly include ordinary quarks.
Ignoring discussion of these effective interactions, only
concentrate on their contribution to bosonic part EWCL coefficients
is not self-consistent and efficient. Further one special feature of
topcolor-assisted technicolor model is its arrangements on the
interactions among ordinary quarks, especially for top and bottom
quark mass splitting and the top pions resulted from top quark
condensation through topcolor interactions, therefore dealing with
quark interactions is a separated important issue which needs
special care, previous formal derivations from underlying gauge
theory to low energy chiral Lagrangian, no matter QCD and
electroweak theory, all not involve in matter part of chiral
Lagrangian and further initial computation shows that we need some
special techniques to handle top-bottom splitting which are beyond
those techniques developed in this paper, to simplify the
computations and reduce the lengthy formulae,  we will not involve
in discussion of ordinary quark in this paper and would rather
specially focus our attentions on this issue in future works.

The strategy to derive the electroweak chiral Lagrangian from the
schematic topcolor-assisted technicolor model can be formulated as
\begin{eqnarray}
\exp\bigg(iS_{\mathrm{EW}}[W_\mu^a,B_\mu]\bigg)
&=&\int\mathcal{D}\bar{Q}\mathcal{D}Q\mathcal{D}\bar{T}\mathcal{D}T
\mathcal{D}G_{1\mu}^\alpha
\mathcal{D}G_{2\mu}^\alpha\mathcal{D}B_\mu^A\mathcal{D}Z_\mu^\prime\nonumber\\
&&\times\exp\bigg(iS_{\mathrm{SBS}}[G_{1\mu}^\alpha,G_{2\mu}^\alpha,A_{1\mu}^A,A_{2\mu}^A,
W_\mu^a,B_{1\mu},B_{2\mu},\bar{Q},Q,\bar{T},T]\bigg)\nonumber\\
&=&\mathcal{N}[W_\mu^a,B_\mu]\int\mathcal{D}\mu(U)\exp\bigg(iS_{\mathrm{eff}}[U,W_\mu^a,B_\mu]\bigg)\;,
\label{strategy-TC2}
\end{eqnarray}
where $U(x)$ is a dimensionless unitary unimodular matrix field in
the electroweak chiral Lagrangian, and ${\cal D}\mu(U)$ denotes
normalized functional integration measure on $U$. The normalization
factor $\mathcal{N}[W_\mu^a,B_\mu]$ is determined through
requirement that when the gauge coupling $g_{32}$ is switched off
$S_{\mathrm{eff}}[U,W_\mu^a,B_\mu]$ vanishes, this leads the
electroweak gauge fields $W_\mu^a$, $B_\mu$ dependent part of
$\mathcal{N}[W_\mu^a,B_\mu]$ is
\begin{eqnarray}
\mathcal{N}[W_\mu^a,B_\mu]&=&\int\mathcal{D}\bar{Q}\mathcal{D}Q\mathcal{D}\bar{T}\mathcal{D}T
\mathcal{D}G_{1\mu}^\alpha\mathcal{D}B_\mu^A\mathcal{D}Z_\mu^\prime\label{Ndef}\\
&&\times\exp\bigg(iS_{\mathrm{SBS}}[G_{1\mu}^\alpha,0,A_{1\mu}^A,A_{2\mu}^A,
W_\mu^a,B_{1\mu},B_{2\mu},\bar{Q},Q,\bar{T},T]\bigg)\nonumber
\end{eqnarray}
Since there different interactions in present model, in following
several subsections, we discuss them and their contributions to EWCL
separately.
\subsection{Topcolor Symmetry Breaking: the Contribution of
$SU(3)_{\rm TC1}$}

It can be shown below, by Schwinger-Dyson analysis, that the
$SU(3)_{\rm TC1}$ interaction induces the techniquark condensate
$\langle\bar{Q}Q\rangle\neq0$, which will trigger the topcolor
symmetry breaking $SU(3)_1\times SU(3)_2\times U(1)_{Y_1}\times
U(1)_{Y_2}\rightarrow SU(3)_c\times U(1)_Y$ at the scale $\Lambda=1$
TeV. This typically leaves a degenerate, massive color octet of
``colorons", $B_\mu^A$, and a singlet heavy $Z_\mu^\prime$ in the
coset space $[SU(3)_1\times SU(3)_2\times U(1)_{Y_1}\times
U(1)_{Y_2}]/[SU(3)_c\times U(1)_Y]$. The gluon $A_\mu^A$ and coloron
$B_\mu^A$ (the SM $U(1)_Y$ field $B_\mu$ and the $U(1)^\prime$ field
$Z_\mu^\prime$) are defined by orthogonal rotations with mixing
angle $\theta$ ($\theta^\prime$):
\begin{subequations}
\begin{eqnarray}
&&\begin{pmatrix}A_{1\mu}^A &
A_{2\mu}^A\end{pmatrix}=\begin{pmatrix}B_\mu^A &
A_\mu^A\end{pmatrix}
\begin{pmatrix}\cos\theta & -\sin\theta\\ \sin\theta &
\cos\theta\end{pmatrix}\;,\label{A1A2-AB}\\
&&\begin{pmatrix}B_{1\mu} &
B_{2\mu}\end{pmatrix}=\begin{pmatrix}Z_\mu^\prime &
B_\mu\end{pmatrix}
\begin{pmatrix}\cos\theta^\prime & -\sin\theta^\prime\\ \sin\theta^\prime &
\cos\theta^\prime\end{pmatrix}\;,\label{B1B2-BZpri}
\end{eqnarray}
\end{subequations}
which lead to
\begin{subequations}
\begin{eqnarray}
&&-h_1\frac{\lambda^A}{2}\slashed{A}_1^AP_L-h_2\frac{\lambda^A}{2}\slashed{A}_2^AP_R
=-g_3\frac{\lambda^A}{2}\slashed{A}^A-g_3(\cot\theta P_L-\tan\theta
P_R)\frac{\lambda^A}{2}\slashed{B}^A\;,\label{TC2-A1A2-AB}\\
&&-q_1\frac{1}{2}\slashed{B}_1P_L-q_2\frac{1}{2}\slashed{B}_2P_R
=-g_1\frac{1}{2}\slashed{B}-g_1(\cot\theta^\prime
P_L-\tan\theta^\prime
P_R)\frac{1}{2}\slashed{Z}^\prime\;,\label{TC2-B1B2-BZpri}
\end{eqnarray}
\end{subequations}
with
\begin{subequations}
\begin{eqnarray}
&&g_3\equiv h_1\sin\theta=h_2\cos\theta\;,\label{g3-h1-h2}\\
&&g_1\equiv
q_1\sin\theta^\prime=q_2\cos\theta^\prime\;.\label{g1-q1-q2}
\end{eqnarray}
\end{subequations}
As a first step  , we formally integrate out the
$SU(3)_{\mathrm{TC1}}$ technigluons $G_{1\mu}^\alpha$ in
Eq.\eqref{strategy-TC2} by introducing full $n$-point Green's
function of the $G_{1\mu}^\alpha$-field
$G_{\mu_1\ldots\mu_n}^{\alpha_1\ldots\alpha_n}$
\begin{eqnarray}
&&\hspace{-0.5cm}\exp\bigg(iS_{\mathrm{EW}}[W_\mu^a,B_\mu]\bigg)\nonumber\\
&=&\exp\bigg[i\int d^4x(-\frac{1}{4}W_{\mu\nu}^a W^{a\mu\nu})\bigg]
\int\mathcal{D}\bar{T}\mathcal{D}T\mathcal{D}G_{2\mu}^\alpha\mathcal{D}B_\mu^A\mathcal{D}Z_\mu^\prime
\exp\bigg[i\int
d^4x(-\frac{1}{4}F_{2\mu\nu}^\alpha
F_2^{\alpha\mu\nu}\nonumber\\
&&-\frac{1}{4}A_{1\mu\nu}^A A_1^{A\mu\nu}-\frac{1}{4}A_{2\mu\nu}^A
A_2^{A\mu\nu}-\frac{1}{4}B_{1\mu\nu}
B_1^{\mu\nu}-\frac{1}{4}B_{2\mu\nu}
B_2^{\mu\nu})+iS_{\mathrm{TC1}}[A_{1\mu}^A,A_{2\mu}^A,B_{1\mu},B_{2\mu}]\nonumber\\
&&+i\int
d^4x~\bar{T}\big[i\slashed{\partial}-g_{32}r_2^\alpha\slashed{G}_2^\alpha
-g_2\frac{\tau^a}{2}\slashed{W}^aP_L-q_2\frac{1}{6}\slashed{B}_2P_L
-q_2(\frac{1}{6}+\frac{\tau^3}{2})\slashed{B}_2P_R\big]T
\bigg]\;,\label{strategy-TC2-action2}
\end{eqnarray}
where
\begin{eqnarray}
&&\hspace{-0.5cm}\exp\bigg(iS_{\mathrm{TC1}}[A_{1\mu}^A,A_{2\mu}^A,B_{1\mu},B_{2\mu}]\bigg)\nonumber\\
&\equiv&\int\mathcal{D}\bar{Q}\mathcal{D}Q\exp\bigg[i\int
d^4x\bar{Q}\big(i\slashed{\partial}
-h_1\frac{\lambda^A}{2}\slashed{A}_1^AP_L-h_2\frac{\lambda^A}{2}\slashed{A}_2^AP_R
-q_1\frac{1}{2}\slashed{B}_1P_L-q_2\frac{1}{2}\slashed{B}_2P_R\big)Q\nonumber\\
&&+\sum_{n=2}^\infty\int d^4x_1\ldots
d^4x_n\frac{(-ig_{31})^n}{n!}G_{\mu_1\ldots\mu_n}^{\alpha_1\ldots\alpha_n}(x_1,\ldots,x_n)
J_{1\alpha_1}^{\mu_1}(x_1)\ldots
J_{1\alpha_n}^{\mu_n}(x_n)\bigg]\;.\label{action-TC1}
\end{eqnarray}
$J_{1\alpha}^\mu(x)\equiv\bar{Q}(x)r_{1}^\alpha\gamma^\mu Q(x)$ is
effective source.

Since the total functional derivative of the integrand in
Eq.\eqref{action-TC1} with respect to $\bar{Q}^\sigma(x)$ is zero,
(here and henceforth the suffixes $\sigma$ and $\rho$  are short
notations for Lorentz spinor, techniflavor and technicolor indices,)
{\it i.e.},
\begin{eqnarray}
0&=&\int\mathcal{D}\bar{Q}\mathcal{D}Q\frac{\delta}{\delta\bar{Q}^\sigma(x)}
\exp\bigg[\int d^4x(\bar{Q}I+\bar{I}Q)+i\int
d^4x\bar{Q}\big(i\slashed{\partial}
-h_1\frac{\lambda^A}{2}\slashed{A}_1^AP_L-h_2\frac{\lambda^A}{2}\slashed{A}_2^AP_R\nonumber\\
&&-q_1\frac{1}{2}\slashed{B}_1P_L-q_2\frac{1}{2}\slashed{B}_2P_R\big)Q
+\sum_{n=2}^\infty\int d^4x_1\ldots
d^4x_n\frac{(-ig_{31})^n}{n!}G_{\mu_1\ldots\mu_n}^{\alpha_1\ldots\alpha_n}(x_1,\ldots,x_n)
\nonumber\\
&&\times J_{1\alpha_1}^{\mu_1}(x_1)\ldots
J_{1\alpha_n}^{\mu_n}(x_n)\bigg]\;,
\end{eqnarray}
where $I(x)$ and $\bar{I}(x)$ are the external sources for,
respectively, $\bar{Q}(x)$ and $Q(x)$, then continue the similar
procedure from (\ref{eq-to-SDE1}) to (\ref{eq-to-SDE4}), by {\it
neglecting higher-point Green's functions} and {\it taking
factorization approximation}, we obtain
\begin{eqnarray}
0&=&\delta_{\sigma\rho}\delta(x-y)+i\bigg[i\slashed{\partial}_x
-h_1\frac{\lambda^A}{2}\slashed{A}_1^A(x)P_L-h_2\frac{\lambda^A}{2}\slashed{A}_2^A(x)P_R
-q_1\frac{1}{2}\slashed{B}_1(x)P_L\nonumber\\
&&-q_2\frac{1}{2}\slashed{B}_2(x)P_R\bigg]_{\sigma\tau}\langle
Q^\tau(x)\bar{Q}^\rho(y)\rangle-g_{31}^2\int
d^4x_2G_{\mu_1\mu_2}^{\alpha_1\alpha_2}(x,x_2)\nonumber\\
&&\times(r_1^{\alpha_1}\gamma^{\mu_1})_{\sigma\tau}
(r_1^{\alpha_2}\gamma^{\mu_2})_{\gamma\delta}
\big\langle\bar{Q}^\rho(y)Q^\delta(x_2)\big\rangle
\big\langle\bar{Q}^\gamma(x_2)Q^\tau(x)\big\rangle\;,\label{eq-to-TC2SDE4}
\end{eqnarray}
where $\langle\mathcal{O}(x)\big\rangle\equiv\left.
\big\langle\!\big\langle\mathcal{O}(x)\big\rangle\!\big\rangle_I/
\langle\langle\,1\,\rangle\rangle_I\right|_{I=\bar{I}=0}$ and we
have defined the notation $\langle\langle\,\cdots\rangle\rangle_I$
in this section by
\begin{eqnarray}
\big\langle\!\big\langle\mathcal{O}(x)\big\rangle\!\big\rangle_I
&\equiv&\int\mathcal{D}\bar{Q}\mathcal{D}Q\,\mathcal{O}(x)\,\exp\bigg[\int
d^4x(\bar{Q}I+\bar{I}Q)+i\int d^4x\bar{Q}\big(i\slashed{\partial}
-h_1\frac{\lambda^A}{2}\slashed{A}_1^AP_L\nonumber\\
&&-h_2\frac{\lambda^A}{2}\slashed{A}_2^AP_R
-q_1\frac{1}{2}\slashed{B}_1P_L-q_2\frac{1}{2}\slashed{B}_2P_R\big)Q
+\sum_{n=2}^\infty\int d^4x_1\ldots d^4x_n\frac{(-ig_{31})^n}{n!}\nonumber\\
&&\times
G_{\mu_1\ldots\mu_n}^{\alpha_1\ldots\alpha_n}(x_1,\ldots,x_n)J_{1\alpha_1}^{\mu_1}(x_1)\ldots
J_{1\alpha_n}^{\mu_n}(x_n)\bigg]\;.
\end{eqnarray}
Denote the technifermion propagator
$S^{\sigma\rho}(x,x^\prime)\equiv \langle
Q^\sigma(x)\bar{Q}^\rho(x^\prime)\rangle$, Eq.\eqref{eq-to-TC2SDE4}
can be written as SDE for techniquark propagator,
\begin{eqnarray}
0&=&S_{\sigma\rho}^{-1}(x,y)+i\bigg[i\slashed{\partial}_x
-h_1\frac{\lambda^A}{2}\slashed{A}_1^A(x)P_L-h_2\frac{\lambda^A}{2}\slashed{A}_2^A(x)P_R
-q_1\frac{1}{2}\slashed{B}_1(x)P_L-q_2\frac{1}{2}\slashed{B}_2(x)P_R\bigg]_{\sigma\rho}
\nonumber\\
&&\times\delta(x-y)-g_{31}^2G_{\mu_1\mu_2}^{\alpha_1\alpha_2}(x,y)
\bigg[r_1^{\alpha_1}\gamma^{\mu_1}S(x,y)r_1^{\alpha_2}\gamma^{\mu_2}
\bigg]_{\sigma\rho}\;.\label{eq-TC2SDE1}
\end{eqnarray}
By defining techniquark self energy $\Sigma$ as
\begin{eqnarray}
i\Sigma_{\sigma\rho}(x,y)&\equiv& S^{-1}_{\sigma\rho}(x,y)
+i\bigg[i\slashed{\partial}_x
-h_1\frac{\lambda^A}{2}\slashed{A}_1^A(x)P_L-h_2\frac{\lambda^A}{2}\slashed{A}_2^A(x)P_R
-q_1\frac{1}{2}\slashed{B}_1(x)P_L
\nonumber\\
&&-q_2\frac{1}{2}\slashed{B}_2(x)P_R\bigg]_{\sigma\rho}\delta(x-y)\;,
\end{eqnarray}
the SDE \eqref{eq-TC2SDE1} can be written as
\begin{eqnarray}
i\Sigma_{\sigma\rho}(x,y)=g_{31}^2G_{\mu_1\mu_2}^{\alpha_1\alpha_2}(x,y)
\bigg[r_1^{\alpha_1}\gamma^{\mu_1}S(x,y)r_1^{\alpha_2}\gamma^{\mu_2}
\bigg]_{\sigma\rho}\;.\label{eq-TC2SDE2}
\end{eqnarray}
Moreover, from the fact that technigluon propagator is diagonal in
the adjoint representation space of $SU(3)_{\rm TC1}$, {\it i.e.},
$G_{\mu\nu}^{\alpha\beta}(x,y)=\delta^{\alpha\beta}G_{\mu\nu}(x,y)$,
and techniquark propagator $\langle Q\bar{Q}\rangle$ is diagonal in
the fundamental representation space of $SU(3)_{\rm TC1}$, and also
$(r_1^\alpha r_1^\alpha)_{ab}=C_2(3)\delta_{ab}$,
Eq.\eqref{eq-TC2SDE2} is diagonal in indices $a,b$ and diagonal part
becomes
\begin{eqnarray}
i\Sigma_{\eta\zeta}^{ij}(x,y)=C_2(3)g_{31}^2G_{\mu_1\mu_2}(x,y)
[\gamma^{\mu_1}S(x,y)
\gamma^{\mu_2}]_{\eta\zeta}^{ij}\;,\label{eq-TC2SDE3}
\end{eqnarray}
where $\{i,~j\}$, and $\{\eta,~\zeta\}$ are, respectively,
techniflavor, and Lorentz spinor indices; and the Casimir operator
$C_2(3)=(3^2-1)/(2\times3)=4/3$.

\subsubsection{The Gap Equation}

We first consider the case of
$A_{1\mu}^A=A_{2\mu}^A=B_{1\mu}=B_{2\mu}=0$. The $SU(3)_{\rm TC1}$
technigluon propagator in Landau gauge is
\begin{eqnarray}
G_{\mu\nu}^{\alpha\beta}(x,y)
=\delta^{\alpha\beta}\int\frac{d^4p}{(2\pi)^4}e^{-ip(x-y)}\frac{i}{-p^2[1+\Pi(-p^2)]}\bigg(g_{\mu\nu}-\frac{p_\mu
p_\nu}{p^2}\bigg)\;, \label{TC2eq-mom-technigluon}
\end{eqnarray}
In the case of $A_{1\mu}^A=A_{2\mu}^A=B_{1\mu}=B_{2\mu}=0$, the
$SU(3)_{\rm TC1}$ techniquark self energy and propagator are
respectively
\begin{eqnarray}
\Sigma_{\eta\zeta}^{ij}(x,y)
=\int\frac{d^4p}{(2\pi)^4}e^{-ip(x-y)}\,\Sigma_{\eta\zeta}^{ij}(-p^2)\hspace{1cm}
S_{\eta\zeta}^{ij}(x,y)=\int\frac{d^4p}{(2\pi)^4}e^{-ip(x-y)}S_{\eta\zeta}^{ij}(p)\;,~~~
\end{eqnarray}
with
$S_{\eta\zeta}^{ij}(p)=i\{1/(\slashed{p}-\Sigma(-p^2)]\}_{\eta\zeta}^{ij}$.
Substituting above equations into the SDE \eqref{eq-TC2SDE3}, we
have
\begin{eqnarray}
\Sigma_{\eta\zeta}^{ij}(-p^2)
=\int\frac{d^4q}{(2\pi)^4}\frac{-C_2(3)g_{31}^2}{(p-q)^2[1+\Pi(-(p-q)^2)]}
\bigg[g_{\mu\nu}-\frac{(p-q)_\mu (p-q)_\nu}{(p-q)^2}\bigg]
\bigg[\gamma^\mu\frac{i}{\slashed{q}-\Sigma(-q^2)}\gamma^\nu\bigg]_{\eta\zeta}^{ij}~~
\label{eq-TC2SDE4}
\end{eqnarray}
As discussions of dynamical computation prescription for one-doublet
technicolor model, above equation will lead
$\Sigma_{\eta\zeta}^{ij}(-p^2)=\delta^{ij}\delta_{\eta\zeta}\Sigma_\mathrm{TC}(p^2_E)$
and in Euclidean space $\Sigma_\mathrm{TC}(p^2_E)$ satisfy
\begin{eqnarray}
\Sigma_{\rm TC1}(p_E^2)=3C_2(3)\int\frac{d^4q_E}{4\pi^3}
\frac{\alpha_{31}[(p_E-q_E)^2]}{(p_E-q_E)^2} \frac{\Sigma_{\rm
TC1}(q_E^2)}{q_E^2+\Sigma^2_{\rm TC1}(q_E^2)}\;.\label{eq-TC2SDE8a}
\end{eqnarray}
The corresponding techniquark condensate
$\langle\bar{Q}^kQ^j\rangle$ with $k$ and $j$ techniflavor indices,
\begin{eqnarray}
\langle\bar{Q}^k(x)Q^j(x)\rangle
=-12\delta^{jk}\int\frac{d^4p_E}{(2\pi)^4}\frac{\Sigma_\mathrm{TC}(p_E^2)}{p_E^2
+\Sigma_\mathrm{TC}^2(p_E^2)}\;,
\end{eqnarray}
where ${\rm tr}_{lc}$ is the trace with respect to Lorentz,
technicolor indices. Nonzero techniquark self energy can give a
nontrivial diagonal condensate $\langle\bar{Q}Q\rangle\neq0$, which
spontaneously breaks $SU(3)_1\times SU(3)_2\times U(1)_{Y_1}\times
U(1)_{Y_2}\rightarrow SU(3)_c\times U(1)_Y$.

In the following we consider the effects of the nonzero electroweak
gauge fields $A_{1\mu}^A$, $A_{2\mu}^A$, $B_{1\mu}$ and $B_{2\mu}$.
The SDE \eqref{eq-TC2SDE3} is explicitly
\begin{eqnarray}
\Sigma(x,y)&=&C_2(3)g_{31}^2G_{\mu\nu}(x,y)\gamma^\mu
\bigg[\bigg(i\slashed{\partial}_x
-h_1\frac{\lambda^A}{2}\slashed{A}_1^A(x)P_L-h_2\frac{\lambda^A}{2}\slashed{A}_2^A(x)P_R
-q_1\frac{1}{2}\slashed{B}_1(x)P_L\nonumber\\
&&-q_2\frac{1}{2}\slashed{B}_2(x)P_R\bigg)\delta(x-y)
-\Sigma(x,y)\bigg]^{-1}\gamma^\nu\;,\label{eq-TC2SDE-sources}
\end{eqnarray}
where the techniflavor and Lorentz spinor indices of the techniquark
self energy are implicitly contained.

Suppose the function $\Sigma_{\rm TC1}(-p^2)$ is a solution of the
SDE in the case $A_{1\mu}^A=A_{2\mu}^A=B_{1\mu}=B_{2\mu}=0$, that
is, it satisfies the equation
\begin{eqnarray}
\Sigma_{\rm TC1}(-p^2)&=&C_2(3)g_{31}^2\int\frac{d^4q}{(2\pi)^4}
G_{\mu\nu}(q^2)\gamma^\mu
\frac{1}{\slashed{q}+\slashed{p}-\Sigma_{\rm
TC1}[-(q+p)^2]}\gamma^\nu\;,\label{eq-sol-TC2SDE1}
\end{eqnarray}
 Replacing the variable $p$ by $p+\Delta$ in
Eq.\eqref{eq-sol-TC2SDE1} and subsequently integrating over $p$ with
the weight $e^{-ip(x-y)}$, we obtain, as long as $\Delta$ is
commutative with $\partial_x$ and Dirac matrices,
\begin{eqnarray}
&&\Sigma_{\rm TC1}[(\partial_x-i\Delta)^2]\delta(x-y)\nonumber\\
&&=C_2(3)g_{31}^2G_{\mu\nu}(x,y)\gamma^\mu\frac{1}{i\slashed{\partial}_x
+\slashed{\Delta}-\Sigma_{\rm
TC1}[-(i\partial_x+\Delta)^2]}\delta(x-y)\gamma^\nu\;.\label{eq-sol-TC2SDE2}
\end{eqnarray}
Even if $\Delta$ is noncommutative with $\partial_x$ and Dirac
matrices, the above equation holds as the lowest order
approximation, for the commutator $[\slashed{\partial}, \Delta]$ is
higher order of momentum than $\Delta$ itself. Now if we take
$\Delta$ to be
$-h_1\frac{\lambda^A}{2}A_1^AP_L-h_2\frac{\lambda^A}{2}A_2^AP_R
-q_1\frac{1}{2}B_1P_L-q_2\frac{1}{2}B_2P_R$, {\it ignoring its
noncommutative property with $\partial_x$ and Dirac matrices},
Eq.\eqref{eq-sol-TC2SDE2} is just the SDE \eqref{eq-TC2SDE-sources}
in the case
$A_{1\mu}^A\neq0,~A_{2\mu}^A\neq0,~B_{1\mu}\neq0,~\mbox{and}~B_{2\mu}\neq0$.
Thus, $\Sigma_{\rm
TC1}[(\partial_\mu^x+ih_1\frac{\lambda^A}{2}A_{1\mu}^AP_L
+ih_2\frac{\lambda^A}{2}A_{2\mu}^AP_R
+iq_1\frac{1}{2}B_{1\mu}P_L+iq_2\frac{1}{2}B_{2\mu}P_R)^2]\delta(x-y)$,
which is $SU(3)_1\times SU(3)_2\times U(1)_{Y_1}\times U(1)_{Y_2}$
covariant, can be regarded as the lowest-order solution of
Eq.\eqref{eq-TC2SDE-sources}. From Eqs.(\ref{TC2-A1A2-AB},
\ref{TC2-B1B2-BZpri}), we can write the covariant derivative of
$SU(3)_1\times SU(3)_2\times U(1)_{Y_1}\times U(1)_{Y_2}$ as
\begin{eqnarray}
\nabla_\mu&\equiv&\partial_\mu+ih_1\frac{\lambda^A}{2}A_{1\mu}^AP_L
+ih_2\frac{\lambda^A}{2}A_{2\mu}^AP_R
+iq_1\frac{1}{2}B_{1\mu}P_L+iq_2\frac{1}{2}B_{2\mu}P_R\\
&=&\partial_\mu+ig_3\frac{\lambda^A}{2}A_\mu^A+ig_1\frac{1}{2}B_\mu+ig_3(\cot\theta
P_L-\tan\theta P_R)\frac{\lambda^A}{2}B_\mu^A+ig_1(\cot\theta^\prime
P_L-\tan\theta^\prime)\frac{1}{2}Z_\mu^\prime\;,\nonumber
\end{eqnarray}
where $A_\mu^A$ and $B_\mu$ are the gauge fields of the unbroken
symmetry group $SU(3)_c\times U(1)_Y$. To further simplify the
calculations, we can {\it just keep this $SU(3)_c\times U(1)_Y$
covariance of the self energy}, that is, we can replace $\nabla_\mu$
by the covariant derivative of $SU(3)_c\times U(1)_Y$,
\begin{eqnarray}
\overline{\nabla}_\mu\equiv\partial_\mu+ig_3\frac{\lambda^A}{2}A_\mu^A+ig_1\frac{1}{2}B_\mu
\end{eqnarray}
inside the techniquark self energy. Thus, if the function
$\Sigma(\partial_x^2)\delta(x-y)$ is the self-energy solution of the
SDE in the case $A_{1\mu}^A=A_{2\mu}^A=B_{1\mu}=B_{2\mu}=0$, we can
replace its argument $\partial_x$ by the $SU(3)_c\times U(1)_Y$
covariant derivative $\overline{\nabla}_x$, {\it i.e.},
$\Sigma(\overline{\nabla}_x^2)\delta(x-y)$, as an approximate
solution the SDE in the case
$A_{1\mu}^A\neq0,~A_{2\mu}^A\neq0,~B_{1\mu}\neq0,~\mbox{and}~B_{2\mu}\neq0$.

Now we are ready to integrate out the techniquarks $Q$ and
$\bar{Q}$. The exponential terms on the right-hand side of
Eq.\eqref{action-TC1} can be written explicitly as
\begin{eqnarray}
&&\hspace{-0.5cm}\sum_{n=2}^\infty\int d^4x_1\ldots
d^4x_n\frac{(-ig_{31})^n}{n!}G_{\mu_1\ldots\mu_n}^{\alpha_1\ldots\alpha_n}(x_1,\ldots,x_n)
J_{1,\alpha_1}^{\mu_1}(x_1)\ldots J_{1,\alpha_n}^{\mu_n}(x_n)\nonumber\\
&\approx&\int d^4xd^4x^\prime
\bar{Q}^\sigma(x)\Pi_{\sigma\rho}(x,x^\prime)Q^\rho(x^\prime)\;,\label{TC2-mean-field}
\end{eqnarray}
where in the last equality we have taken the approximation of {\it
replacing the summation over $2n$-fermion interactions with parts of
them by their vacuum expectation values}, that is,
\begin{eqnarray}
\Pi_{\sigma\rho}(x,x^\prime)&=&\sum_{n=2}^\infty\Pi_{\sigma\rho}^{(n)}(x,x^\prime)\;,\\
\Pi_{\sigma\rho}^{(n)}(x,x^\prime)&=&n\times\int d^4x_2\ldots
d^4x_{n-1}\frac{(-ig_{31})^n}{n!}
G_{\mu_1\ldots\mu_n}^{\alpha_1\ldots\alpha_n}(x,x_2\ldots,x_{n-1},x^\prime)
\bigg\langle(r_1^{\alpha_1}\gamma^{\mu_1})_{\sigma\sigma_1}Q^{\sigma_1}(x)\nonumber\\
&&\times\bar{Q}(x_2) r_1^{\alpha_2}\gamma^{\mu_2}Q(x_2)\ldots
\bar{Q}(x_{n-1})
r_1^{\alpha_{n-1}}\gamma^{\mu_{n-1}}Q(x_{n-1})\bar{Q}^{\rho_n}(x^\prime)
(r_1^{\alpha_n}\gamma^{\mu_n})_{\rho_n\rho}\bigg\rangle\;,
\end{eqnarray}
where the factor $n$ comes from $n$ different choices of unaveraged
$\bar{Q}Q$, and the lowest term of which is
\begin{eqnarray}
\Pi_{\sigma\rho}^{(2)}(x,x^\prime)&=&2\cdot\frac{(-ig_{31})^2}{2!}
G_{\mu_1\mu_2}^{\alpha_1\alpha_2}(x,x^\prime)
\bigg\langle(r_1^{\alpha_1}\gamma^{\mu_1})_{\sigma\sigma_1}Q^{\sigma_1}(x)
\bar{Q}^{\rho_2}(x^\prime)
(r_1^{\alpha_2}\gamma^{\mu_2})_{\rho_2\rho}\bigg\rangle\nonumber\\
&=&-g_{31}^2G_{\mu_1\mu_2}^{\alpha_1\alpha_2}(x,x^\prime)
\bigg[r_1^{\alpha_1}\gamma^{\mu_1}S(x,y)
r_1^{\alpha_2}\gamma^{\mu_2}\bigg]_{\sigma\rho}\;.\label{eq-TC2-Pi-2}
\end{eqnarray}
Comparing Eq.\eqref{eq-TC2-Pi-2} with Eq.\eqref{eq-TC2SDE2}, we have
\begin{eqnarray}
i\Pi_{\sigma\rho}^{(2)}(x,x^\prime)=\Sigma_{\sigma\rho}(x,x^\prime)
\approx\Sigma_{\sigma\rho}(\overline{\nabla}_x^2)\delta(x-y)\;.
\label{eq-TC2-Pi-Sigma}
\end{eqnarray}
Substituting Eq.\eqref{TC2-mean-field} into Eq.\eqref{action-TC1},
we obtain
\begin{eqnarray}
&&\hspace{-0.5cm}\exp\bigg(iS_{\mathrm{TC1}}[A_{1\mu}^A,A_{2\mu}^A,B_{1\mu},B_{2\mu}]\bigg)\nonumber\\
&\approx&\mathrm{Det}\bigg[i\slashed{\partial}
-g_3\frac{\lambda^A}{2}\slashed{A}^A-g_3(\cot\theta P_L-\tan\theta
P_R)\frac{\lambda^A}{2}\slashed{B}^A-g_1\frac{1}{2}\slashed{B}-g_1(\cot\theta^\prime
P_L-\tan\theta^\prime
P_R)\frac{1}{2}\slashed{Z}^\prime\nonumber\\
&&-\Sigma_{\rm TC1}(\overline{\nabla}^2)\bigg]\;,
\end{eqnarray}
where we have taken further approximation of {\it keeping only the
lowest order, {\it i.e.} $\Pi_{\sigma\rho}^{(2)}(x,x^\prime)$, of
$\Pi_{\sigma\rho}(x,x^\prime)$}. With all these approximations, we
have
\begin{eqnarray}
iS_{\mathrm{TC1}}[A_{\mu}^A,B_{\mu}^A,B_{\mu},Z_{\mu}^\prime]
&=&\mathrm{Tr}\log\bigg[i\slashed{\partial}
-g_3\frac{\lambda^A}{2}\slashed{A}^A-g_3(\cot\theta P_L-\tan\theta
P_R)\frac{\lambda^A}{2}\slashed{B}^A-g_1\frac{1}{2}\slashed{B}\nonumber\\
&&-g_1(\cot\theta^\prime P_L-\tan\theta^\prime
P_R)\frac{1}{2}\slashed{Z}^\prime-\Sigma_{\rm
TC1}(\overline{\nabla}^2)\bigg]\;,\label{action2-TC1}
\end{eqnarray}
 Since we know QCD-induced condensate is too
weak to give sufficiently large masses of $W$ and $Z$ bosons and
thus it is negligible when we consider the main cause responsible
for the electroweak symmetry breaking which imply that ordinary QCD
gluon fields has very little effects on our technicolor and
electroweak interactions, therefore for simplicity, we ignore them
by just vanishing gluon field $A_\mu^A=0$. In next two
subsubsections, we perform low energy expansion and explicitly
expand above action up to order of $p^4$.

\subsubsection{Low Energy Expansion for $iS_{\mathrm{TC1}}[0,B^A_\mu,B_{\mu},Z_\mu^\prime]$}

We have
\begin{eqnarray}
iS_{\mathrm{TC1}}[0,B^A_\mu,B_{\mu},Z_\mu^\prime]&=&\mathrm{Tr}\log[i\slashed{\partial}
+\slashed{v}_1+\slashed{a}_1\gamma_5-\Sigma_{\rm TC1}(\overline{\nabla}^2)]\nonumber\\
&=&i\int d^4x(F_0^{\rm
TC1})^2\mathrm{tr}[a^2_1(x)]+S_{\mathrm{TC1}}^{(4)}[0,B^A_\mu,B_{\mu},Z_\mu^\prime]+O(p^6)\;,
\label{actionTC1-2}
\end{eqnarray}
where the parameter $F_0^{\rm TC1}$ is depend on the techniquark
self energy $\Sigma_{\rm TC1}$. The fields $v_\mu$ and $a_\mu$ are
identified with
\begin{subequations}
\label{def-v-a}
\begin{eqnarray}
&&v_{1,\mu}\equiv-\frac{g_3}{2}(\cot\theta-\tan\theta)\frac{\lambda^A}{2}B^A_\mu
-g_1\frac{1}{2}B_\mu-\frac{g_1}{4}(\cot\theta^\prime
-\tan\theta^\prime )Z^\prime_\mu\\
&&a_{1,\mu}\equiv-\frac{g_3}{2}(\cot\theta+\tan\theta)\frac{\lambda^A}{2}B^A_\mu
-\frac{g_1}{4}(\cot\theta^\prime+\tan\theta^\prime)Z^\prime_\mu\;.
\end{eqnarray}
\end{subequations}
Substituting Eqs.\eqref{def-v-a} into Eq.\eqref{actionTC1-2}, we
obtain, at the order of $p^2$,
\begin{eqnarray}
&&S_{\mathrm{TC1}}^{(2)}[0,B^A_\mu,B_{\mu},Z_\mu^\prime]\nonumber\\
&&=\frac{(F^{\rm TC1}_0)^2}{16}\int
d^4x[2g_3^2(\cot\theta+\tan\theta)^2B^A_{\mu}B^{A,,\mu}+3g_1^2(\cot\theta^\prime+\tan\theta^\prime)^2Z^{\prime
2}]\;.\label{actionTC1-2-1}
\end{eqnarray}

Now we come to consider the $p^4$ order effective action. It can be
divided into two parts
\begin{eqnarray}
&&S_{\mathrm{TC1}}^{(4)}[0,B^A_\mu,B_{\mu},Z_\mu^\prime]
=S_{\mathrm{TC1}}^{(4d)}[0,B^A_\mu,B_{\mu},Z_\mu^\prime]
+S_{\mathrm{TC1}}^{(4c)}[0,B^A_\mu,B_{\mu},Z_\mu^\prime]\label{actionTC1-4-1}\\
&&iS_{\mathrm{TC1}}^{(4d)}[0,B^A_\mu,B_{\mu},Z_\mu^\prime]=
\mathrm{Tr}\log[i\slashed{\partial}+\slashed{v}_1+\slashed{a}_1\gamma_5]\\
&&iS_{\mathrm{TC1}}^{(4c)}[0,B^A_\mu,B_{\mu},Z_\mu^\prime]
=\mathrm{Tr}\log[i\slashed{\partial}+\slashed{v}_1+\slashed{a}_1\gamma_5-
\Sigma_{\rm
TC1}(\overline{\nabla}^2)]-\mathrm{Tr}\log[i\slashed{\partial}+\slashed{v}_1+\slashed{a}_1\gamma_5]\nonumber
\end{eqnarray}
$S_{\mathrm{TC1}}^{(4d)}[0,B^A_\mu,B_{\mu},Z_\mu^\prime]$ is
divergent part of action, which can calculated by following standard
formula
\begin{eqnarray}
&&i\mathrm{Tr}\log[i\slashed{\partial}+\slashed{l}P_L+\slashed{r}P_R]
=-\frac{1}{2}\mathcal{K}\int
d^4x~\mathrm{tr}[r^{\mu\nu}r_{\mu\nu}+l^{\mu\nu}l_{\mu\nu}]\label{TrlogExp}\\
&&r_{\mu\nu}=\partial_{\mu}r_{\nu}-\partial_{\nu}r_{\mu}-i(r_{\mu}r_{\nu}-r_{\nu}r_{\mu})\hspace{2cm}
l_{\mu\nu}=\partial_{\mu}l_{\nu}-\partial_{\nu}l_{\mu}-i(l_{\mu}l_{\nu}-l_{\nu}l_{\mu})\nonumber\\
&&\mathcal{K}=-\frac{1}{48\pi^2}(\log\frac{\kappa^2}{\Lambda^2}+\gamma)\label{kappaDef}
\end{eqnarray}
with $\mathcal{K}$ a divergent constant depend on the ratio between
ultraviolet cutoff $\Lambda$ and infrared cutoff $\kappa$ of the
theory. Identify
\begin{eqnarray}
r_\mu&=&v_{1,\mu}+a_{1,\mu}=-g_3\cot\theta\frac{\lambda^A}{2}B^A_\mu
-g_1\frac{1}{2}B_\mu-\frac{g_1}{2}\cot\theta^\prime Z^\prime_\mu\\
l_\mu&=&v_{1,\mu}-a_{1,\mu}=g_3\tan\theta\frac{\lambda^A}{2}B^A_\mu
-g_1\frac{1}{2}B_\mu+\frac{g_1}{2}\tan\theta^\prime Z^\prime_\mu
\end{eqnarray}
With these preparations,
\begin{eqnarray}
S_{\mathrm{TC1}}^{(4d)}[0,B^A_\mu,B_{\mu},Z_{c,\mu}^\prime]&=&
-\frac{1}{2}\mathcal{K}\int d^4x~\bigg[\frac{g_3^2}{2}(\cot^2\theta
B^A_{r,\mu\nu}B_r^{A,\mu\nu}+\tan^2\theta
B^A_{l,\mu\nu}B_l^{A,\mu\nu})+\frac{3g_1^2}{2}B_{\mu\nu}B^{\mu\nu}\nonumber\\
&&+\frac{3g_1^2}{4}(\cot^2\theta^\prime+\tan^2\theta^\prime)
Z^\prime_{\mu\nu}Z^{\prime,\mu\nu}+\frac{3g_1^2}{2}(\cot\theta^\prime-\tan\theta^\prime)B_{\mu\nu}Z^{\prime,\mu\nu}
\bigg]\nonumber
\end{eqnarray}
with
\begin{eqnarray}
&&\hspace{-0.5cm}B^A_{r,\mu\nu}=\partial_{\mu}B^A_{\nu}-\partial_{\nu}B^A_{\mu}-g_3\cot\theta
f^{ABC}B^B_{\mu}B^C_{\nu} \hspace{1cm}
B^A_{l,\mu\nu}=\partial_{\mu}B^A_{\nu}-\partial_{\nu}B^A_{\mu}+g_3\tan\theta f^{ABC}B^B_{\mu}B^C_{\nu}\nonumber\\
&&\hspace{-0.5cm}B_{\mu\nu}=\partial_{\mu}B_{\nu}-\partial_{\nu}B_{\mu}\hspace{2cm}
Z^\prime_{\mu\nu}=\partial_{\mu}Z^\prime_{\nu}-\partial_{\nu}Z^\prime_{\mu}
\end{eqnarray}
$S_{\mathrm{TC1}}^{(4c)}[0,B^A_\mu,B_{\mu},Z_\mu^\prime]$ is
convergent part of action, which can calculated by following
standard formula
\begin{eqnarray}
S_{\mathrm{TC1}}^{(4c)}[0,B^A_\mu,B_{\mu},Z_\mu^\prime] &=&\int
d^4x~\mathrm{tr}[-\mathcal{K}_1^{\rm TC1,\Sigma\neq 0}(d_\mu
a^\mu_1)^2-\mathcal{K}_2^{\rm TC1,\Sigma\neq 0}(d_\mu
a_{1,\nu}-d_\nu a_{1,\mu})^2+\mathcal{K}_3^{\rm TC1,\Sigma\neq 0}(a^2_1)^2\nonumber\\
&&+\mathcal{K}_4^{\rm TC1,\Sigma\neq 0}(a_{1,\mu}
a_{1,\nu})^2-\mathcal{K}_{13}^{{\rm TC1},\Sigma\neq
0}V_{1,\mu\nu}V_1^{\mu\nu}+i\mathcal{K}_{14}^{{\rm TC1},\Sigma\neq
0}V_{1,\mu\nu}a_1^{\mu}a_1^{\nu}]\;,
\end{eqnarray}
with $V_{1,\mu\nu}\equiv
\partial_{\mu}v_{1,\nu}-\partial_{\nu}v_{1,\mu}-i(v_{1,\mu}v_{1,\nu}-v_{1,\nu}v_{1,\mu})$
 and
 $d_{\mu}a_{1,\nu}\equiv\partial_{\mu}a_{1,\nu}-i(v_{1,\mu}a_{1,\nu}-a_{1,\nu}v_{1,\mu})$.
 $S_{\mathrm{TC1}}^{(4c)}[0,B^A_\mu,B_{\mu},Z_\mu^\prime]$ can be
further divided into four parts
\begin{eqnarray}
S_{\mathrm{TC1}}^{(4c)}[0,B^A_\mu,B_{\mu},Z_\mu^\prime]&=&
S_{\mathrm{TC1}}^{(4c,B^A)}[B^A]+S_{\mathrm{TC1}}^{(4c,B)}[B]+S_{\mathrm{TC1}}^{(4c,Z')}[Z']\nonumber\\
&&+S_{\mathrm{TC1}}^{(4c,B^AZ')}[B^A,Z']
+S_{\mathrm{TC1}}^{(4c,BZ')}[B,Z']
\end{eqnarray}
The detail form of $S_{\mathrm{TC1}}^{(4c,B^A)}[B^A]$,
$S_{\mathrm{TC1}}^{(4c,B)}[B]$, $S_{\mathrm{TC1}}^{(4c,Z')}[Z']$,
$S_{\mathrm{TC1}}^{(4c,B^AZ')}[B^A,Z']$ and
$S_{\mathrm{TC1}}^{(4c,BZ')}[B,Z']$ are given in
(\ref{B1})-(\ref{B5}) respectively. Since TC1 interaction is $SU(3)$
gauge interaction which is same as QCD interaction and the quark
number $N_f$ are all equal to three \footnote{This is in fact an
approximation in which we have ignored possible effects on the
running of TC1 gauge coupling constant from ordinary color gauge
fields and coloron fields.}, as we discussed before, due to scale
invariance we have $\mathcal{K}_i^{{\rm TC1},\Sigma\neq
0},~~i=2,3,4,13,14$ are equal to those of QCD values within our
approximations
\begin{eqnarray}
\mathcal{K}_i^{{\rm TC1},\Sigma\neq 0}=\mathcal{K}_i^{\Sigma\neq
0}\hspace{2cm}i=2,3,4,13,14
\end{eqnarray}
Use relation given in Ref.\cite{Wang02}
\begin{eqnarray}
&&\hspace{-1cm}H_1=-\frac{1}{4}(\mathcal{K}_2^{\Sigma\neq
0}+\mathcal{K}_{13}^{\Sigma\neq
0})\hspace{1cm}L_{10}=\frac{1}{2}(\mathcal{K}_2^{\Sigma\neq
0}-\mathcal{K}_{13}^{\Sigma\neq
0})\hspace{1cm}L_9=\frac{1}{8}(4\mathcal{K}_{13}^{\Sigma\neq
0}-\mathcal{K}_{14}^{\Sigma\neq 0})\\
&&\hspace{-1cm}L_1=\frac{1}{32}(\mathcal{K}_4^{\Sigma\neq
0}+2\mathcal{K}_{13}^{\Sigma\neq 0}-\mathcal{K}_{14}^{\Sigma\neq
0})\hspace{1cm}L_3=\frac{1}{16}(\mathcal{K}_3^{\Sigma\neq
0}-2\mathcal{K}_4^{\Sigma\neq 0}-6\mathcal{K}_{13}^{\Sigma\neq
0}+3\mathcal{K}_{14}^{\Sigma\neq 0})
\end{eqnarray}
In Table III, we list down original QCD calculation result given in
Ref.\cite{Wang02}, the value for $H_1$ in the original paper is a
divergent constant therefore not given its value, now the divergent
part is already extracted out by
$S_{\mathrm{TC1}}^{(4d)}[0,B^A_\mu,B_{\mu},Z_\mu^\prime]$, $H_1$
here is a convergent quantity which can be obtained in original
formula for $H_1$ by subtracting out its divergent part caused by
terms with $\Sigma=0$.

\begin{table}[h]
{\small{\bf TABLE III}. The obtained nonzero values of the $O(p^4)$
coefficients $H_1,L_{10},L_9,L_2,L_1,L_3$ for topcolor-assisted
technicolor model.\\ The $F_0^{\rm TC1}$ and $\Lambda_{TC1}$ are in
units of TeV and coefficients are in units of $10^{-3}$.}

\vspace*{0.3cm}\begin{tabular}{c c|c c c c c c}\hline\hline
$F_0^{\rm TC1}$&$\Lambda_{\rm TC1}$ &$H_1$& $L_{10}$ & $L_9$ &
$L_2$ & $L_1$ & $L_3$ \\
\hline 1&5.21&43.0&-7.04&5.06&2.19&1.10&-7.81\\
 \hline\hline
\end{tabular}
\end{table}
We finally obtain
\begin{eqnarray}
&&\mathcal{K}_2^{{\rm TC1},\Sigma\neq
0}=L_{10}-2H_1\hspace{2cm}\mathcal{K}_3^{{\rm TC1},\Sigma\neq
0}=64L_1+16L_3+8L_9+2L_{10}+4H_1\nonumber\\
&&\mathcal{K}_4^{{\rm TC1},\Sigma\neq 0}=32L_1-8L_9-2L_{10}-4H_1\nonumber\\
&&\mathcal{K}_{13}^{{\rm TC1},\Sigma\neq
0}=-L_{10}-2H_1\hspace{2cm}\mathcal{K}_{14}^{{\rm TC1},\Sigma\neq
0}=-4L_{10}-8L_9-8H_1\label{KTC1}
\end{eqnarray}
\subsection{Electroweak Symmetry Breaking: the Contribution of
$SU(3)_{\rm TC2}$}

Likewise, it is easily check that the $SU(3)_{\mathrm{TC2}}$
interaction does induce the techniquark condensate
$\langle\bar{T}T\rangle\neq0$, which triggers the electroweak
symmetry breaking $SU(2)_L\times U(1)_{Y}\rightarrow U(1)_{\rm EM}$.
Integrating out the $SU(3)_{\mathrm{TC2}}$ technigluons
$G_{2\mu}^\alpha$ and the techniquarks $T$ and $\bar{T}$,
Eq.\eqref{strategy-TC2-action2} can be written as
\begin{eqnarray}
&&\hspace{-0.5cm}\exp\bigg(iS_{\mathrm{EW}}[W_\mu^a,B_\mu]\bigg)\nonumber\\
&=&\exp\bigg[i\int d^4x(-\frac{1}{4}W_{\mu\nu}^a W^{a\mu\nu})\bigg]
\int\mathcal{D}B_\mu^A\mathcal{D}Z_\mu^\prime \exp\bigg[i\int
d^4x(-\frac{1}{4}A_{1\mu\nu}^A
A_1^{A\mu\nu}\label{strategy-TC2-action3}\\
&&-\frac{1}{4}A_{2\mu\nu}^A A_2^{A\mu\nu}-\frac{1}{4}B_{1\mu\nu}
B_1^{\mu\nu}-\frac{1}{4}B_{2\mu\nu}
B_2^{\mu\nu})+iS_{\mathrm{TC1}}[A_{\mu}^A,B_{\mu}^A,B_{\mu},Z_{\mu}^\prime]
+iS_{\mathrm{TC2}}[W_\mu^a,B_{2\mu}] \bigg]\;,\nonumber
\end{eqnarray}
where $S_{\mathrm{TC1}}[A_{\mu}^A,B_{\mu}^A,B_{\mu},Z_{\mu}^\prime]$
has been given in Eq.\eqref{action2-TC1} for its general form and
expanded up to order of $p^2$ in (\ref{actionTC1-2-1}) and $p^4$ in
(\ref{actionTC1-4-1}), and $S_{\mathrm{TC2}}[W_\mu^a,B_{2\mu}]$ is
given by
\begin{eqnarray}
&&\hspace{-0.5cm}\exp\bigg(iS_{\mathrm{TC2}}[W_\mu^a,B_{2\mu}]\bigg)\label{action-TC2}\\
&=&\int\mathcal{D}\bar{T}\mathcal{D}T
\mathcal{D}G_{2\mu}^\alpha\exp\bigg[i\int
d^4x\bigg(-\frac{1}{4}F_{2\mu\nu}^\alpha
F_2^{\alpha\mu\nu}+\bar{T}\big[i\slashed{\partial}-g_{32}r_2^\alpha\slashed{G}_2^\alpha
+\slashed{l}_2P_L+\slashed{r}_2P_R\big]T\bigg)\bigg]\nonumber
\end{eqnarray}
with
$l_{2,\mu}\equiv-g_2\frac{\tau^a}{2}W_\mu^a-q_2\frac{1}{6}B_{2\mu}$
and $r_{2,\mu}\equiv-q_2(\frac{1}{6}+\frac{\tau^3}{2})B_{2\mu}$.

By means of the Gasser-Leutwyler's prescription presented in section
II, the functional integration \eqref{action-TC2} can be related to
the QCD-type chiral Lagrangian by
\begin{eqnarray}
&&\frac{\int\mathcal{D}\bar{T}\mathcal{D}T
\mathcal{D}G_{2\mu}^\alpha\exp\bigg[i\int
d^4x\bigg(-\frac{1}{4}F_{2\mu\nu}^\alpha
F_2^{\alpha\mu\nu}+\bar{T}\big[i\slashed{\partial}-g_{32}r_2^\alpha\slashed{G}_2^\alpha
+\slashed{l}_2P_L+\slashed{r}_2P_R\big]T\bigg)\bigg]}
{\int\mathcal{D}\bar{T}\mathcal{D}T \exp\big\{i\int
d^4x\bar{T}[i\slashed{\partial}
+\slashed{l}_2P_L+\slashed{r}_2P_R]T\big\}}\nonumber\\
&&=\int{\cal D}\mu(\tilde{U})\exp\{iS_{\mbox{\scriptsize TC2-induced
eff}}[\tilde{U},l_{2,\mu},r_{2,\mu}]\}\;,\label{TC2inducedeffDEF}
\end{eqnarray}
with the $SU(3)_{\mathrm{TC2}}$-induced chiral effective action
\begin{eqnarray}
S_{\mbox{\scriptsize TC2-induced
eff}}[\tilde{U},l_{2,\mu},r_{2,\mu}]&=&\int d^4x
\bigg[\frac{(F_0^{\rm TC2})^2}{4} {\rm
tr}[(\nabla^\mu\tilde{U}^\dag)(\nabla_\mu\tilde{U})]+ L_1^{\rm
TC2}[{\rm
tr}(\nabla^\mu\tilde{U}^\dag\nabla_\mu\tilde{U})]^2\nonumber\\&&+L_2^{\rm
TC2}{\rm tr}[\nabla_\mu\tilde{U}^{\dag}\nabla_\nu\tilde{U}] {\rm
tr}[\nabla^\mu\tilde{U}^{\dag}\nabla^\nu\tilde{U}]+L_3^{\rm TC2}{\rm
tr}[(\nabla^\mu\tilde{U}^\dag\nabla_\mu\tilde{U})^2]\nonumber\\
&&-iL_9^{\rm TC2}{\rm
tr}[F_{\mu\nu}^R\nabla^\mu\tilde{U}\nabla^\nu\tilde{U}^{\dag}
+F_{\mu\nu}^L\nabla^\mu\tilde{U}^\dag\nabla^\nu\tilde{U}]+L_{10}^{\rm
TC2}{\rm
tr}[\tilde{U}^{\dag}F_{\mu\nu}^R\tilde{U}F^{L,\mu\nu}]\nonumber\\
&& +H_1^{\rm TC2}{\rm
tr}[F_{\mu\nu}^RF^{R,\mu\nu}+F_{\mu\nu}^LF^{L,\mu\nu}]\bigg]\;,\label{TC1action-QCD-type}
\end{eqnarray}
where
\begin{eqnarray}
&&\nabla_\mu\tilde{U}\equiv\partial_\mu\tilde{U}-ir_{2,\mu}\tilde{U}+i\tilde{U}l_{2,\mu}\;,
\qquad\nabla_{\mu}\tilde{U}^\dag=-\tilde{U}^\dag(\nabla_{\mu}\tilde{U})\tilde{U}^\dag
=\partial_\mu\tilde{U}^\dag-il_{2,\mu}\tilde{U}^\dag+i\tilde{U}^\dag
r_{2,\mu}\;,\nonumber\\
&&F_{\mu\nu}^R\equiv i[\partial_\mu-ir_{2,\mu},
\partial_\nu-ir_{2,\nu}]\;, \qquad F_{\mu\nu}^L\equiv
i[\partial_\mu-il_{2,\mu},
\partial_\nu-il_{2,\nu}]\;.\label{TC2eq-U-tilde-definition}
\end{eqnarray}
The coefficients $F_0^{\rm TC2}$, $L_1^{\rm TC2}$, $L_2^{\rm TC2}$,
$L_3^{\rm TC2}$, $L_{10}^{\rm TC2}$, $H_1^{\rm TC2}$ arise from
$SU(3)_{\mathrm{TC2}}$ dynamics. These coefficients relates to the
$\mathcal{K}_i^{\rm TC2}$ coefficients as that appeared in one
doublet technicolor model as
\begin{eqnarray}
&& H_1^{\rm TC2}=-\frac{\mathcal{K}_{2}^{{\rm
TC2},\Sigma'}+\mathcal{K}_{13}^{{\rm TC2},\Sigma'}}{4}\qquad
L_{10}^{\rm TC2}=\frac{\mathcal{K}_{2}^{{\rm
TC2},\Sigma\neq0}-\mathcal{K}_{13}^{{\rm TC2},\Sigma\neq0}}{2}\nonumber\\
&& L_9^{\rm TC2}=\frac{\mathcal{K}_{13}^{{\rm
TC2},\Sigma\neq0}}{2}-\frac{\mathcal{K}_{14}^{{\rm
TC2},\Sigma\neq0}}{8}\qquad L_2^{\rm
TC2}=\frac{\mathcal{K}_{4}^{{\rm
TC2},\Sigma\neq0}+2\mathcal{K}_{13}^{{\rm
TC2},\Sigma\neq0}-\mathcal{K}_{14}^{{\rm
TC2},\Sigma\neq0}}{16}\nonumber\\
&&L_1^{\rm TC2}+\frac{L_3^{\rm TC2}}{2}=\frac{\mathcal{K}_{3}^{{\rm
TC2},\Sigma\neq0}-\mathcal{K}_{4}^{{\rm
TC2},\Sigma\neq0}-4\mathcal{K}_{13}^{{\rm TC2},\Sigma\neq0}
+2\mathcal{K}_{14}^{{\rm TC2},\Sigma\neq0}}{32}
\end{eqnarray}
$\mathcal{K}_i^{\rm TC2}$ coefficients with superscript TC2 denote
present TC2 interaction, they are functions of technifermion $T$
self energy $\Sigma_{\rm TC2}(p^2)$ and detail expressions are
already written down in (36) of Ref.\cite{Wang02} with the
replacement of $N_c\rightarrow 3$ and subtract out their
$\Sigma_{\rm TC2}(p^2)=0$ parts. Since TC2 interactions among
techiquark doublet T is $SU(3)$ which is same as one doublet
technicolor model discussed before except $\Lambda_{\rm TC2}$ may be
different as $\Lambda_{\rm TC}$ of one doublet technicolor model,
but as we discussed before $\mathcal{K}_i^{\rm
TC2},~~i=1,2,3,4,13,14$ are independent of $\Lambda_{\rm TC2}$,
therefore our $\mathcal{K}_i^{\rm TC2},~~i=1,2,3,4,13,14$ are same
as those obtained in one doublet technicolor model. This results
present $L_i^{\rm TC2},~~i=1,2,3,9,10$ and $H_1^{\rm TC2}$
coefficients are same as those in one doublet technicolor model,in
Table VI, we list down the numerical calculation result in which the
method is already mentioned in previous section and except the
result for $H_1^{\rm TC2}$, all others are already used in Table I.

\begin{table}[h]
{\small{\bf TABLE VI}. The obtained nonzero values of the $O(p^4)$
coefficients $H_1^{\rm TC2},L_{10}^{\rm TC2},L_9^{\rm TC2},L_2^{\rm
TC2},L_1^{\rm TC2},L_3^{\rm TC2}$ for topcolor-assisted technicolor
model.\\ The $F_0^{\rm TC2}$ and $\Lambda_{TC2}$ are in units of TeV
and coefficients are in units of $10^{-3}$.}

\vspace*{0.3cm}\begin{tabular}{c c|c c c c c c}\hline\hline
$F_0^{\rm TC2}$&$\Lambda_{\rm TC2}$ &$H_1^{\rm TC2}=H_1^{\rm 1D}$&
$L_{10}^{\rm TC2}=L_{10}^{\rm 1D}$ & $L_9^{\rm TC2}=L_9^{\rm 1D}$ &
$L_2^{\rm TC2}=L_2^{\rm 1D}$ & $L_1^{\rm TC2}=L_1^{\rm 1D}$ & $L_3^{\rm TC2}=L_3^{\rm 1D}$ \\
\hline 0.25&1.34&43.0&-6.90&4.87&2.02&1.01&-7.40\\
 \hline\hline
\end{tabular}
\end{table}
Similar as one-doublet case for $\tilde{U}$ is a $2\times 2$ unitary
matrix, and thus the $L_1^{\rm TC2}$ term and the $L_3^{\rm TC2}$
term are linearly related,
\begin{eqnarray}
L_3^{\rm TC2}{\rm
tr}[(\nabla^\mu\tilde{U}^\dag\nabla_\mu\tilde{U})^2]
&=&\frac{L_3^{\rm TC2}}{2}\bigg\{{\rm
tr}\big[\tilde{U}^\dag(\nabla^\mu\tilde{U})\tilde{U}^\dag(\nabla^\mu\tilde{U})\big]\bigg\}^2
\end{eqnarray}
 Comparing
Eqs.(\ref{TC2eq-U-tilde-definition}) with standard covariant
derivative given in Ref.\cite{EWCL}, we need to recognize
\begin{eqnarray}
\tilde{U}^\dag=U\;,\qquad \nabla_\mu\tilde{U}^\dag=\tilde{D}_\mu
U\equiv\partial_\mu
U+ig_2\frac{\tau^a}{2}W_\mu^aU-Uiq_2\frac{\tau^3}{2}B_{2\mu}\;.\label{TC2eq-U-tilde2U}
\end{eqnarray}
And $F_{\mu\nu}^R=-q_2(\frac{1}{6}+\frac{\tau^3}{2})B_{2\mu\nu}$,~
$F_{\mu\nu}^L=-g_2\frac{\tau^a}{2}W_{\mu\nu}^a-q_2\frac{1}{6}B_{2\mu\nu}$
with $B_{2\mu\nu}\equiv\partial_\mu B_{2\nu}-\partial_\nu B_{2\mu}$
is the $U(1)_{Y_2}$ gauge field strength tensor.

Substituting above equations back into
Eq.(\ref{TC1action-QCD-type}), we obtain
\begin{eqnarray}
S_{\mbox{\scriptsize TC2-induced eff}}[U,W,B_2]&=&\int d^4x
\bigg[-\frac{(F_0^{\rm TC2})^2}{4} {\rm tr}(\tilde{X}_\mu
\tilde{X}^\mu)+(L_1^{\rm 1D}+\frac{L_3^{\rm 1D}}{2})[{\rm
tr}(\tilde{X}_\mu
\tilde{X}^\mu)]^2\nonumber\\
&&+L_2^{\rm 1D}[{\rm tr}(\tilde{X}_\mu
\tilde{X}_\nu)]^2-i\frac{L_9^{\rm 1D}}{2}q_2B_{2\mu\nu}{\rm
tr}(\tau^3\tilde{X}^\mu \tilde{X}^\nu)-iL_9^{\rm 1D}{\rm
tr}(\overline{W}_{\mu\nu}\tilde{X}^\mu
\tilde{X}^\nu)\nonumber\\
&&+\frac{L_{10}^{\rm 1D}}{2}q_2B_{2\mu\nu}{\rm
tr}(\tau^3\overline{W}^{\mu\nu})+\frac{1}{18}(L_{10}^{\rm 1D}+11H_1^{\rm 1D})q_2^2B_{2\mu\nu}B_2^{\mu\nu}\nonumber\\
&&+H_1^{\rm 1D}{\rm
tr}(\overline{W}_{\mu\nu}\overline{W}^{\mu\nu})\bigg]\;,\label{TC1action-QCD-type2}
\end{eqnarray}
where $\tilde{X}_\mu$ is defined by
\begin{eqnarray}
\tilde{X}_\mu\equiv U^\dag(\tilde{D}_\mu U)\;.\label{def-X-tilde}
\end{eqnarray}
With (\ref{def-X-tilde}) and from Eqs.\eqref{TC2eq-U-tilde2U},
\eqref{B1B2-BZpri} and \eqref{g1-q1-q2}, we obtain
\begin{eqnarray}
\tilde{X}_\mu=X_\mu+ig_1\tan\theta^\prime
Z_{\mu}^\prime\frac{\tau^3}{2}\;.\label{X-tilde-X}
\end{eqnarray}
Substituting Eq.\eqref{X-tilde-X} into
Eq.\eqref{TC1action-QCD-type2}, we obtain, at the order of $p^2$,
\begin{eqnarray}
&&S_{\mbox{\scriptsize TC2-induced
eff}}^{(2)}[U,W^a,B\cos\theta^\prime-Z^\prime\sin\theta^\prime]\nonumber\\
&&=\frac{(F_0^{\rm TC2})^2}{4}\int d^4x \bigg[-{\rm tr}(X_\mu X^\mu)
-ig_1\tan\theta^\prime Z_{\mu}^\prime\mathrm{tr}(\tau^3X^\mu)+
\frac{g_1^2}{2}\tan^2\theta^\prime Z^{\prime
2}\bigg]\;.\label{TC1action-QCD-type2-2-2}
\end{eqnarray}
Similarly detail algebra gives
\begin{eqnarray}
&&\hspace{-0.5cm}iS_{\mbox{\scriptsize TC2-induced
eff}}^{(4)}[U,W^a,B\cos\theta^\prime-Z^\prime\sin\theta^\prime]\nonumber\\
&=&i\int d^4x\bigg[(L_1^{\rm 1D}+\frac{L_3^{\rm 1D}}{2})[{\rm
tr}(X_{\mu}X^\mu)]^2+L_2^{\rm 1D}[{\rm tr}(X_{\mu}X_\nu)]^2
-iL_9^{\rm 1D}{\rm tr}(\overline{W}_{\mu\nu}X^{\mu}X^\nu)
\nonumber\\
&&-i\frac{L_9^{\rm 1D}}{2}g_1B_{\mu\nu}{\rm tr}(\tau^3 X^\mu X^\nu)
+\frac{L_{10}^{\rm 1D}}{2}g_1B_{\mu\nu}{\rm
tr}(\tau^3\overline{W}^{\mu\nu}) +\frac{1}{18}(L_{10}^{\rm
1D}+11H_1^{\rm 1D})g_1^2B_{\mu\nu}B^{\mu\nu}\nonumber\\
&&+H_1^{\rm 1D}{\rm tr}(\overline{W}_{\mu\nu}\overline{W}^{\mu\nu})
-(L_1^{\rm 1D}+\frac{L_3^{\rm 1D}}{2})g_1^2\tan^2\theta^\prime
Z^{\prime,2}(\mathrm{tr}X^2)\label{TC2action4-QCD-3}\\
&&+(L_1^{\rm 1D}+L_2^{\rm 1D}+\frac{L_3^{\rm
1D}}{2})[\frac{1}{4}g_1^4\tan^4\theta^\prime Z^{\prime,4}
-ig_1^3\tan^3\theta^\prime
Z^{\prime,2}Z^{\prime,\mu}\mathrm{tr}(X_{\mu}\tau^3)]\nonumber\\
&&-\frac{1}{2}L_2^{\rm 1D}g_1^2\tan^2\theta^\prime [Z^{\prime
2}\mathrm{tr}(X_{\nu}\tau^3)\mathrm{tr}(X^{\nu}\tau^3)+
Z^{\prime\mu}Z^{\prime\nu}\mathrm{tr}(X_{\mu}\tau^3)\mathrm{tr}(X_{\nu}\tau^3)]\nonumber\\
&&-(L_1^{\rm 1D}+\frac{L_3^{\rm 1D}}{2})g_1^2\tan^2\theta^\prime
Z^{\prime,\mu}Z^{\prime,\nu}\mathrm{tr}(X_{\mu}\tau^3)\mathrm(X_{\nu}\tau^3)
-L_2^{\rm 1D}g_1^2\tan^2\theta^\prime
Z^{\prime,\mu}Z^{\prime,\nu}\mathrm{tr}(X_{\mu}X_{\nu})
\nonumber\\
&&+2i(L_1^{\rm 1D}+\frac{L_3^{\rm 1D}}{2})g_1\tan\theta^\prime
Z_\mu^{\prime}\mathrm{tr}(X^{\mu}\tau^3)(\mathrm{tr}X^2) +2iL_2^{\rm
1D}g_1\tan\theta^\prime
Z_\mu^{\prime}\mathrm{tr}(X_{\nu}\tau^3)\mathrm{tr}(X^{\mu}X^{\nu})\nonumber\\
&&+i\frac{L_9^{\rm 1D}}{2}g_1\tan\theta^\prime Z^\prime_{\mu\nu}{\rm
tr}(\tau^3 X^\mu X^\nu) +\frac{1}{2}g_1L_9^{\rm
1D}\tan\theta^\prime[{\rm tr}(\overline{W}_{\mu\nu}X^\mu\tau^3
)Z_\nu^\prime +{\rm
tr}(\overline{W}_{\mu\nu}\tau^3X^\nu)Z_\mu^\prime]\nonumber\\
&&-\frac{L_{10}^{\rm 1D}}{2}g_1\tan\theta^\prime Z^\prime_{\mu\nu}
{\rm tr}(\tau^3\overline{W}^{\mu\nu}) +\frac{1}{18}(L_{10}^{\rm
1D}+11H_1^{\rm 1D})[g_1^2\tan^2\theta^\prime
Z^\prime_{\mu\nu}Z^{\prime,\mu\nu}-2g_1^2\tan\theta^\prime
Z^\prime_{\mu\nu}B^{\mu\nu}]\bigg]\;.~~\nonumber
\end{eqnarray}
Thus, from Eqs.(\ref{action-TC2}--\ref{def-X-tilde}) we obtain
\begin{eqnarray}
&&\hspace{-1cm}iS_{\mathrm{TC2}}[W_\mu^a,B_{2\mu}]
=\mathrm{Tr}\ln\big[i\slashed{\partial}-g_2\frac{\tau^a}{2}\slashed{W}^aP_L
-q_2\frac{1}{6}\slashed{B}_2P_L-q_2(\frac{1}{6}
+\frac{\tau^3}{2})\slashed{B}_2P_R\big]\label{action-TC2-2}\\
&&\hspace{2cm}\times \log\int{\cal
D}\mu(U)\exp\bigg(iS_{\mbox{\scriptsize TC2-induced
eff}}^{(2)}[U,W,B_2]+iS_{\mbox{\scriptsize TC2-induced
eff}}^{(4)}[U,W,B_2]\bigg)\;.~~~~\nonumber
\end{eqnarray}
We still left to compute
$\mathrm{Tr}\ln\big[i\slashed{\partial}-g_2\frac{\tau^a}{2}\slashed{W}^aP_L
-q_2\frac{1}{6}\slashed{B}_2P_L-q_2(\frac{1}{6}
+\frac{\tau^3}{2})\slashed{B}_2P_R\big]$, which is at least order of
$p^4$. We can write
$\mathrm{Tr}\ln\big[i\slashed{\partial}-g_2\frac{\tau^a}{2}\slashed{W}^aP_L
-q_2\frac{1}{6}\slashed{B}_2P_L-q_2(\frac{1}{6}
+\frac{\tau^3}{2})\slashed{B}_2P_R\big]$ as
\begin{eqnarray}
&&\hspace{-0.5cm}\mathrm{Tr}\ln\big[i\slashed{\partial}-g_2\frac{\tau^a}{2}\slashed{W}^aP_L
-q_2\frac{1}{6}\slashed{B}_2P_L-q_2(\frac{1}{6}
+\frac{\tau^3}{2})\slashed{B}_2P_R\big]=
\mathrm{Tr}\log[i\slashed{\partial}+\slashed{l}_2P_L+\slashed{r}_2P_R]~~~~\\
&&\hspace{-0.5cm}l_2^{\mu}=-g_2\frac{\tau^a}{2}W^{a,\mu}
-q_2\frac{1}{6}B_2^\mu=-g_2\frac{\tau^a}{2}W^{a,\mu} -\frac{1}{6}
(g_1B^{\mu}-g_1\tan\theta^\prime Z^{\prime,\mu})\nonumber\\
&&\hspace{-0.5cm}r_2^{\mu}=-q_2(\frac{1}{6}
+\frac{\tau^3}{2})B_2^\mu=-(\frac{1}{6} +\frac{\tau^3}{2})
(g_1B^{\mu}-g_1\tan\theta^\prime Z^{\prime,\mu})\nonumber
\end{eqnarray}
Then with help of (\ref{TrlogExp}), computation gives
\begin{eqnarray}
&&\hspace{-0.5cm}i\mathrm{Tr}\ln\big[i\slashed{\partial}-g_2\frac{\tau^a}{2}\slashed{W}^aP_L
-q_2\frac{1}{6}\slashed{B}_2P_L-q_2(\frac{1}{6}
+\frac{\tau^3}{2})\slashed{B}_2P_R\big]\\
&&\hspace{-0.5cm} =-\frac{1}{2}\mathcal{K}\int
d^4x~\bigg[\frac{11}{18}g_1^2
(B_{\mu\nu}B^{\mu\nu}+\tan^2\theta^\prime
Z^\prime_{\mu\nu}Z^{\prime,\mu\nu}-2\tan\theta^\prime
Z^\prime_{\mu\nu}B^{\mu\nu})+\frac{1}{2}g_2^2W^a_{\mu\nu}W^{a,\mu\nu}\bigg]\nonumber
\end{eqnarray}

Substituting Eq.\eqref{action-TC2-2} into
Eq.\eqref{strategy-TC2-action3} and then comparing it with the last
line of Eq.\eqref{strategy-TC2}, we have
\begin{eqnarray}
&&\hspace{-0.5cm}\mathcal{N}[W_\mu^a,B_\mu]\exp\bigg(iS_{\mathrm{eff}}[U,W_\mu^a,B_\mu]\bigg)\nonumber\\
&=&\exp\bigg[i\int d^4x(-\frac{1}{4}W_{\mu\nu}^a W^{a\mu\nu})\bigg]
\int\mathcal{D}B_\mu^A\mathcal{D}Z_\mu^\prime \exp\bigg[i\int
d^4x(-\frac{1}{4}A_{1\mu\nu}^A
A_1^{A\mu\nu}-\frac{1}{4}A_{2\mu\nu}^A A_2^{A\mu\nu}\nonumber\\
&&-\frac{1}{4}B_{1\mu\nu} B_1^{\mu\nu}-\frac{1}{4}B_{2\mu\nu}
B_2^{\mu\nu})+\mathrm{Tr}\ln\big[i\slashed{\partial}-g_2\frac{\tau^a}{2}\slashed{W}^aP_L
-q_2\frac{1}{6}\slashed{B}_2P_L-q_2(\frac{1}{6}
+\frac{\tau^3}{2})\slashed{B}_2P_R\big]\nonumber\\
&&+iS_{\mathrm{TC1}}[A_{\mu}^A,B_{\mu}^A,B_{\mu},Z_{\mu}^\prime]
+iS_{\mbox{\scriptsize TC2-induced eff}}[U,W,B_2]
\bigg]_{A_\mu^A=0}\;,\label{strategy-TC2-action4}
\end{eqnarray}
where we have put the $SU(3)_c$ gluon fields $A_\mu^A=0$ on the
right-hand side, for the QCD effects are small here. The
normalization factor from its definition (\ref{Ndef}) can be
calculated similarly as previous procedure, the only difference is
that we switch off TC2 interaction by taking $g_{32}=0$ and it will
result $S_{\mbox{\scriptsize TC2-induced eff}}[U,W,B_2]$ vanishes,
then this leads ignoring term $iS_{\mbox{\scriptsize TC2-induced
eff}}[U,W,B_2]$ in above expression, we get expression for
$\mathcal{N}[W_\mu^a,B_\mu]$.

\subsection{Integrate out of Colorons}

Now, as shown in Eqs.\eqref{strategy-TC2-action4}, the next work is
to integrate out the $SU(3)_c$ octet of colorons, $B_\mu^A$. From
Eqs.\eqref{A1A2-AB} and \eqref{g3-h1-h2}, it is straightforward to
get
\begin{subequations}
\label{TCA1A2-field-strength}
\begin{eqnarray}
&&A_{1\mu\nu}^A=A_{\mu\nu}^A\sin\theta+B_{1,\mu\nu}^A\cos\theta\;,\\
&&A_{2\mu\nu}^A=A_{\mu\nu}^A\cos\theta-B_{2,\mu\nu}^A\sin\theta\;,
\end{eqnarray}
\end{subequations}
where
\begin{eqnarray}
B_{1,\mu\nu}^A&\equiv&\partial_\mu B_\nu^A-\partial_\nu
B_\mu^A+g_3f^{ABC}(\cot\theta
B_{\mu}^BB_\nu^C+B_{\mu}^BA_{\nu}^C+A_{\mu}^BB_{\nu}^C)\\
B_{2,\mu\nu}^A&\equiv&\partial_\mu B_\nu^A-\partial_\nu
B_\mu^A-g_3f^{ABC}(\tan\theta
B_{\mu}^BB_\nu^C+B_{\mu}^BA_{\nu}^C+A_{\mu}^BB_{\nu}^C)
\end{eqnarray}
then Eqs.\eqref{strategy-TC2-action4} become
\begin{eqnarray}
&&\hspace{-0.5cm}\mathcal{N}[W_\mu^a,B_\mu]\exp\bigg(iS_{\mathrm{eff}}[U,W_\mu^a,B_\mu]\bigg)\nonumber\\
&&\hspace{-0.5cm}=\exp\bigg[i\int d^4x(-\frac{1}{4}W_{\mu\nu}^a
W^{a\mu\nu})\bigg]
\int\mathcal{D}B_\mu^A\mathcal{D}Z_\mu^\prime\exp\bigg[iS_{\mathrm{TC1}}[0,B_{\mu}^A,B_{\mu},Z_{\mu}^\prime]
\label{coloron-action}\\
&&+i\int d^4x(-\frac{1}{4}B_{1\mu\nu}^A
B_1^{A\mu\nu}\cos^2\theta-\frac{1}{4}B_{2\mu\nu}^A
B_2^{A\mu\nu}\sin^2\theta-\frac{1}{4}B_{1\mu\nu}
B_1^{\mu\nu}-\frac{1}{4}B_{2\mu\nu}
B_2^{\mu\nu})\nonumber\\
&&+\mathrm{Tr}\ln\big[i\slashed{\partial}-g_2\frac{\tau^a}{2}\slashed{W}^aP_L
-q_2\frac{1}{6}\slashed{B}_2P_L-q_2(\frac{1}{6}
+\frac{\tau^3}{2})\slashed{B}_2P_R\big]+iS_{\mbox{\scriptsize
TC2-induced eff}}[U,W,B_2] \bigg]\;,\nonumber
\end{eqnarray}
Ignoring term $iS_{\mbox{\scriptsize TC2-induced eff}}[U,W,B_2]$ in
above expression, we get expression for
$\mathcal{N}[W_\mu^a,B_\mu]$. In above result, if we denote the
coloron involved part be
$\int\mathcal{D}B_\mu^A~e^{iS_\mathrm{coloron}[B^A,Z']}$, then
\begin{eqnarray}
S_\mathrm{coloron}[B^A,Z']&=&
S_{\mathrm{TC1}}^{4c,B^A}[B^A]+S_{\mathrm{TC1}}^{4c,B^AZ'}[B^A,Z']+
\int~d^4x\bigg(-\frac{1}{4}B_{1\mu\nu}^A B_1^{A\mu\nu}\cos^2\theta
\nonumber\\
&&-\frac{1}{4}B_{2\mu\nu}^A B_2^{A\mu\nu}\sin^2\theta+\frac{(F^{\rm
TC1}_0)^2}{8}g_3^2(\cot\theta+\tan\theta)^2B^A_{\mu}B^{A,,\mu}\nonumber\\
&&-\frac{g_3^2}{4}\mathcal{K}(\cot^2\theta
B^A_{r,\mu\nu}B_r^{A,\mu\nu}+\tan^2\theta
B^A_{l,\mu\nu}B_l^{A,\mu\nu})\bigg)\\
&=&S_\mathrm{coloron}^0[B^A,Z']+S_\mathrm{coloron}^\mathrm{int}[B^A,Z']
\end{eqnarray}
with $S_\mathrm{coloron}^0[B^A,Z']$ be linear and quadratic in
coloron fields and $S_\mathrm{coloron}^\mathrm{int}[B^A,Z']$ be
cubic and quartic in coloron fields, the detail form of them are
given in (\ref{Scoloron0}) and (\ref{ScoloronInt}). Now coloron
fields is not correctly normalized, since the coefficient in front
of kinetic term is not standard $-1/4$. We now introduce normalized
fields $B_{R,\mu}^A$  as
\begin{eqnarray}
&&\hspace{-0.8cm}B_\mu^A=\frac{1}{c}B_{R,\mu}^A\\
&&\hspace{-0.8cm}c^2=1+g_3^2[\frac{1}{2}\mathcal{K}_2^{\rm
TC1,\Sigma\neq
0}(\cot\theta+\tan\theta)^2+\frac{1}{2}\mathcal{K}_{13}^{{\rm
TC1},\Sigma\neq 0}(\cot\theta-\tan\theta)^2
+\mathcal{K}(\cot^2\theta +\tan^2\theta)]\nonumber\\
\label{cdef}
\end{eqnarray}
With them, $S_\mathrm{coloron}^0[B^A,Z']$ in terms of normalized
coloron fields become
\begin{eqnarray}
S_\mathrm{coloron}^0[B^A,Z']=\int
d^4x~\frac{1}{2}B_{R,\mu}^A(x)D_B^{-1,\mu\nu}(Z')B_{R,\nu}^A(x)
\end{eqnarray}
with
\begin{eqnarray}
&&\hspace{-0.5cm}D_B^{-1,\mu\nu}(Z')=D_{B0}^{-1,\mu\nu}+\Delta^{\mu\nu}(Z')\hspace{1cm}
D_{B0}^{-1,\mu\nu}=g^{\mu\nu}(\partial^2+M^2_\mathrm{coloron})
-(1+\lambda_B)\partial^{\mu}\partial^{\nu}~~~~~\\
&&\hspace{-0.5cm}\Delta^{\mu\nu}(Z')=[g^{\mu\nu}(\frac{1}{2}\mathcal{K}_3^{\rm
TC1,\Sigma\neq 0}+\mathcal{K}_4^{\rm TC1,\Sigma\neq 0})
Z^\prime_{\nu'}Z^{\prime,\nu'}+(2\mathcal{K}_3^{\rm TC1,\Sigma\neq
0}+\frac{5}{4}\mathcal{K}_4^{\rm TC1,\Sigma\neq 0}
)Z^{\prime,\mu}Z^{\prime,\nu}]\nonumber\\
&&\hspace{1.5cm}\times \frac{g_1^2g_3^4(\cot\theta+\tan\theta)^2(\cot\theta^\prime+\tan\theta^\prime)^2}{32c^2}\\
 &&\hspace{-0.5cm}M_\mathrm{coloron}=\frac{1}{2}g_3\frac{\cot\theta+\tan\theta}{c}F^{\rm
TC1}_0=\frac{g_3F^{\rm
TC1}_0}{2c\sin\theta\cos\theta}\\
&&\hspace{-0.5cm}\lambda_B=-\frac{1}{4}g_3^2\frac{(\cot\theta+\tan\theta)^2}{c^2}\mathcal{K}_1^{\rm
TC1,\Sigma\neq 0}
\end{eqnarray}
Here we recover the estimation for coloron mass
$M_\mathrm{coloron}\sim g_3\Lambda/(\sin\theta\cos\theta)$ given in
Ref.\cite{Hill02} if we identify $\Lambda=F^{\rm TC1}_0/(2c)$. We
now denote the result action after integration over colorons as
\begin{eqnarray}
\int\mathcal{D}B_\mu^A~e^{iS_\mathrm{coloron}[B^A,Z']}=e^{i\bar{S}_\mathrm{coloron}[Z']}
\label{ColoronOut}
\end{eqnarray}
 $\bar{S}_\mathrm{coloron}[Z']$ are  all
vacuum diagrams with propagator $D_B^{\mu\nu}(Z')$ and vertices
determined by $S_\mathrm{coloron}^\mathrm{int}[B^A,Z']$. The loop
expansion result is
\begin{eqnarray}
i\bar{S}_\mathrm{coloron}[Z']=-\frac{1}{2}\mathrm{Tr}\log
D_B^{-1}(Z')+ \mbox{two or more loop contributions}
\end{eqnarray}
The first term in the r.h.s. of above equation is one loop result,
if we further perform low energy expansion for it and drop out total
derivative terms, we find the contributions from one loop term is
quartically divergent up to order of $p^4$ which will be vanish if
we take dimensional regularization, then up to order of one loop
precision, colorons makes no contributions.

\subsection{Integrate out of $Z^\prime$}

From Eqs.\eqref{B1B2-BZpri} and \eqref{g1-q1-q2} imply
\begin{subequations}
\label{TCB1B2-field-strength}
\begin{eqnarray}
&&B_{1\mu\nu}=B_{\mu\nu}\sin\theta^\prime+(\partial_\mu
Z_\nu^\prime-\partial_\nu Z_\mu^\prime)\cos\theta^\prime\;,\\
&&B_{2\mu\nu}=B_{\mu\nu}\cos\theta^\prime-(\partial_\mu
Z_\nu^\prime-\partial_\nu Z_\mu^\prime)\sin\theta^\prime\;.
\end{eqnarray}
\end{subequations}

Substituting
$B_{1\mu}=B_\mu\sin\theta^\prime+Z_\mu^\prime\cos\theta^\prime$,
 $B_{2\mu}=B_\mu\cos\theta^\prime-Z_\mu^\prime\sin\theta^\prime$
and Eq.\eqref{TCB1B2-field-strength} into the right-hand side of
Eq.\eqref{coloron-action}, combined with (\ref{ColoronOut}), we get
\begin{eqnarray}
&&\hspace{-0.5cm}\mathcal{N}[W_\mu^a,B_\mu]\exp\bigg(iS_{\mathrm{eff}}[U,W_\mu^a,B_\mu]\bigg)\nonumber\\
&&\hspace{-0.5cm}=\exp\bigg[i\int d^4x(-\frac{1}{4}W_{\mu\nu}^a
W^{a\mu\nu})\bigg] \int\mathcal{D}Z_\mu^\prime\exp\bigg[i\int
d^4x\bigg(-\frac{1}{4}B_{\mu\nu} B^{\mu\nu}-\frac{1}{4}(\partial_\mu
Z_\nu^\prime-\partial_\nu
Z_\mu^\prime)^2\nonumber\\
&&+\frac{3g_1^2(F^{\rm
TC1}_0)^2}{16}(\cot\theta^\prime+\tan\theta^\prime)^2Z^{\prime
2}-\frac{1}{2}\mathcal{K}[\frac{3g_1^2}{2}B_{\mu\nu}B^{\mu\nu}
+\frac{3g_1^2}{4}(\cot^2\theta^\prime+\tan^2\theta^\prime)
Z^\prime_{\mu\nu}Z^{\prime,\mu\nu}\nonumber\\ &&
+\frac{3g_1^2}{2}(\cot\theta^\prime-\tan\theta^\prime)B_{\mu\nu}Z^{\prime,\mu\nu}]\bigg)
+i\bar{S}_\mathrm{coloron}[Z']+iS_{\mathrm{TC1}}^{(4c,B)}[B]+iS_{\mathrm{TC1}}^{(4c,Z')}[Z']\nonumber\\
&&+iS_{\mathrm{TC1}}^{(4c,BZ')}[B,Z']+\mathrm{Tr}\ln\big[i\slashed{\partial}-g_2\frac{\tau^a}{2}\slashed{W}^aP_L
-q_2\frac{1}{6}\slashed{B}_2P_L-q_2(\frac{1}{6}
+\frac{\tau^3}{2})\slashed{B}_2P_R\big] \nonumber\\
&&+iS_{\mbox{\scriptsize TC2-induced
eff}}[U,W,B_2]\bigg]\;,\label{Z'-action}
\end{eqnarray}
Ignoring term $iS_{\mbox{\scriptsize TC2-induced eff}}[U,W,B_2]$ in
above expression, we get expression for
$\mathcal{N}[W_\mu^a,B_\mu]$. In above result, if we denote the $Z'$
involved part be $\int\mathcal{D}Z'_\mu~e^{iS_{Z'}[Z',U,W^a,B]}$,
then we will find that $Z'$ field in $S_{Z'}[Z',U,W^a,B]$ is not
correctly normalized, since the coefficient in front of kinetic term
is not standard $-1/4$. We now introduce normalized fields
$Z'_{R,\mu}$  as
\begin{eqnarray}
Z'_\mu&=&\frac{1}{c'}Z'_{R,\mu}\\
c^{\prime,2}&=& 1
+\mathcal{K}\frac{3g_1^2}{2}(\cot^2\theta^\prime+\tan^2\theta^\prime)
+\frac{3g_1^2}{4}[\mathcal{K}_2^{\rm TC1,\Sigma\neq
0}(\cot\theta^\prime+\tan\theta^\prime)^2\label{cpDef}\\
&&+\mathcal{K}_{13}^{{\rm TC1},\Sigma\neq 0}(\cot\theta^\prime
-\tan\theta^\prime
)^2]-\frac{11}{9}g_1^2\mathcal{K}\tan^2\theta^\prime
-\frac{2}{9}(L_{10}^{\rm 1D}+11H_1^{\rm
1D})g_1^2\tan^2\theta^\prime\nonumber
\end{eqnarray}
then
\begin{eqnarray}
S_{Z'}[Z',U,W^a,B]=\int d^4x~[
\frac{1}{2}Z'_{R,\mu}(x)D_Z^{-1,\mu\nu}Z'_{R,\nu}(x)
+Z_R^{\prime,\mu}J_{Z,\mu}+Z_R^2Z_{R,\mu}'J^{\mu}_{3Z}
+g_{4Z}\frac{g_1^4}{c^{\prime 4}}Z_R^{\prime,4}]~~~~\label{SZ'}
\end{eqnarray}
with
\begin{eqnarray}
&&\hspace{-0.5cm}D_Z^{-1,\mu\nu}=g^{\mu\nu}(\partial^2+M^2_{Z'})
-(1+\lambda_Z)\partial^{\mu}\partial^{\nu}+\Delta^{\mu\nu}_Z(X)\label{DZdef}\\
&&\hspace{-0.5cm}M_{Z'}^2=\frac{3g_1^2(F^{\rm TC1}_0)^2}{8c^{\prime
2}}(\cot\theta^\prime+\tan\theta^\prime)^2+\frac{g_1^2(F_0^{\rm
TC2})^2}{4c^{\prime 2}}\tan^2\theta^\prime \label{MZdef}\\
&&\hspace{-0.5cm}\lambda_Z=-\mathcal{K}_1^{\rm TC1,\Sigma\neq
0}\frac{3g_1^2}{8c^{\prime 2}}(\cot\theta^\prime+\tan\theta^\prime)^2\\
&&\hspace{-0.5cm}\Delta^{\mu\nu}_Z(X)=
-g_1^2\frac{\tan^2\theta^\prime}{c^{\prime 2}}\{(2L_1^{\rm
1D}+L_3^{\rm 1D}) g^{\mu\nu}(\mathrm{tr}X^2)+(2L_1^{\rm 1D}+L_3^{\rm
1D})\mathrm{tr}(X_{\mu}\tau^3)\mathrm{tr}(X_{\nu}\tau3)
\nonumber\\
&&\hspace{1.5cm}+2L_2^{\rm 1D}\mathrm{tr}(X_{\mu}X_{\nu})+L_2^{\rm
1D}[g^{\mu\nu}\mathrm{tr}(X_{\nu}\tau^3)\mathrm{tr}(X^{\nu}\tau^3)+
\mathrm{tr}(X_{\mu}\tau^3)\mathrm{tr}(X_{\nu}\tau^3)]\}
\end{eqnarray}
and
\begin{eqnarray}
&&\hspace{-0.5cm}J_Z^\mu=J_{Z0}^\mu+\frac{g_1^2\gamma}{c'}\partial^{\nu}B_{\mu\nu}+\tilde{J}_Z^\mu\label{JZdef}\\
&&\hspace{-0.5cm}J_{Z0}^\mu=-ig_1\frac{(F_0^{\rm
TC2})^2}{4c'}\tan\theta^\prime\mathrm{tr}(\tau^3X^\mu)\label{JZ0def}\\
&&\hspace{-0.5cm}\gamma=-(3\mathcal{K}+\mathcal{K}_{13}^{{\rm
TC1},\Sigma\neq 0})\frac{1}{2}(\cot\theta^\prime -\tan\theta^\prime)
-(11\mathcal{K}+2L_{10}^{\rm 1D}+22H_1^{\rm
1D})\frac{1}{9}\tan\theta^\prime\label{gammaDef}\\
&&\hspace{-0.5cm}\tilde{J}_{Z,\mu}=\frac{2i}{c'}(L_1^{\rm
1D}+\frac{L_3^{\rm 1D}}{2})g_1\tan\theta^\prime
\mathrm{tr}(X_{\mu}\tau^3)(\mathrm{tr}X^2)+\frac{2i}{c'}L_2^{\rm
1D}g_1\tan\theta^\prime
\mathrm{tr}(X_{\nu}\tau^3)(\mathrm{tr}X_{\mu}X_{\nu})\nonumber\\
 &&\hspace{0.9cm}+\frac{1}{2c'}g_1L_9^{\rm 1D}\tan\theta^\prime{\rm
tr}[(\overline{W}_{\mu\nu}\tau^3-\tau^3\overline{W}_{\mu\nu})X^\nu]
+\frac{i}{2c'}L_9^{\rm 1D}g_1\tan\theta^\prime\partial^{\nu}[{\rm
tr}\tau^3(X_\mu X_\nu-X_\nu X_\mu)]\nonumber\\
&&\hspace{0.9cm} -\frac{1}{c'}L_{10}^{\rm
1D}g_1\tan\theta^\prime\partial^{\nu}{\rm
tr}(\tau^3\overline{W}_{\mu\nu})
\nonumber\\
&&\hspace{-0.5cm}g_{4Z}=(\mathcal{K}_3^{\rm TC1,\Sigma\neq
0}+\mathcal{K}_4^{\rm TC1,\Sigma\neq
0})\frac{3}{256}(\cot\theta^\prime+\tan\theta^\prime)^4 +(L_1^{\rm
1D}+L_2^{\rm 1D}+\frac{L_3^{\rm 1D}}{2})\frac{1}{4}\tan^4\theta^\prime\label{g4Zdef}\\
&&\hspace{-0.5cm}J_{3Z}^\mu=-\frac{i}{c^{\prime 3}}(L_1^{\rm
1D}+L_2^{\rm 1D}+\frac{L_3^{\rm 1D}}{2})
g_1^3\tan^3\theta^\prime\mathrm{tr}(X_{\mu}\tau^3)
\end{eqnarray}

We denote the result action after the integration over $Z'$ as
\begin{eqnarray}
\int\mathcal{D}Z'_\mu~e^{iS_{Z'}[Z',U,W^a,B]}=e^{i\bar{S}_{Z'}[U,W^a,B]}
\end{eqnarray}
We can use loop expansion to calculate above integration
\begin{eqnarray}
\bar{S}_{Z'}[U,W^a,B]=S_{Z'}[Z'_c,U,W^a,B]+\mbox{loop terms}
\end{eqnarray}
with classical field $Z'_c$ satisfy
\begin{eqnarray}
\frac{\partial }{\partial
Z'_{c,\mu}(x)}\bigg[S_{Z'}[Z'_c,U,W^a,B]+\mbox{loop terms}\bigg]=0
\end{eqnarray}
With (\ref{SZ'}), the solution is
\begin{eqnarray}
Z_c^{\prime\mu}(x)=-D^{\mu\nu}_ZJ_{Z,\nu}(x)+O(p^3)+\mbox{loop
terms}
\end{eqnarray}
then
\begin{eqnarray}
\bar{S}_{Z'}[U,W^a,B] &=&\int d^4x~[
-\frac{1}{2}J_{Z,\mu}D_Z^{\mu\nu}J_{Z,\nu}
-J_{3Z,\mu'}(D_Z^{\mu'\nu'}J_{Z,\nu'})(D_Z^{\mu\nu}J_{Z,\nu})^2+g_{4Z}\frac{g_1^4}{c^{\prime 4}}(D_Z^{\mu\nu}J_{Z,\nu})^4]\nonumber\\
&&+\mbox{loop terms}\label{SZ'out}
\end{eqnarray}
where
\begin{eqnarray}
D_Z^{-1,\mu\nu}D_{Z,\nu\lambda}=D_Z^{\mu\nu}D_{Z,\nu\lambda}^{-1}=g^\mu_\lambda
\end{eqnarray}
and it is not difficult to show that if we are accurate up to order
of $p^4$, then $p$ order $Z_c'$ solution is enough, all
contributions from $p^3$ order $Z'_c$ are at least belong to order
of $p^6$.

With these results, (\ref{Z'-action}) become
\begin{eqnarray}
&&\hspace{-0.5cm}\mathcal{N}[W_\mu^a,B_\mu]\exp\bigg(iS_{\mathrm{eff}}[U,W_\mu^a,B_\mu]\bigg)\label{Z'out}\\
&&\hspace{-0.5cm}=\exp\bigg[i\bar{S}_{Z'}[U,W^a,B]+iS_{\mathrm{TC1}}^{(4c,B)}[B]
+i\int d^4x\bigg[-\frac{1}{4}W_{\mu\nu}^a
W^{a\mu\nu}-\frac{1}{4}B_{\mu\nu} B^{\mu\nu}
-\mathcal{K}\frac{3g_1^2}{4}B_{\mu\nu}B^{\mu\nu}
\nonumber\\
&&+\frac{1}{2}\mathcal{K}(\frac{11}{18}g_1^2
B_{\mu\nu}B^{\mu\nu}+\frac{1}{2}g_2^2W^a_{\mu\nu}W^{a,\mu\nu})
-\frac{(F_0^{\rm TC2})^2}{4}{\rm tr}(X_\mu X^\mu) +(L_1^{\rm
1D}+\frac{L_3^{\rm 1D}}{2})[{\rm tr}(X_{\mu}X^\mu)]^2
\nonumber\\
&&+L_2^{\rm 1D}[{\rm tr}(X_{\mu}X_\nu)]^2-iL_9^{\rm 1D}{\rm
tr}(\overline{W}_{\mu\nu}X^{\mu}X^\nu) -i\frac{L_9^{\rm
1D}}{2}g_1B_{\mu\nu}{\rm tr}(\tau^3 X^\mu X^\nu) +\frac{L_{10}^{\rm
1D}}{2}g_1B_{\mu\nu}{\rm tr}(\tau^3\overline{W}^{\mu\nu})\nonumber\\
&&+\frac{1}{18}(L_{10}^{\rm 1D}+11H_1^{\rm
1D})g_1^2B_{\mu\nu}B^{\mu\nu} +H_1^{\rm 1D}{\rm
tr}(\overline{W}_{\mu\nu}\overline{W}^{\mu\nu})\bigg]\;,\nonumber
\end{eqnarray}
 Ignoring term with coefficients $F_0^\mathrm{TC2}$, $L_i^\mathrm{1D}$ and
$H_1^\mathrm{1D}$ in above expression, we get expression for
$\mathcal{N}[W_\mu^a,B_\mu]$, with it we finally obtain
 $S_{\mathrm{eff}}[U,W_\mu^a,B_\mu]$
\begin{eqnarray}
S_{\mathrm{eff}}[U,W_\mu^a,B_\mu]&=&\int d^4x~\bigg[
-\frac{(F_0^{\rm TC2})^2}{4}{\rm tr}(X_\mu X^\mu)+(L_1^{\rm
1D}+\frac{L_3^{\rm 1D}}{2})[{\rm tr}(X_{\mu}X^\mu)]^2+L_2^{\rm
1D}[{\rm tr}(X_{\mu}X_\nu)]^2 \nonumber\\
&&-iL_9^{\rm 1D}{\rm
tr}(\overline{W}_{\mu\nu}X^{\mu}X^\nu)-i\frac{L_9^{\rm
1D}}{2}g_1B_{\mu\nu}{\rm tr}(\tau^3 X^\mu X^\nu)+\frac{L_{10}^{\rm
1D}}{2}g_1B_{\mu\nu}{\rm tr}(\tau^3\overline{W}^{\mu\nu})
\nonumber\\
&&+\frac{1}{18}(L_{10}^{\rm 1D}+11H_1^{\rm
1D})g_1^2B_{\mu\nu}B^{\mu\nu}+H_1^{\rm 1D}{\rm
tr}(\overline{W}_{\mu\nu}\overline{W}^{\mu\nu}) \bigg]+\Delta
S_{\mathrm{eff}}[U,W_\mu^a,B_\mu]\label{Seffresult}
\end{eqnarray}
i.e. our result EWCL is equal to standard one-doublet technicolor
model result plus contributions from $Z'$, we denote this $Z'$
contribution part $\Delta S_{\mathrm{eff}}[U,W_\mu^a,B_\mu]$,
\begin{eqnarray}
\Delta S_{\mathrm{eff}}[U,W_\mu^a,B_\mu]=
 \bar{S}_{Z'}[U,W^a,B]-\bar{S}_{Z'}[U,W^a,B]|_{
F_0^\mathrm{TC2}=0,L_i^\mathrm{1D}=H_1^\mathrm{1D}=0}\;.
\end{eqnarray}
Correspondingly, EWCL coefficients for topcolor assisted technicolor
model  can also be divided into two parts
\begin{eqnarray}
f^2=(F_0^\mathrm{TC2})^2
\hspace{1cm}\beta_1=\Delta\beta_1\hspace{1cm}\alpha_i=\alpha_i|_\mathrm{one~doulet}+\Delta\alpha_i
~~i=1,2,\ldots,14~~~~
\end{eqnarray}
in which $\alpha_i|_\mathrm{one~doulet}~~i=1,2,\ldots,14$ are
coefficients from one-doublet technicolor model, their values are
given in (\ref{Gprescript}) and Table I. $\Delta\beta_1$ and
$\Delta\alpha_i ~~i=1,2,\ldots,14$ are contributions from $Z'$ and
ordinary quarks, since we do not consider ordinary quarks in this
work, so in the next part of this paper, we calculate $Z'$
contributions.

 With help of (\ref{SZ'out}), (\ref{DZdef}) and
(\ref{JZdef})
\begin{eqnarray}
\Delta S_{\mathrm{eff}}[U,W_\mu^a,B_\mu]&=&\int
d^4x~[-\frac{1}{2}J_{Z0,\mu}D_Z^{\mu\nu}J_{Z0,\nu}
-\frac{1}{M_{Z'}^2}
J_{Z0,\mu}(\tilde{J}^\mu_Z+\frac{g_1^2\gamma}{c'}\partial_{\nu}B^{\mu\nu})\nonumber\\
&&-\frac{1}{M_{Z'}^6}J_{3Z,\mu}J_{Z0}^{\mu}J_{Z0}^2
+\frac{g_{4Z}g_1^4}{c^{\prime 4}M_{Z'}^8}J_{Z0}^4]\label{DeltaSZ'}
\end{eqnarray}
With help of following algebra relations,
\begin{eqnarray}
&&\partial_\mu\mathrm{tr}[\tau^3X^\mu]=0\nonumber\\
 &&\mathrm{tr}[\tau^3(\partial_\mu X_\nu-\partial_\nu X_\mu)]=
-2\mathrm{tr}(\tau^3X_\mu X_\nu)
+i\mathrm{tr}(\tau^3\overline{W}_{\mu\nu})-ig_1B_{\mu\nu}\nonumber\\
&&{\rm tr}(\tau^3X_\mu X_\nu){\rm tr}(\tau^3X^\mu
X^\nu)\\
&&=[{\rm tr}(X_\mu X_\nu)]^2-[{\rm tr}(X_\mu X^\mu)]^2-{\rm
tr}(X_\mu X_\nu){\rm tr}(\tau^3X^\mu){\rm tr}(\tau^3X^\nu)+{\rm
tr}(X_\mu X^\mu)[{\rm tr}(\tau^3X_\nu)]^2\nonumber\\
&&\mathrm{tr}(TA)\mathrm{tr}(TBC)+\mathrm{tr}(TB)\mathrm{tr}(TCA)+\mathrm{tr}(TC)\mathrm{tr}(TAB)
=2\mathrm{tr}(ABC)\nonumber
\end{eqnarray}
where $\mathrm{tr}A=\mathrm{tr}B=\mathrm{tr}C=0$ and $T^2=1$. We can
show (\ref{DeltaSZ'}) leads to the form of standard EWCL, further
combined with (\ref{Seffresult}) and (\ref{JZ0def}), we can read out
$p^2$ coefficient
\begin{eqnarray}
\beta_1=\frac{g_1^2(F_0^{\rm TC2})^2}{8c^{\prime
2}M_{Z'}^2}\tan^2\theta^\prime=\frac{(F_0^{\rm TC2})^2} {3(F^{\rm
TC1}_0)^2(\cot^2\theta^\prime+1)^2+2(F_0^{\rm
TC2})^2}\label{beta1-0}
\end{eqnarray}
which imply a positive $\beta_1$ which is further bounded above.
With the fact that $f=F_0^{\rm TC2}=250$GeV and original model
requirement $F_0^{\rm TC1}=1$TeV, we find
\begin{eqnarray}
2\beta_1=\frac{1} {24(\cot^2\theta^\prime+1)^2+1}\;.
\end{eqnarray}
Combine with $\alpha T=2\beta_1$ given in Ref.\cite{EWCL}, we obtain
result that topcolor-assisted technicolor model produce positive and
bounded above $T$ parameter! The upper limit of $\beta_1$ is $1/50$
which corresponds to upper limit of $T$ parameter $1/(25\alpha)\sim
5.1$. From (\ref{beta1-0}), we know $\beta_1$ coefficient is
uniquely determined by parameter $\theta'$, therefore instead of
using $\theta'$ as the input parameter of the theory, we can further
use $\beta_1$ or $T=2\beta_1/\alpha$ as the parameter of the theory.

\begin{figure}[t]
\caption{The $S$ and $U$ parameters for topcolor assisted
technicolor model. $F_0^\mathrm{TC2}=250$ GeV, the $T$ parameter and
$M_{Z'}=\{0.5,~1,~2\}$ TeV are as input parameters of the model.}
\label{fig-S-U} \centering
\begin{minipage}[t]{\textwidth}
\subfloat[$S$]{%
    \label{fig-TC2-S}
    \includegraphics[scale=0.6]{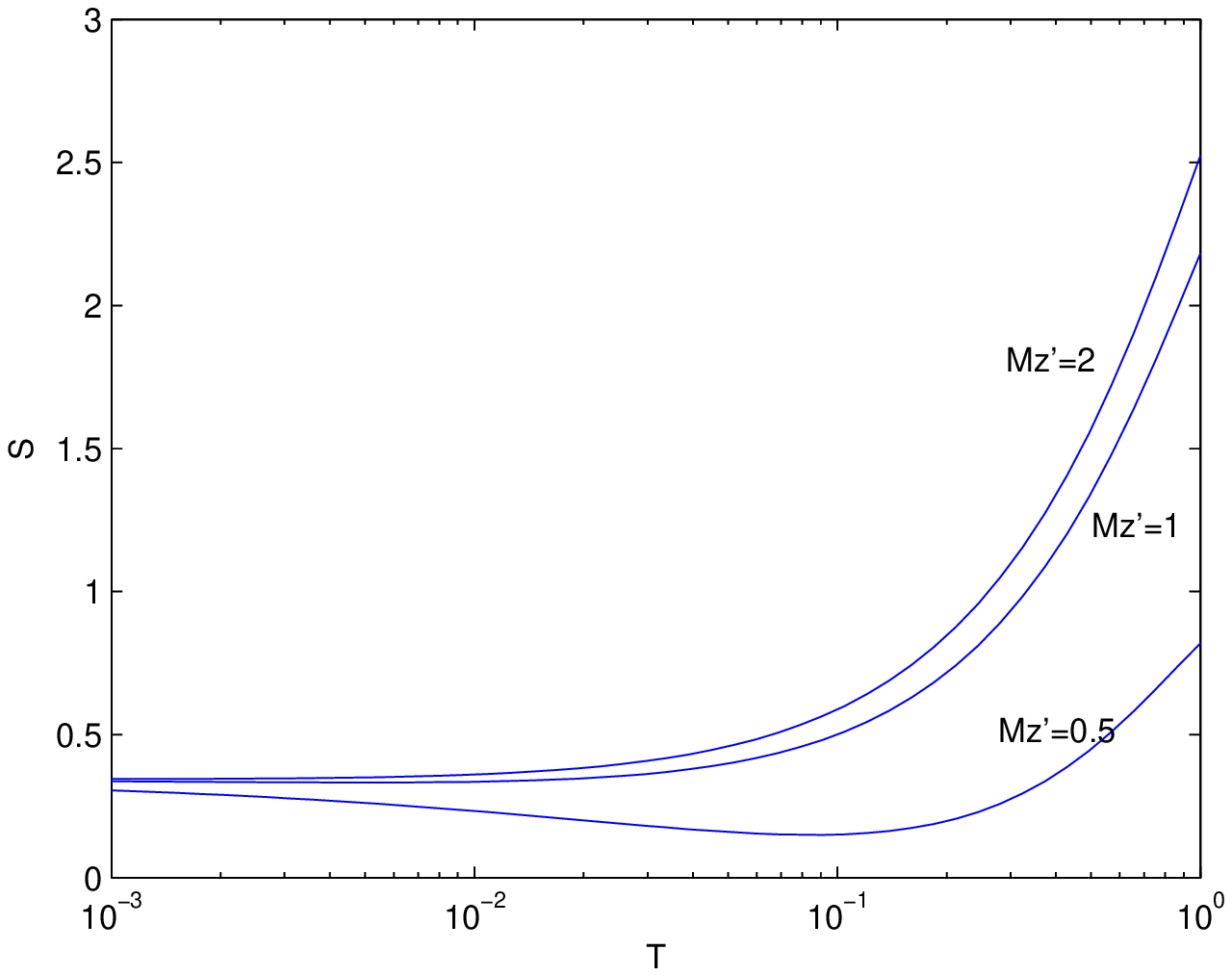}}
\subfloat[$U$]{%
    \label{fig-TC2-U}
    \includegraphics[scale=0.6]{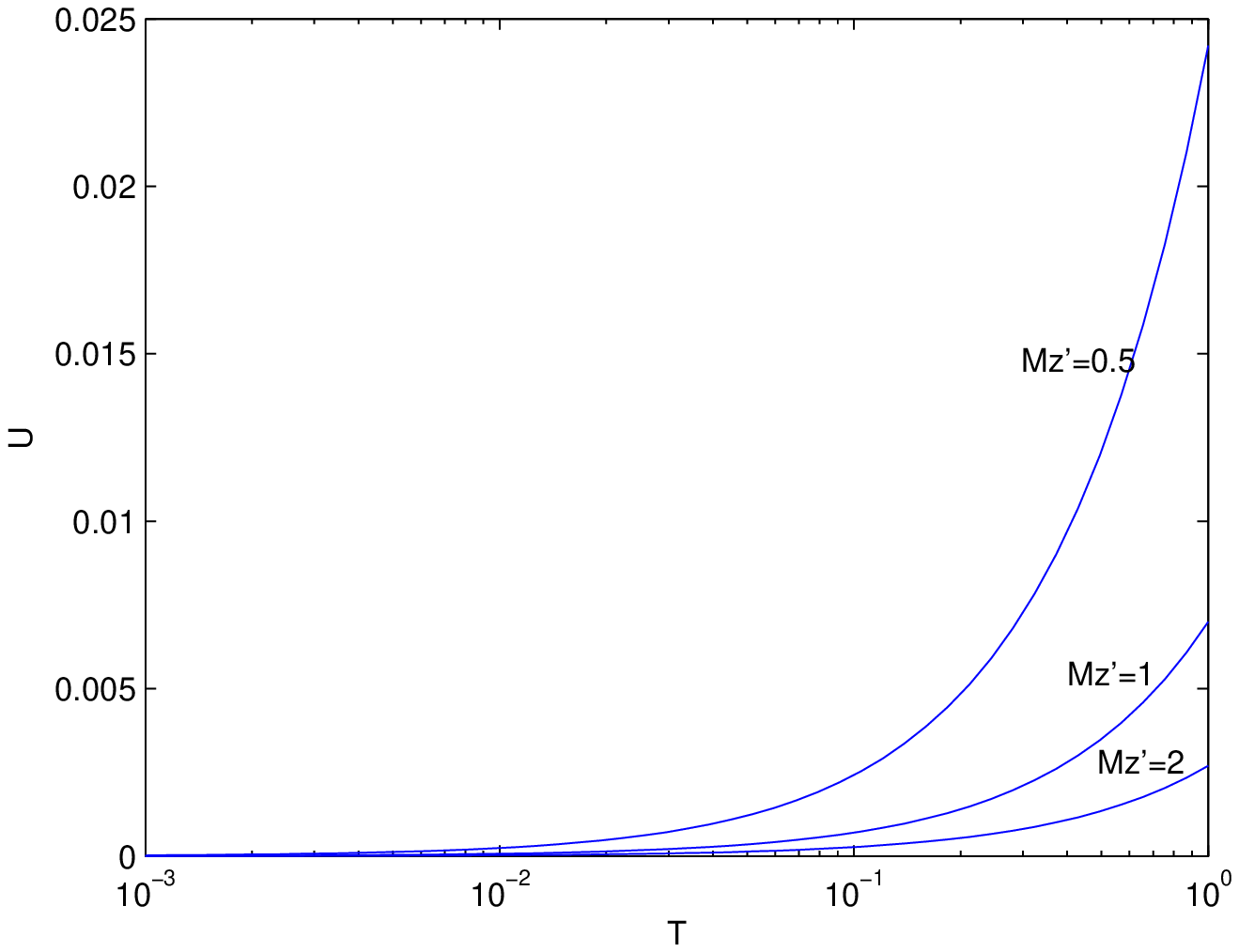}}
\end{minipage}
\end{figure}

\begin{figure}[t]
\caption{Nonzero EWCL coefficients $\alpha_i$($i=1,2,\ldots,10$) for
the topcolor-assisted technicolor model up to order of
$\mathcal{O}(p^4)$. $F_0^\mathrm{TC2}=250$ GeV, the $T$ parameter
and $M_{Z'}=\{0.5,~1,~2\}$ TeV are as input parameters of the
model.}\label{fig-alpha-i}\centering
\begin{minipage}[t]{\textwidth}
\subfloat[$\alpha_1$]{%
    \label{fig-alpha-1}
    \includegraphics[scale=0.6]{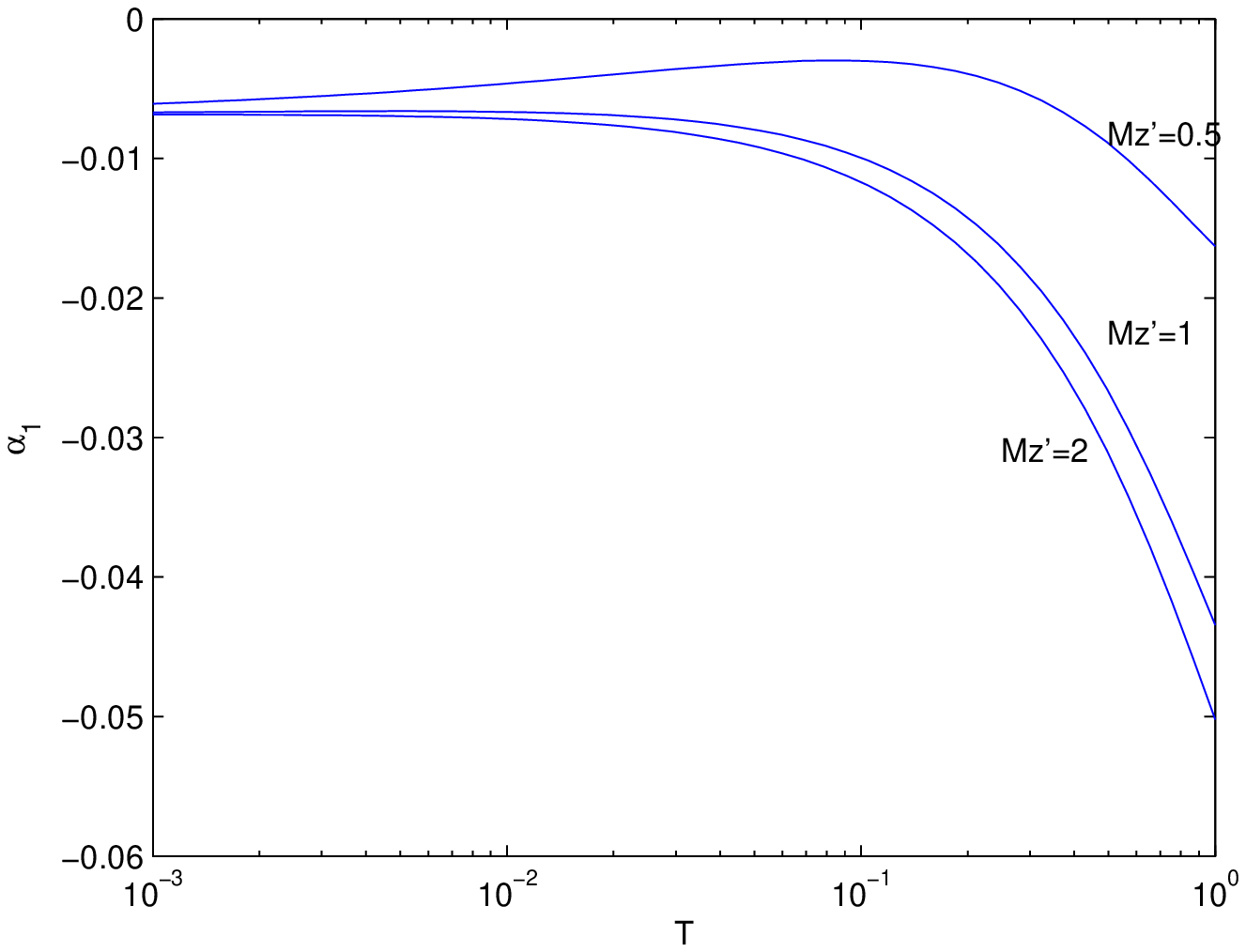}}
\subfloat[$\alpha_2$]{%
    \label{fig-alpha-2}
    \includegraphics[scale=0.6]{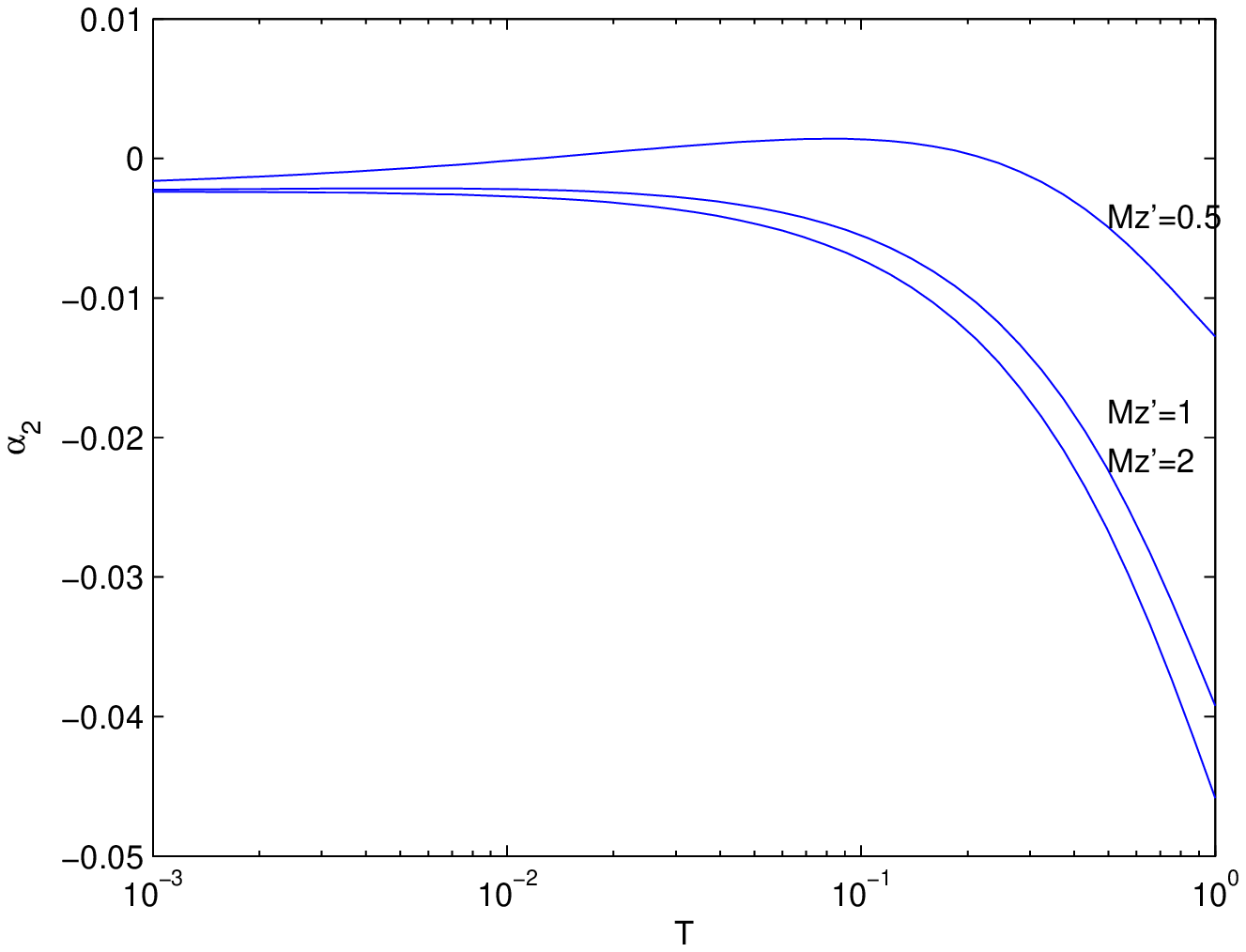}}
\end{minipage}
\begin{minipage}[t]{\textwidth}
\subfloat[$\alpha_3$]{%
    \label{fig-alpha-3}
    \includegraphics[scale=0.6]{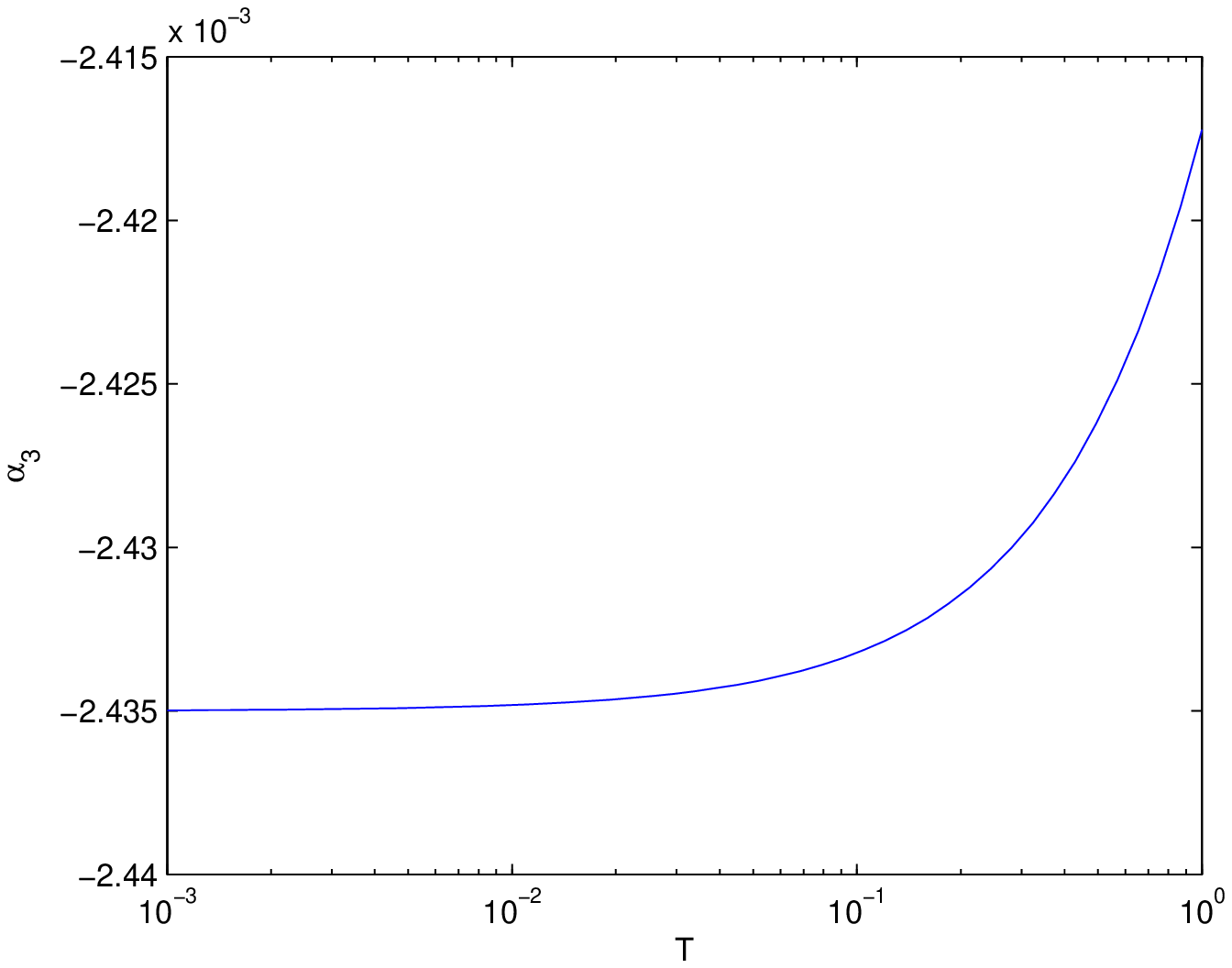}}
\subfloat[$\alpha_4$]{%
    \label{fig-alpha-4}
    \includegraphics[scale=0.6]{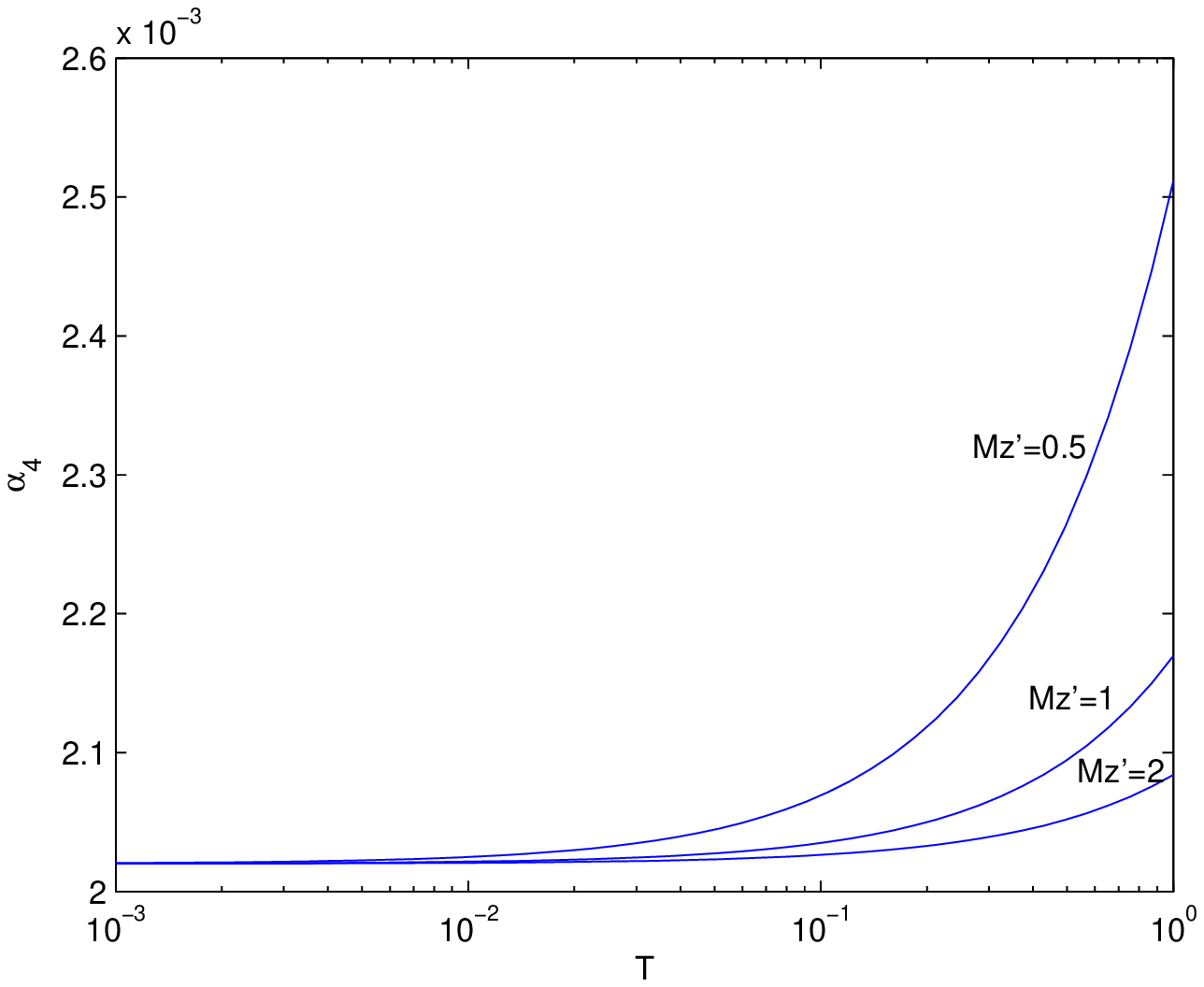}}
\end{minipage}\end{figure}

The $p^4$ order coefficients can be read out from derived EWCL
(\ref{Seffresult}) and (\ref{DeltaSZ'}), we list down the results as
following,
\begin{eqnarray}
&&\alpha_1=(1-2\beta_1)L_{10}^\mathrm{1D}+\frac{(F_0^{\rm
TC2})^2}{2M_{Z'}^2}\beta_1-2\gamma\beta_1\cot\theta^\prime\nonumber\\
&&\alpha_2=-\frac{1}{2}(1-2\beta_1)L_9^\mathrm{1D}-\frac{(F_0^{\rm
TC2})^2}{2M_{Z'}^2}\beta_1-2\gamma\beta_1\cot\theta^\prime
\nonumber\\
&&\alpha_3=-\frac{1}{2}(1-2\beta_1)L_9^\mathrm{1D}\hspace{2cm}
\alpha_4=L_2^\mathrm{1D}+\frac{(F_0^{\rm
TC2})^2}{2M_{Z'}^2}\beta_1+2\beta_1L_9^\mathrm{1D}\nonumber\\
&&\alpha_5=
L_1^\mathrm{1D}+\frac{L_3^\mathrm{1D}}{2}-\frac{(F_0^{\rm
TC2})^2}{2M_{Z'}^2}\beta_1-2\beta_1L_9^\mathrm{1D}\nonumber\\
&&\alpha_6=-\frac{(F_0^{\rm
TC2})^2}{2M_{Z'}^2}\beta_1+4\beta_1^2L_2^{\rm 1D}-4\beta_1(L_2^{\rm
1D}+\frac{L_9^{\rm
1D}}{2})\nonumber\\
&&\alpha_7=\beta_1\frac{(F_0^{\rm
TC2})^2}{2M_{Z'}^2}+2(\beta_1^2-\beta_1)(2L_1^{\rm 1D}+L_3^{\rm
1D})+2\beta_1L_9^{\rm 1D}\nonumber\\
&& \alpha_8=-\beta_1\frac{(F_0^{\rm
TC2})^2}{2M_{Z'}^2}+\beta_1L_{10}^\mathrm{1D}\nonumber\\
&&\alpha_9=-\beta_1\frac{(F_0^{\rm
TC2})^2}{2M_{Z'}^2}+2\beta_1(-L_9^{\rm
1D}+L_{10}^\mathrm{1D})\label{p4result}\\
&&\alpha_{10}=(4\beta_1^2-8\beta_1^3)(L_1^{\rm
1D}+L_2^{\rm 1D}+\frac{L_3^{\rm 1D}}{2})+ 16\beta_1^4g_{4Z}\cot^4\theta'\nonumber\\
&&\alpha_{11}=\alpha_{12}=\alpha_{13}=\alpha_{14}=0\nonumber
\end{eqnarray}
Several features of this result are:
\begin{enumerate}
\item Except the part of one-doublet technicolor model result, all
corrections from $Z'$ particle are at least proportional to
$\beta_1$ which vanish if the mixing disappear by $\theta'=0$.
\item Since $L_{10}^\mathrm{1D}<0$, therefore (\ref{p4result}) tells
us $\alpha_8$ is negative and then $U=-16\pi\alpha_8$ is always
positive in this model.
\item Except $\alpha_1$, $\alpha_2$ and $\alpha_{10}$, all other
coefficients are determined by one-doublet technicolor model
coefficients given in Table VI and two other parameter $\beta_1$ and
$F_0^\mathrm{TC2}/M_{Z'}$.
\item $\alpha_{10}$ further depend on parameter $g_{4Z}$ which from
(\ref{g4Zdef}) further depend on $\mathcal{K}_3^{\rm TC1,\Sigma\neq
0}+\mathcal{K}_4^{\rm TC1,\Sigma\neq 0}$ which are already given by
(\ref{KTC1}) and Table III.
\item $\alpha_1$ and $\alpha_2$ further depend on $\gamma$ which
from (\ref{gammaDef}) further rely on an extra parameter
$\mathcal{K}$. We can combine (\ref{beta1-0}) and (\ref{cpDef})
together to fix $\mathcal{K}$,
\begin{eqnarray}
\frac{(F_0^{\rm TC2})^2}{8\beta_1M_{Z'}^2}\tan^2\theta^\prime &=&
\frac{1}{g_1^2}
+\mathcal{K}(\frac{3}{2}\cot^2\theta^\prime+\frac{5}{18}\tan^2\theta^\prime)
+\frac{3}{4}[\mathcal{K}_2^{\rm TC1,\Sigma\neq
0}(\cot\theta^\prime+\tan\theta^\prime)^2\nonumber\\
&&+\mathcal{K}_{13}^{{\rm TC1},\Sigma\neq 0}(\cot\theta^\prime
-\tan\theta^\prime )^2]-\frac{2}{9}(L_{10}^{\rm 1D}+11H_1^{\rm
1D})\tan^2\theta^\prime\nonumber
\end{eqnarray}
Once $\mathcal{K}$ is fixed, with help of (\ref{kappaDef}), we can
determine the ratio of infrared cutoff $\kappa$ and ultraviolet
cutoff $\Lambda$, in Fig.\ref{fig-cutoff}(a), we draw the
$\kappa/\Lambda$ as function of $T$ and $M_{Z'}$, we find our
calculation do produce very large hierarchy and we further find not
all $T$ and $M_{Z'}$ region is available if we consider the natural
criteria $\Lambda>\kappa$. This criteria leads constraints that as
long as $Z'$ mass become large, the allowed range for $T$ parameter
become smaller and smaller approaching to zero, for example,
$T<0.37$ for $M_{Z'}=0.5$TeV, $T<0.0223$ for $M_{Z'}=1$TeV and
$T<0.004$ for $M_{Z'}=2$TeV. In Fig.\ref{fig-cutoff}(b), we draw
$Z'$ mass as function of $T$ parameter and $\kappa/\Lambda$. The
line of $\kappa/\Lambda=1$ give the upper bound of $Z'$ mass. The
upper bound of $Z'$ mass depend on value of $T$ parameter, the
smaller the $T$, the larger the upper bound of $M_{Z'}$.

\begin{figure}[t]
\caption{(a). The ratio of infrared cutoff and ultraviolet cutoff
$\kappa/\Lambda$ as function of $T$ parameter and $Z'$ mass in unit
of TeV.  (b). $Z'$ mass in unit of TeV as function of $T$ parameter
and $\kappa/\Lambda$.} \label{fig-cutoff} \centering
\begin{minipage}[t]{\textwidth}
\subfloat[$\kappa/\Lambda$]{%
    \label{fig-kappa-Lambda}
    \includegraphics[scale=0.6]{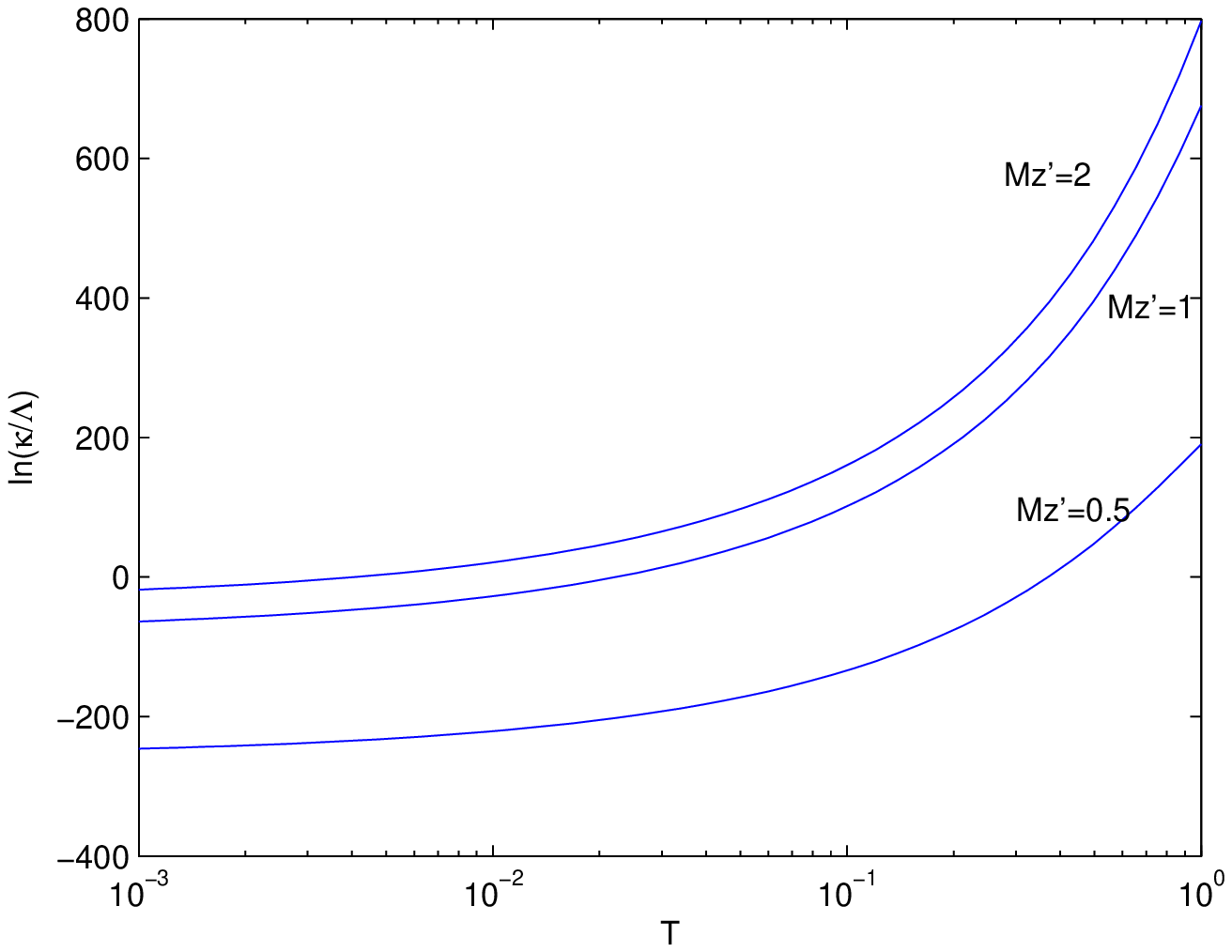}}
\subfloat[$M_{Z'}$]{%
    \label{fig-MZ'}
    \includegraphics[scale=0.6]{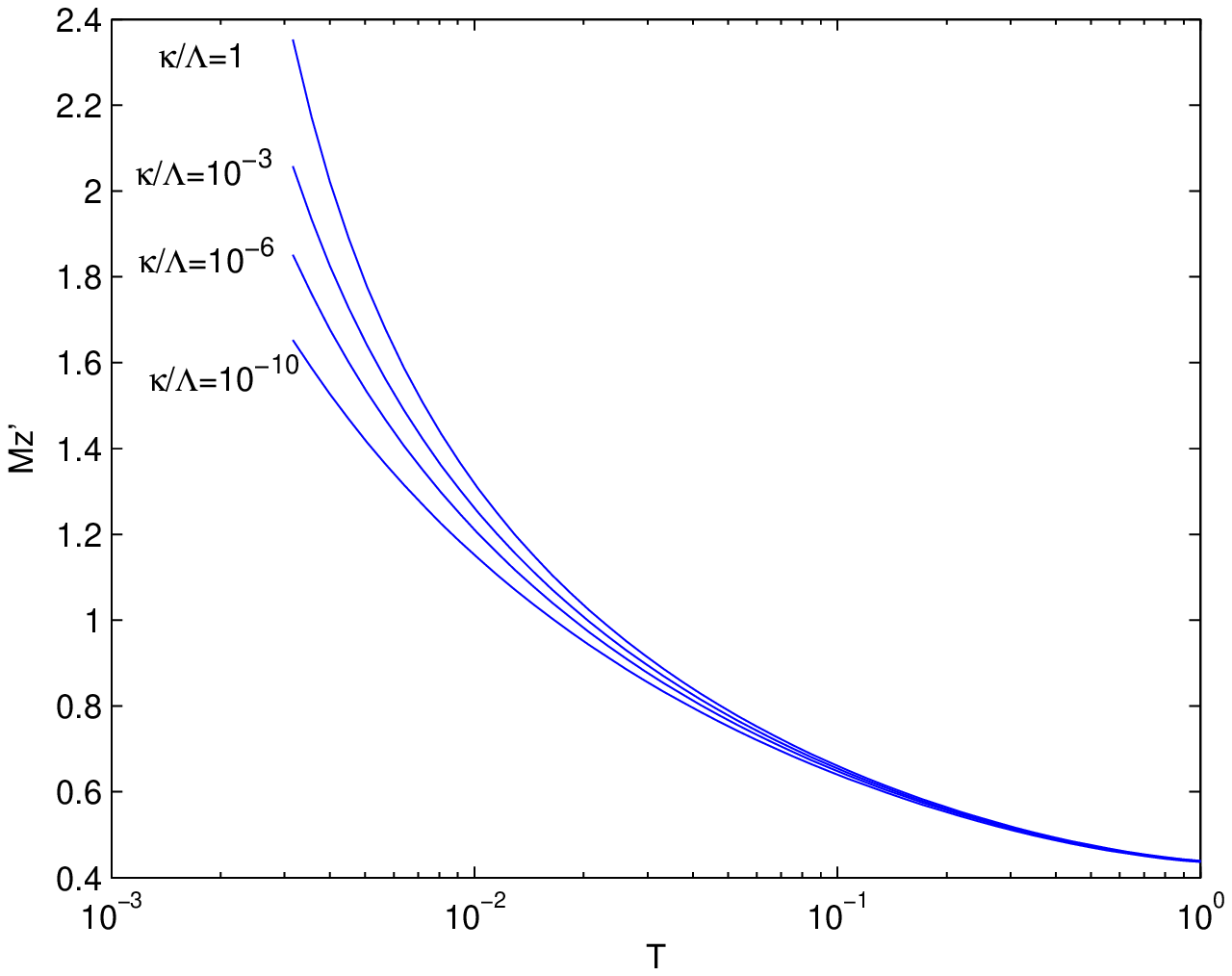}}
\end{minipage}
\end{figure}

\item For fixed $M_{Z'}$, there exists a special $\theta'$ value which
maximizes $\alpha_1$. The parameter, $S=-16\pi\alpha_1$, is of
special importance in new physics search, in Fig.\ref{fig-MinS}, we
draw a graph of minimal $S$ parameter with different $T$ parameter.
We see that if the $Z'$ mass is low enough, say $M_{Z'}<0.441$ TeV
or $T>0.176$, $S$ will become negative.
\begin{figure}[t]
\caption{The dashed line is the minimal $S$ parameter in topcolor
assisted technicolor model for different $T$. The solid lines are
the isolines for different choices of $Z'$ mass in units of TeV.}
\label{fig-MinS} \centering
\begin{minipage}[t]{\textwidth}
    \includegraphics[scale=0.6]{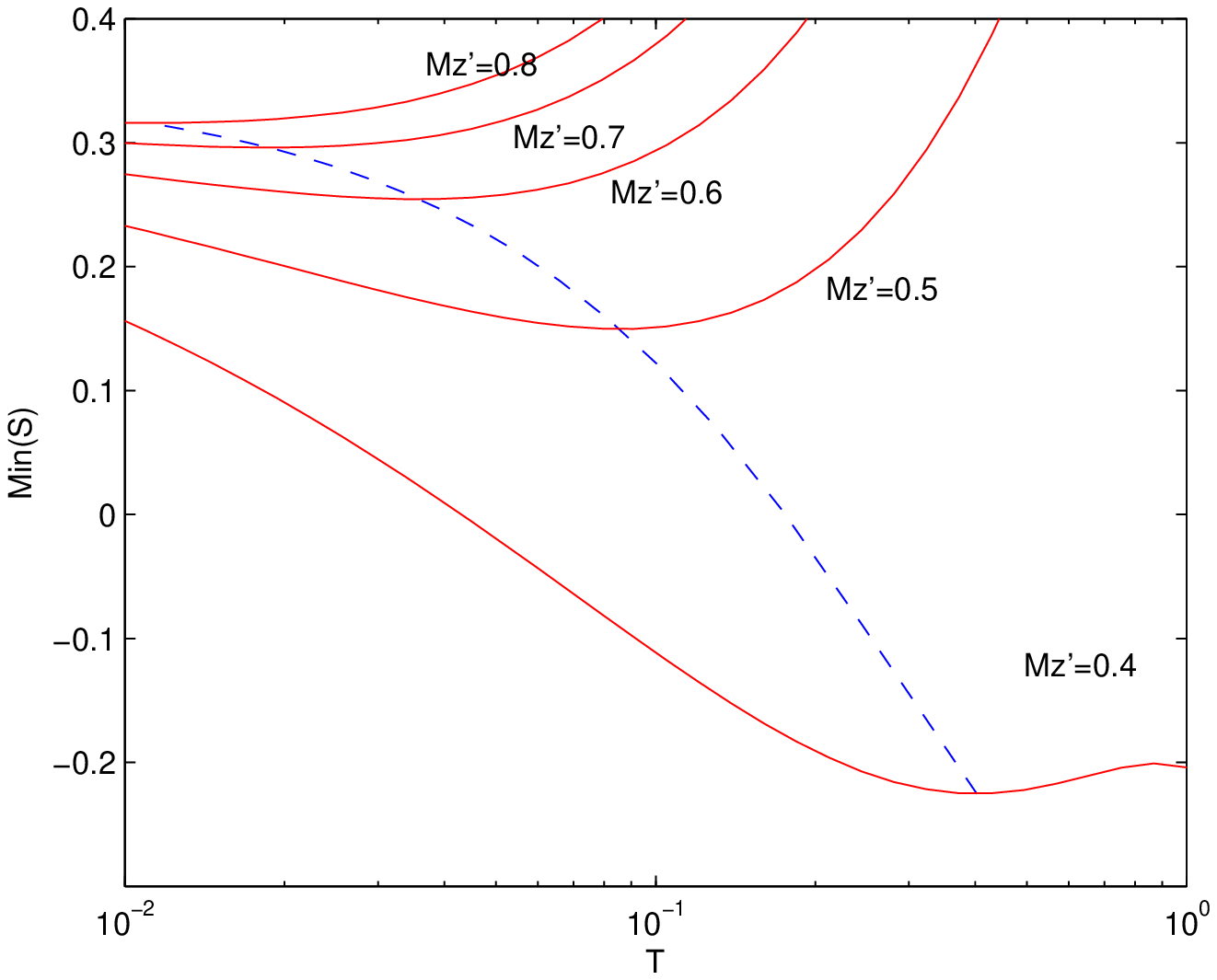}
\end{minipage}
\end{figure}
\end{enumerate}
Since we already know $F_0^\mathrm{TC2}=250$GeV, therefore all EWCL
coefficients depend on two physical parameters $\beta_1$ and
$M_{Z'}$. Combined with $\alpha T=2\beta_1$, we can use the present
experimental result for the $T$ parameter to fix $\beta_1$. In
Fig.\ref{fig-S-U}, we draw graphs for the $S$ and $U$ parameters in
terms of the $T$ parameter.
We take three typical $Z'$ masses $M_{Z'}=0.5,~1,~2$ TeV for
references. In Fig.\ref{fig-alpha-i}, we draw graphs for all $p^4$
order nonzero coefficients in terms of the $T$ parameter. Where for
$\alpha_3$ and $\alpha_{10}$, we only draw one line for each of
them, since they are independent of the $Z'$ mass.

\addtocounter{figure}{-1}
\begin{figure}[t]
\caption{}\centering
\begin{minipage}[t]{\textwidth}
\subfloat[$\alpha_5$]{%
    \label{fig-alpha-5}
    \includegraphics[scale=0.6]{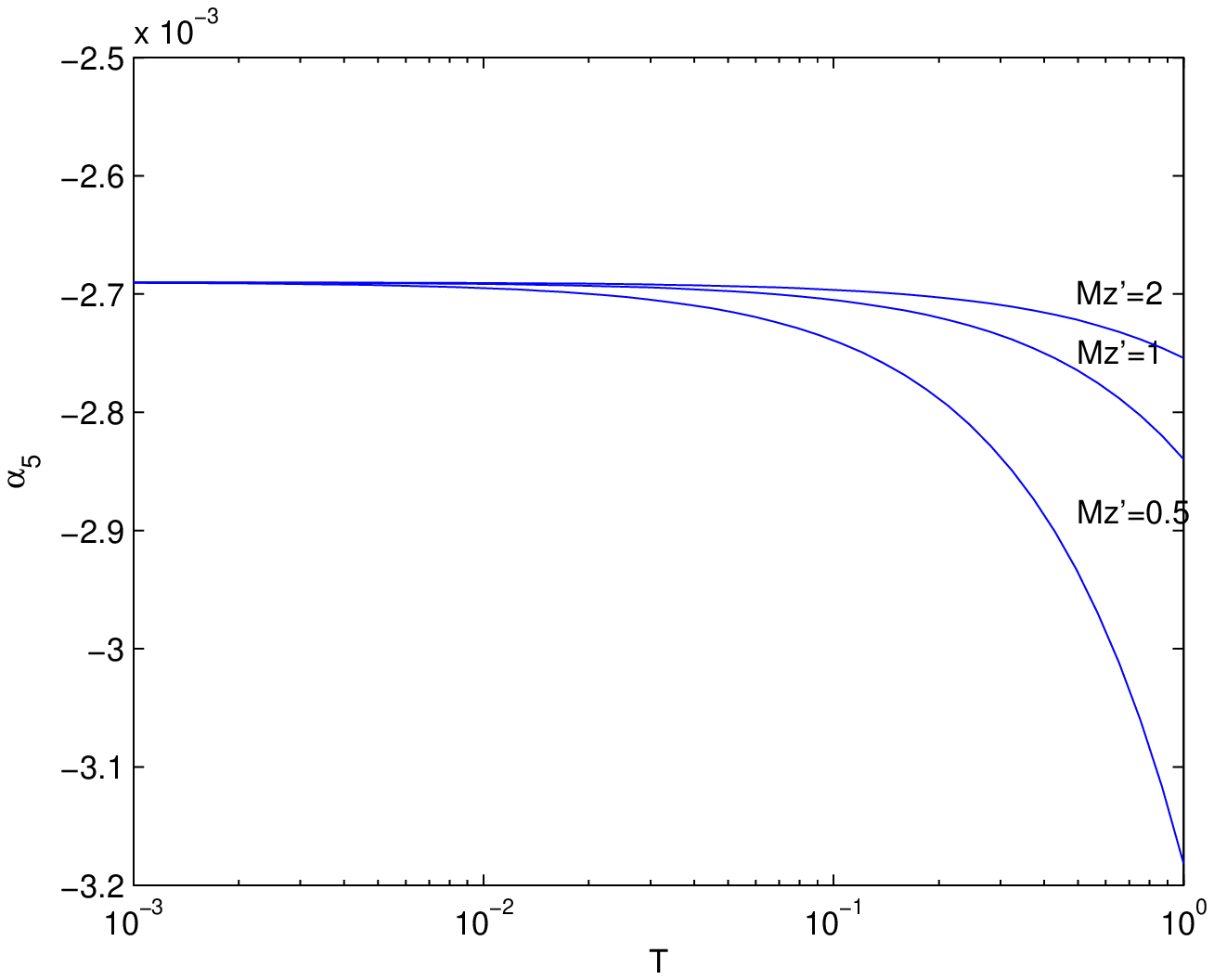}}
\subfloat[$\alpha_6$]{%
    \label{fig-alpha-6}
    \includegraphics[scale=0.6]{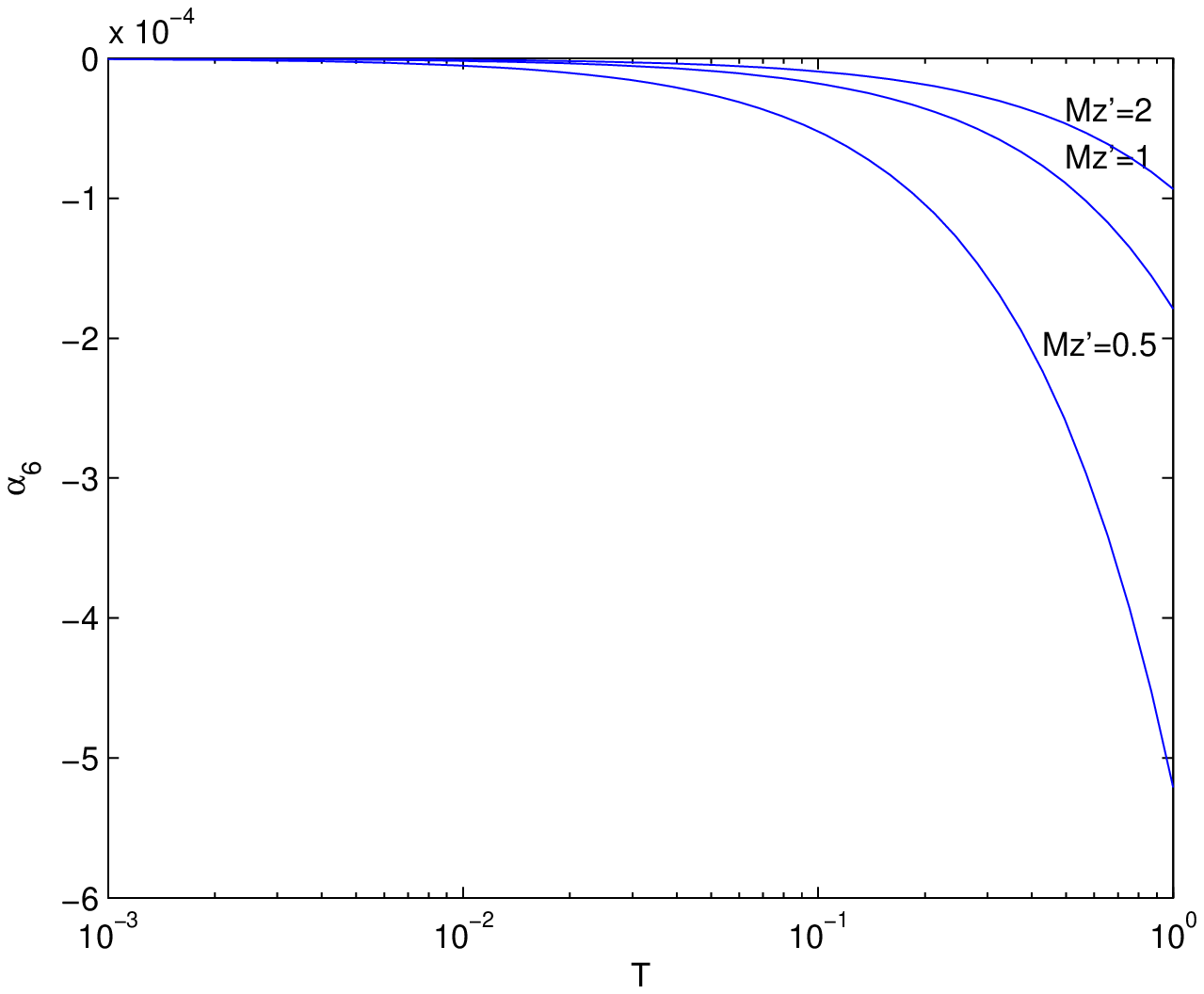}}
\end{minipage}
\begin{minipage}[t]{\textwidth}
\subfloat[$\alpha_7$]{%
    \label{fig-alpha-7}
    \includegraphics[scale=0.6]{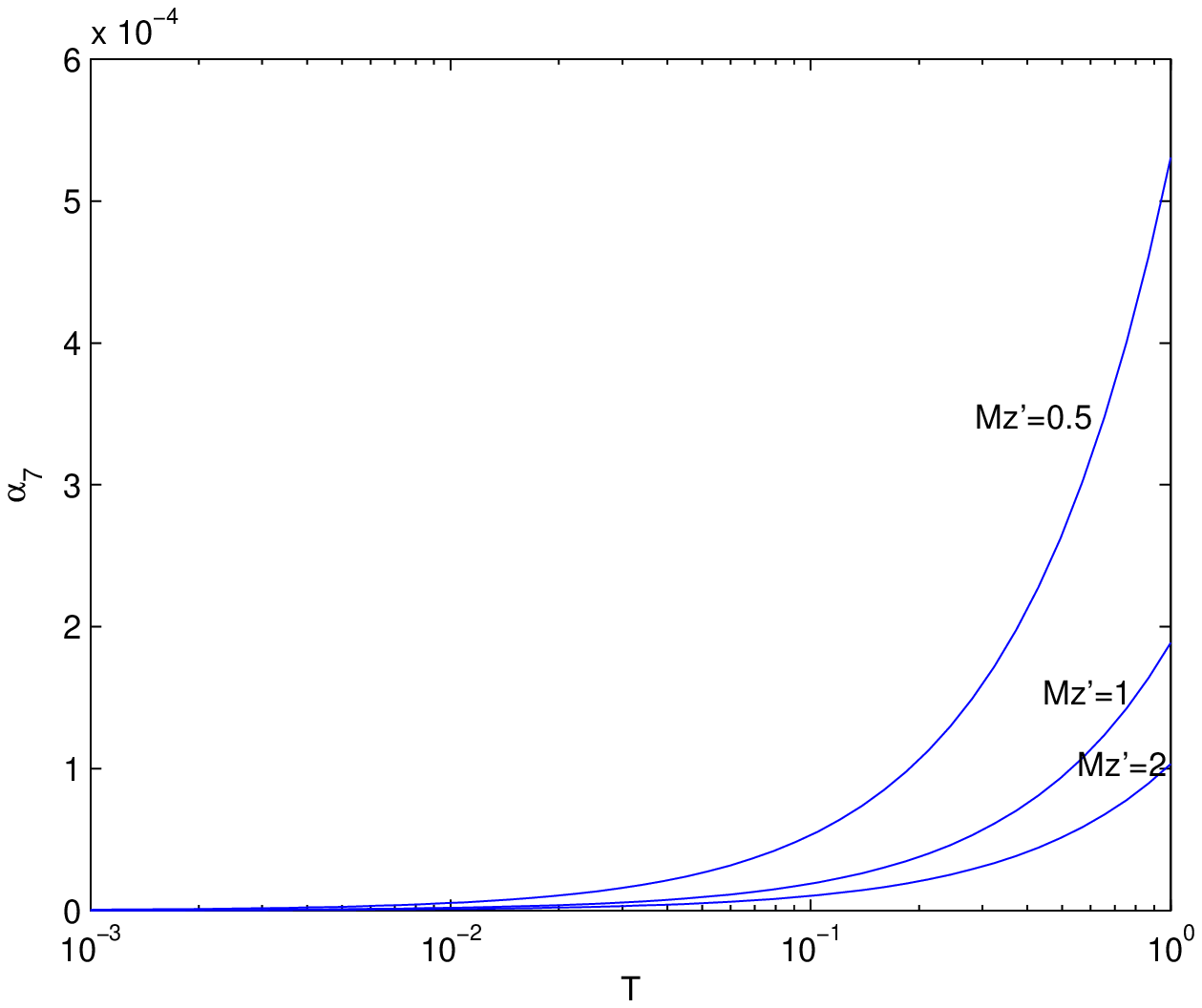}}
\subfloat[$\alpha_8$]{%
    \label{fig-alpha-8}
    \includegraphics[scale=0.6]{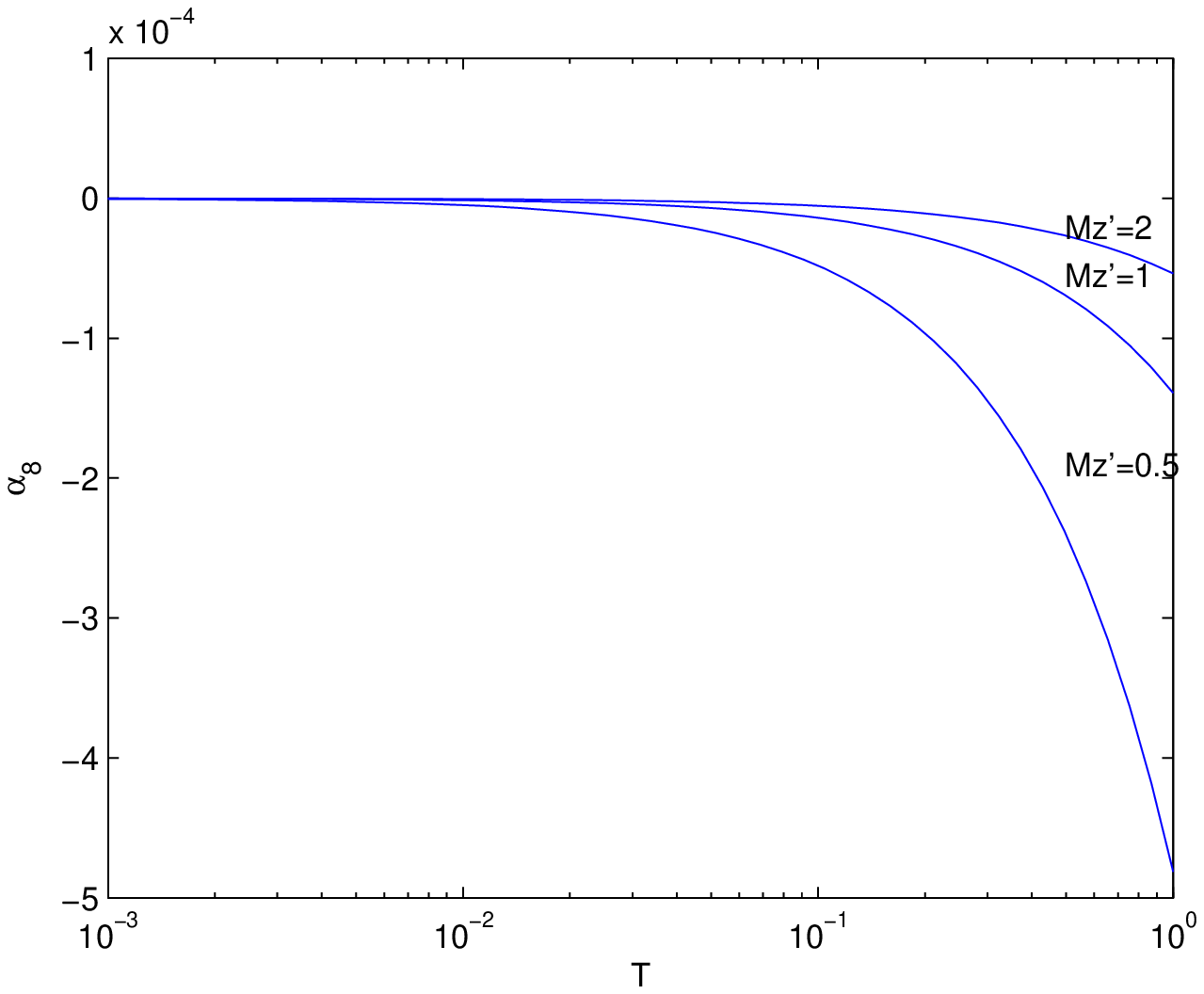}}
\end{minipage}\end{figure}

\addtocounter{figure}{-1}
\begin{figure}[t]
\caption{}\centering
\begin{minipage}[t]{\textwidth}
\subfloat[$\alpha_9$]{%
    \label{fig-alpha-9}
    \includegraphics[scale=0.6]{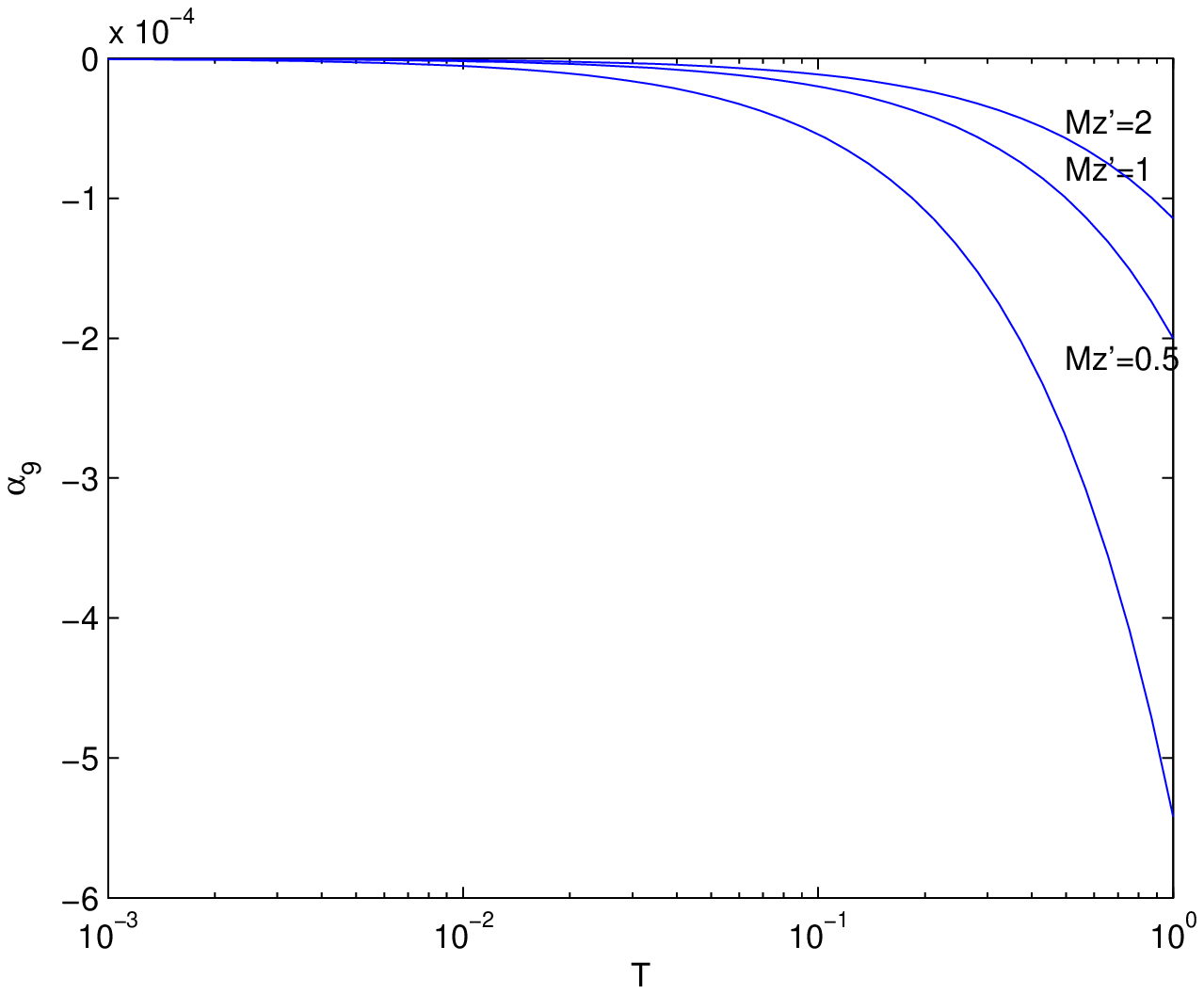}}
\subfloat[$\alpha_{10}$]{%
    \label{fig-alpha-10}
    \includegraphics[scale=0.6]{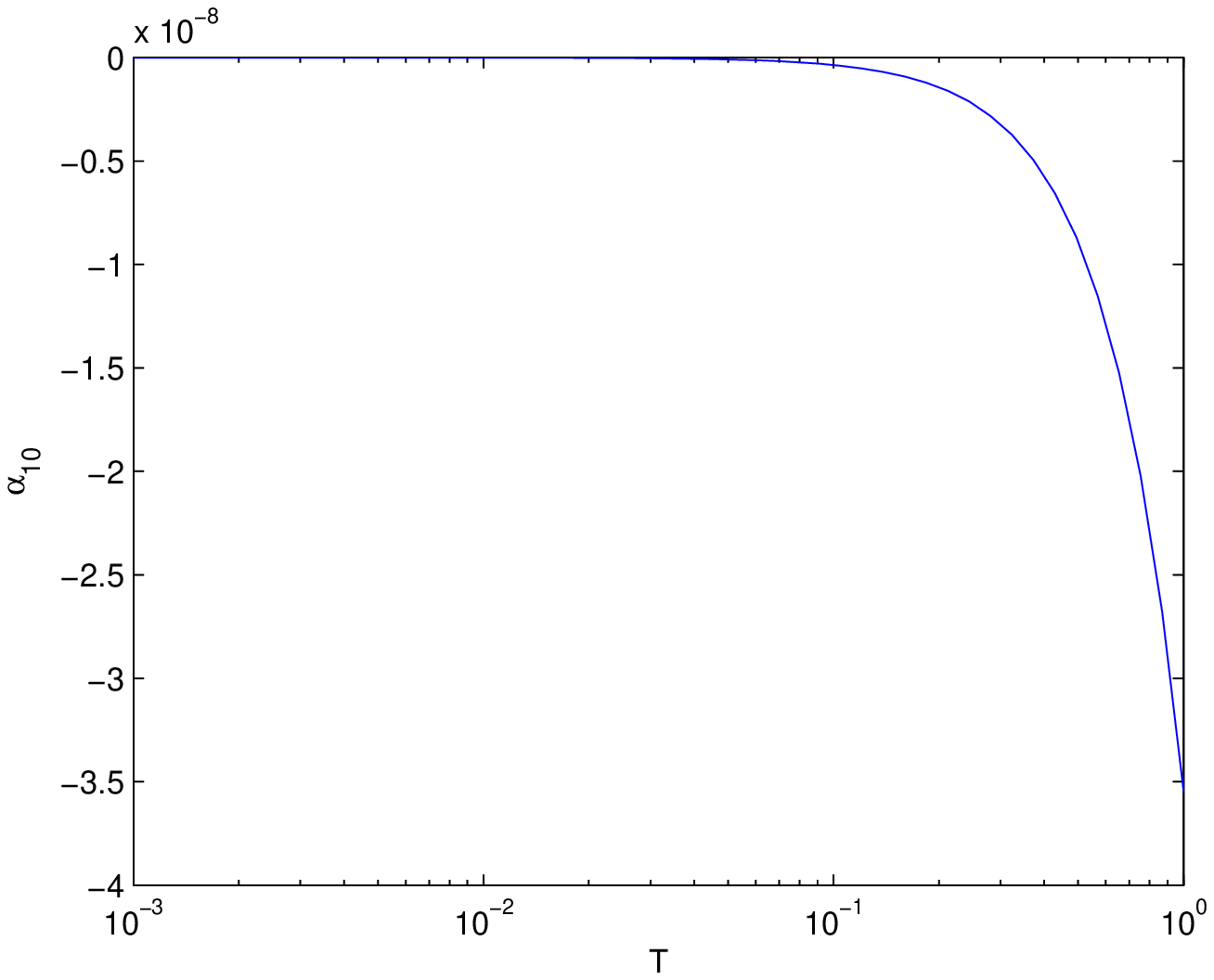}}
\end{minipage}
\end{figure}

\section{Conclusion}

In this paper, we have set up a formulation to perform the dynamical
computation of the bosonic part of EWCL for the one-doublet and
topcolor-assisted technicolor models. The one-doublet technicolor
model as the earliest and simplest dynamical symmetry breaking model
are taken as the trial model to test our formulation. We find our
formulation recovers standard scaling-up results.  The
topcolor-assisted technicolor model is the main model we handle in
this paper. We have computed its TC1 dynamics in detail and verify
the dynamical symmetry breaking of the theory, TC1 interaction will
induce effective interactions among colorons and $Z'$ which are
characterized by a divergent constant $\mathcal{K}$, a dimensional
constant $F_0^\mathrm{TC1}$ and a series of dimensionless QCD
constants $L_1,L_3,L_9,L_{10},H_1$. For TC2 dynamics, it will induce
effective interaction for $Z'$, electroweak gauge fields and their
goldstone bosons. Due to its similarity with QCD, we use
Gasser-Leutwyler prescription to describe its low energy effects in
terms of low energy effective Lagrangian with a divergent constant
$\mathcal{K}$, dimensional constant $F_0^\mathrm{TC2}$ and a series
of dimensionless constants
$L_1^\mathrm{1D},L_2^\mathrm{1D},L_3^\mathrm{1D},L_9^\mathrm{1D},
L_{10}^\mathrm{1D},H_1^\mathrm{1D}$ the same as those in one-doublet
technicolor model. Due to dynamics similarities between TC2 and QCD,
TC2 interaction make a direct contributions to EWCL coefficients a
part which is the same as that of one-doublet technicolor model.
Further corrections are from effective interactions among colorons,
$Z'$ and ordinary quarks induced by TC1 and TC2 interactions. We
have shown that colorons make no contributions to EWCL coefficients
within the approximations we have made in this paper, while ordinary
quark are ignored in this paper for future investigations. In fact,
for some special EWCL coefficients, such as $S=-16\pi\alpha_1$,
$\alpha T=\beta_1$ and $U=-16\pi\alpha_8$ parameters, general
fermion contributions to them are already calculated
\cite{Sparameter}, $S,T,U$ and triple-gauge-vertices from a heavy
non-degenerate fermion doublet has been estimated in
ref.\cite{STU,Tanabashi}. One can based on these general results to
estimate possible contributions to some of EWCL coefficients. For
topcolor-assisted technicolor model in this paper, the main work is
to estimate the effects of $Z'$ particle. Our computation shows that
contributions from $Z'$ particle are at least proportional to
$\beta_1$ and then vanish if $\beta_1$ is zero. One typical feature
of the model is the positivity and bounded above of $\beta_1$
parameter which means the $T$ parameter must vary in the range
$0\sim 1/(25\alpha)$  and the positive $U$ parameter. If we consider
the natural criteria $\Lambda>\kappa$ which will further constraints
the allowed range for $T$ parameter approaching to zero as long as
$Z'$ mass become large, for example, $T<0.37$ for $M_{Z'}=0.5$TeV,
$T<0.0223$ for $M_{Z'}=1$TeV and $T<0.004$ for $M_{Z'}=2$TeV. For
$S$ parameter, it can be either positive and negative depending on
the $Z'$ mass is large or small. As long as $M_{Z'}<0.441$ TeV or
$T>0.176$, we may find negative $S$. There exist a upper bound for
the mass of $Z'$ which is depend on value of $T$ parameter, the
smaller the $T$, the larger the upper bound of $M_{Z'}$. Except
$U(1)_Y$ coupling $g_1$ and coefficients determined in one-doublet
technicolor model and QCD, all EWCL coefficients rely on
experimental $T$ parameter and coloron mass $M_{Z'}$. We have taken
typical values of $M_{Z'}$ and vary $T$ parameter to estimate all
EWCL coefficients up to order of $p^4$. Further works on matter part
of EWCL and computing EWCL coefficients for other dynamical symmetry
breaking new physics models are in progress and will be reported
elsewhere.

\section*{Acknowledgments}

We would like to thank Y.~P.~Kuang, H.~J.~He and J.~K.~Parry for
helpful discussions. This work was  supported by National  Science
Foundation of China (NSFC) under Grant No. 10435040 and Specialized
Research Fund for the Doctoral Program of High Education of China.

\appendix
\section{Two Equivalent EWCL Formalisms}

The electroweak chiral Lagrangian is constructed using a
dimensionless unitary unimodular $2\times2$ matrix field $U(x)$,
 In Ref.\cite{EWCL}, the electroweak chiral Lagrangian has
been constructed with the building blocks which are $SU(2)_L$
covariant and $U(1)_Y$ invariant as:
\begin{eqnarray}
T\equiv U\tau^3U^\dag\;,\qquad V_\mu\equiv(D_\mu U)U^\dag\;, \qquad
g_1B_{\mu\nu}\;,\qquad g_2W_{\mu\nu}\equiv
g_2\frac{\tau^a}{2}W_{\mu\nu}^a\;.
\end{eqnarray}
Alternatively, we reformulate the electroweak chiral Lagrangian
equivalently with $SU(2)_L$ invariant and $U(1)_Y$ covariant
building blocks as:
\begin{eqnarray}
\tau^3\;,\qquad X_\mu\equiv U^\dag(D_\mu U)\;,\qquad
g_1B_{\mu\nu}\;,\qquad\overline{W}_{\mu\nu}\equiv U^\dag
g_2W_{\mu\nu}U\;,
\end{eqnarray}
among which, $\tau^3$ and $g_1B_{\mu\nu}$ are both $SU(2)_L$ and
$U(1)_Y$ invariant, while $X_\mu$ and $\overline{W}_{\mu\nu}$ are
bilinearly $U(1)_Y$ covariant. This second formulation is largely
used throughout this paper. We list down the corresponding relations
of the two formalisms.

\begin{table}[ht]\caption{The Symmetry Breaking Sector of the Electroweak Chiral Lagrangian}
\label{table-ewcl}
\begin{tabular}{*{3}{|l}|}
\hline

& Formulation I & Formulation II\\

\hline \hline

${\cal L}^{(2)}$ & $\frac{1}{4}f^2{\rm tr}[(D_\mu U^\dag) (D^\mu
U)]=-\frac{1}{4}f^2{\rm tr}(V_\mu V^\mu)$ &
$-\frac{1}{4}f^2{\rm tr}(X_\mu X^\mu)$\\

\hline

${\cal L}^{(2)\prime}$ & $\frac{1}{4}\beta_1f^2[{\rm tr}(TV_\mu)]^2$
& $\frac{1}{4}\beta_1f^2[{\rm tr}(\tau^3
X_\mu)]^2$\\

\hline

${\cal L}_1$ & $\frac{1}{2}\alpha_1g_2g_1 B_{\mu\nu}{\rm
tr}(TW^{\mu\nu})$ & $\frac{1}{2}\alpha_1g_1 B_{\mu\nu}{\rm
tr}(\tau^3\overline{W}^{\mu\nu})$\\

\hline

${\cal L}_2$ & $\frac{1}{2}i\alpha_2g_1 B_{\mu\nu}{\rm
tr}(T[V^\mu,V^\nu])$ & $i\alpha_2g_1 B_{\mu\nu}{\rm
tr}(\tau^3X^\mu X^\nu)$\\

\hline

${\cal L}_3$ & $i\alpha_3g_2{\rm tr}(W_{\mu\nu}[V^\mu,V^\nu])$ &
$2i\alpha_3{\rm
tr}(\overline{W}_{\mu\nu}X^\mu X^\nu)$\\

\hline

${\cal L}_4$ & $\alpha_4[{\rm tr}(V_\mu V_\nu)]^2$ &
$\alpha_4[{\rm tr}(X_\mu X_\nu)]^2$\\

\hline

${\cal L}_5$ & $\alpha_5[{\rm tr}(V_\mu V^\mu)]^2$ &
$\alpha_5[{\rm tr}(X_\mu X^\mu)]^2$\\

\hline

${\cal L}_6$ & $\alpha_6{\rm tr}(V_\mu V_\nu){\rm tr}(TV^\mu){\rm
tr}(TV^\nu)$ & $\alpha_6{\rm tr}(X_\mu X_\nu){\rm tr}(\tau^3
X^\mu){\rm
tr}(\tau^3 X^\nu)$\\

\hline

${\cal L}_7$ & $\alpha_7{\rm tr}(V_\mu V^\mu){\rm tr}(TV_\nu){\rm
tr}(TV^\nu)$ & $\alpha_7{\rm tr}(X_\mu X^\mu){\rm
tr}(\tau^3X_\nu){\rm
tr}(\tau^3X^\nu)$\\

\hline

${\cal L}_8$ & $\frac{1}{4}\alpha_8g_2^2[{\rm tr}(TW_{\mu\nu})]^2$ &
$\frac{1}{4}\alpha_8[{\rm
tr}(\tau^3\overline{W}_{\mu\nu})]^2$\\

\hline

${\cal L}_9$ & $\frac{1}{2}i\alpha_9g_2{\rm tr}(TW_{\mu\nu}){\rm
tr}(T[V^\mu,V^\nu])$ & $i\alpha_9{\rm
tr}(\tau^3\overline{W}_{\mu\nu}){\rm tr}(\tau^3X^\mu X^\nu)$\\

\hline

${\cal L}_{10}$ & $\frac{1}{2}\alpha_{10}[{\rm tr}(TV_\mu){\rm
tr}(TV_\nu)]^2$ & $\frac{1}{2}\alpha_{10}[{\rm tr}(\tau^3X_\mu){\rm
tr}(\tau^3X_\nu)]^2$\\

\hline

${\cal L}_{11}$ & $\alpha_{11}g_2\epsilon^{\mu\nu\rho\lambda}{\rm
tr}(TV_\mu){\rm tr}(V_\nu W_{\rho\lambda})$ &
$\alpha_{11}\epsilon^{\mu\nu\rho\lambda}{\rm
tr}(\tau^3X_\mu){\rm tr}(X_\nu \overline{W}_{\rho\lambda})$\\

\hline

${\cal L}_{12}$ & $\alpha_{12}g_2{\rm tr}(TV_\mu){\rm tr}(V_\nu
W^{\mu\nu})$ & $\alpha_{12}{\rm tr}(\tau^3X_\mu){\rm tr}(X_\nu
\overline{W}^{\mu\nu})$\\

\hline

${\cal L}_{13}$ &
$\alpha_{13}g_2g_1\epsilon^{\mu\nu\rho\sigma}B_{\mu\nu}{\rm
tr}(TW_{\rho\sigma})$ &
$\alpha_{13}\epsilon^{\mu\nu\rho\sigma}g_1B_{\mu\nu}{\rm
tr}(\tau^3\overline{W}_{\rho\sigma})$\\

\hline

${\cal L}_{14}$ & $\alpha_{14}g_2^2\epsilon^{\mu\nu\rho\sigma}{\rm
tr}(TW_{\mu\nu}){\rm tr}(TW_{\rho\sigma})$ &
$\alpha_{14}\epsilon^{\mu\nu\rho\sigma}{\rm
tr}(\tau^3\overline{W}_{\mu\nu}){\rm
tr}(\tau^3\overline{W}_{\rho\sigma})$\\

\hline
\end{tabular}
\end{table}
\section{$S_{\mathrm{TC1}}^{(4c,B^A)}[B^A]$,
 $S_{\mathrm{TC1}}^{(4c,B)}[B]$,
 $S_{\mathrm{TC1}}^{(4c,Z')}$,
 $S_{\mathrm{TC1}}^{(4c,B^AZ')}[B^A,Z']$,
$S_{\mathrm{TC1}}^{(4c,BZ')}[B,Z']$}

\begin{eqnarray}
S_{\mathrm{TC1}}^{(4c,B^A)}[B^A]&=&\int
d^4x\bigg[-\mathcal{K}_1^{\rm TC1,\Sigma\neq
0}\frac{g_3^2}{8}(\cot\theta+\tan\theta)^2(\partial_{\mu}B^{A,\mu})^2\nonumber\\
&&-\mathcal{K}_2^{\rm TC1,\Sigma\neq
0}\frac{g_3^2}{8}(\cot\theta+\tan\theta)^2B^A_{a,\mu\nu}B_a^{A,\mu\nu}\nonumber\\
&&+\mathcal{K}_3^{\rm TC1,\Sigma\neq
0}[\frac{g_3^4}{192}(\cot\theta+\tan\theta)^4(B^A_{\mu}B^{A,\mu})^2
+\frac{g_3^4}{128}(\cot\theta+\tan\theta)^4(d^{ABC}B^B_{\mu}B^{C,\mu})^2]\nonumber\\
&&+\mathcal{K}_4^{\rm TC1,\Sigma\neq 0}\{
\frac{g_3^4}{192}(\cot\theta+\tan\theta)^4(B^A_{\mu}B^A_{\nu})^2\nonumber\\
&&+\frac{g_3^4}{128}(\cot\theta+\tan\theta)^4[(if^{ABC}+d^{ABC})B^B_{\mu}B^C_{\nu}]^2\}\nonumber\\
&&-\mathcal{K}_{13}^{{\rm TC1},\Sigma\neq
0}\frac{g_3^2}{8}(\cot\theta-\tan\theta)^2B^A_{v,\mu\nu}B_v^{A,\mu\nu}\nonumber\\
&&+\mathcal{K}_{14}^{{\rm TC1},\Sigma\neq 0}
\frac{g_3^3}{32}(\cot\theta-\tan\theta)(\cot\theta+\tan\theta)^2B_v^{A,\mu\nu}f^{ABC}
B^B_{\mu}B^C_{\nu}\bigg]\;,\label{B1}
\end{eqnarray}
\begin{eqnarray}
S_{\mathrm{TC1}}^{(4c,B)}[B]&=&\int d^4x\bigg[
-\mathcal{K}_{13}^{{\rm TC1},\Sigma\neq
0}\frac{3g_1^2}{4}B_{\mu\nu}B^{\mu\nu}\bigg]\;,\label{B2}
\end{eqnarray}
\begin{eqnarray}
S_{\mathrm{TC1}}^{(4c,Z')}&=&\int d^4x~\bigg[ -\mathcal{K}_1^{\rm
TC1,\Sigma\neq
0}\frac{3g_1^2}{16}(\cot\theta^\prime+\tan\theta^\prime)^2(\partial_{\mu}Z^{\prime,\mu})^2
\nonumber\\
&&+(\mathcal{K}_3^{\rm TC1,\Sigma\neq 0}+\mathcal{K}_4^{\rm
TC1,\Sigma\neq
0})\frac{3g_1^4}{256}(\cot\theta^\prime+\tan\theta^\prime)^4(Z^\prime_{\mu}Z^{\prime,\mu})^2\nonumber\\
&&-\frac{3g_1^2}{16}[\mathcal{K}_2^{\rm TC1,\Sigma\neq
0}(\cot\theta^\prime+\tan\theta^\prime)^2+\mathcal{K}_{13}^{{\rm
TC1},\Sigma\neq 0}(\cot\theta^\prime -\tan\theta^\prime
)^2]Z^\prime_{\mu\nu}Z^{\prime,\mu\nu}\bigg]\;,\label{B3}
\end{eqnarray}
\begin{eqnarray}
S_{\mathrm{TC1}}^{(4c,B^AZ')}[B^A,Z']&=&\int d^4x~\bigg[
\mathcal{K}_3^{\rm TC1,\Sigma\neq 0}
[\frac{g_1^2g_3^2}{64}(\cot\theta+\tan\theta)^2(\cot\theta^\prime+\tan\theta^\prime)^2
B^A_{\mu}B^{A,\mu}Z^\prime_{\nu}Z^{\prime,\nu}\nonumber\\
&&+\frac{g_1^2g_3^2}{32}(\cot\theta+\tan\theta)^2(\cot\theta^\prime+\tan\theta^\prime)^2(B^A_{\mu}Z^{\prime,\mu})^2\nonumber\\
&&+\frac{g_1g_3^3}{32}(\cot\theta+\tan\theta)^3(\cot\theta^\prime+\tan\theta^\prime)d^{ABC}B^B_{\mu}B^{C,\mu}
B^A_{\mu}Z^{\prime,\mu}]\nonumber\\
&&+\mathcal{K}_4^{\rm TC1,\Sigma\neq 0}\{
\frac{g_1^2g_3^2}{256}(\cot\theta+\tan\theta)^2(\cot\theta^\prime+\tan\theta^\prime)^2(B^A_{\mu}Z^{\prime,\mu})^2\nonumber\\
&&+\frac{g_1^2g_3^2}{64}(\cot\theta+\tan\theta)^2(\cot\theta^\prime+\tan\theta^\prime)^2
[B^A_{\mu}B^{A,\mu}Z^\prime_{\nu}Z^{\prime,\nu}+(B^A_{\mu}Z^{\prime,\mu})^2]\nonumber\\
&&+\frac{g_1g_3^3}{32}(\cot\theta+\tan\theta)^3(\cot\theta^\prime+\tan\theta^\prime)d^{ABC}B^B_{\mu}B^C_{\nu}
B^{A,\mu}Z^{\prime,\nu}\}\bigg]\;,\label{B4}
\end{eqnarray}
\begin{eqnarray}
S_{\mathrm{TC1}}^{(4c,BZ')}[B,Z']&=&\int d^4x\bigg[
-\mathcal{K}_{13}^{{\rm TC1},\Sigma\neq
0}\frac{g_1^2}{4}(\cot\theta^\prime -\tan\theta^\prime
)B_{\mu\nu}Z^{\prime,\mu\nu}\bigg]\;,\label{B5}
\end{eqnarray}with
$B^A_{v,\mu\nu}=\partial_{\mu}B^A_{\nu}-\partial_{\nu}B^A_{\mu}-\frac{g_3}{2}(\cot\theta-\tan\theta)
f^{ABC}B^B_{\mu}B^C_{\nu}$.
\section{$S_\mathrm{coloron}^0[B^A,Z']$ and $S_\mathrm{coloron}^\mathrm{int}[B^A,Z']$}

\begin{eqnarray}
&&\hspace{-0.5cm}S_\mathrm{coloron}^0[B^A,Z']\nonumber\\
&&\hspace{-0.5cm}=\int d^4x\bigg[\bigg(g^{\mu\nu}\{\frac{(F^{\rm
TC1}_0)^2}{8}g_3^2(\cot\theta+\tan\theta)^2+[\mathcal{K}_3^{\rm
TC1,\Sigma\neq 0}
\frac{g_1^2g_3^2}{64}(\cot\theta+\tan\theta)^2(\cot\theta^\prime+\tan\theta^\prime)^2\nonumber\\
&&+\mathcal{K}_4^{\rm TC1,\Sigma\neq
0}\frac{g_1^2g_3^2}{64}(\cot\theta+\tan\theta)^2(\cot\theta^\prime+\tan\theta^\prime)^2]
Z^\prime_{\lambda}Z^{\prime,\lambda}\}\nonumber\\
&& +[\mathcal{K}_3^{\rm TC1,\Sigma\neq 0}
\frac{g_1^2g_3^2}{32}(\cot\theta+\tan\theta)^2(\cot\theta^\prime+\tan\theta^\prime)^2\nonumber\\
&&+\mathcal{K}_4^{\rm TC1,\Sigma\neq 0}
\frac{5g_1^2g_3^2}{256}(\cot\theta+\tan\theta)^2(\cot\theta^\prime+\tan\theta^\prime)^2
]Z^{\prime,\mu}Z^{\prime,\nu}
\bigg)B^A_{\mu}B_{A,\nu}\nonumber\\
&&-\mathcal{K}_1^{\rm TC1,\Sigma\neq
0}\frac{g_3^2}{8}(\cot\theta+\tan\theta)^2(\partial_{\mu}B^{A,\mu})^2
\nonumber\\
&&[-\mathcal{K}_2^{\rm TC1,\Sigma\neq
0}\frac{g_3^2}{8}(\cot\theta+\tan\theta)^2-\mathcal{K}_{13}^{{\rm
TC1},\Sigma\neq
0}\frac{g_3^2}{8}(\cot\theta-\tan\theta)^2\nonumber\\
&&-\frac{1}{4}-\frac{g_3^2}{4}\mathcal{K}(\cot^2\theta
+\tan^2\theta)](\partial_{\mu}B_{\nu}^A-\partial_{\nu}B_{\mu}^A)(\partial^{\mu}B^{A,\nu}-\partial^{\nu}B^{A,\mu})
\bigg]\label{Scoloron0}
\end{eqnarray}
and
\begin{eqnarray}
&&\hspace{-0.5cm}S_\mathrm{coloron}^\mathrm{int}[B^A,Z']\label{ScoloronInt}\\
&&\hspace{-0.5cm}=\int d^4x\bigg[-\mathcal{K}_2^{\rm TC1,\Sigma\neq
0}\frac{g_3^2}{8}(\cot\theta+\tan\theta)^2 [
(\partial_{\mu}B^A_{\nu}-\partial_{\nu}B^A_{\mu})2g_3(-\cot\theta+\tan\theta)
f^{ABC}B^{B,\mu}B^{C,\nu}\nonumber\\
&&+g_3^2(-\cot\theta+\tan\theta)^2f^{ABC}B^B_{\mu}B^C_{\nu}f^{AB'C'}B^{B',\mu}B^{C',\nu}]
\nonumber\\
&&+\mathcal{K}_3^{\rm TC1,\Sigma\neq
0}[\frac{g_3^4}{192}(\cot\theta+\tan\theta)^4(B^A_{\mu}B^{A,\mu})^2
+\frac{g_3^4}{128}(\cot\theta+\tan\theta)^4(d^{ABC}B^B_{\mu}B^{C,\mu})^2]\nonumber\\
&&+\mathcal{K}_4^{\rm TC1,\Sigma\neq 0}\{
\frac{g_3^4}{192}(\cot\theta+\tan\theta)^4(B^A_{\mu}B^A_{\nu})^2
+\frac{g_3^4}{128}(\cot\theta+\tan\theta)^4[(if^{ABC}+d^{ABC})B^B_{\mu}B^C_{\nu}]^2\}\nonumber\\
&&-\mathcal{K}_{13}^{{\rm TC1},\Sigma\neq
0}\frac{g_3^2}{8}(\cot\theta-\tan\theta)^2 [
(\partial_{\mu}B^A_{\nu}-\partial_{\nu}B^A_{\mu})g_3(-\cot\theta+\tan\theta)
f^{ABC}B^{B,\mu}B^{C,\nu}\nonumber\\
&&+\frac{1}{4}g_3^2(-\cot\theta+\tan\theta)^2f^{ABC}B^B_{\mu}B^C_{\nu}f^{AB'C'}B^{B',\mu}B^{C',\nu}]
+\mathcal{K}_{14}^{{\rm TC1},\Sigma\neq 0}
\frac{g_3^3}{32}(\cot\theta-\tan\theta)(\cot\theta+\tan\theta)^2\nonumber\\
&&\times[\partial^{\mu}B^{A,\nu}-\partial^{\nu}B^{A,\mu}+\frac{g_3}{2}(-\cot\theta+\tan\theta)
f^{ABC}B^{B,\mu}B^{C,\nu}]f^{ABC}
B^B_{\mu}B^C_{\nu}\nonumber\\
&&+\mathcal{K}_3^{\rm TC1,\Sigma\neq 0}
\frac{g_1g_3^3}{32}(\cot\theta+\tan\theta)^3(\cot\theta^\prime+\tan\theta^\prime)d^{ABC}B^B_{\mu}B^{C,\mu}
B^A_{\nu}Z^{\prime,\nu}\nonumber\\
&&+\mathcal{K}_4^{\rm TC1,\Sigma\neq
0}\frac{g_1g_3^3}{32}(\cot\theta+\tan\theta)^3(\cot\theta^\prime+\tan\theta^\prime)d^{ABC}B^B_{\mu}B^C_{\nu}
B^{A,\mu}Z^{\prime,\nu}\nonumber\\
&&-\frac{1}{4}\cos^2\theta[2g_3(\partial_{\mu}B^A_{\nu}-\partial_{\nu}B^A_{\mu})\cot\theta
f^{ABC}B^{B,\mu}B^{C,\nu}+g_3^2\cot^2\theta
f^{ABC}B^{B,\mu}B^{C,\nu}f^{AB'C'}B^{B'}_{\mu}B^{C'}_{\nu}]\nonumber\\
&&-\frac{1}{4}\sin^2\theta
[-2g_3(\partial_{\mu}B^A_{\nu}-\partial_{\nu}B^A_{\mu})\tan\theta
f^{ABC}B^{B,\mu}B^{C,\nu}+g_3^2\tan^2\theta
f^{ABC}B^{B,\mu}B^{C,\nu}f^{AB'C'}B^{B'}_{\mu}B^{C'}_{\nu}]
\nonumber\\
&&-\frac{g_3^2}{4}\mathcal{K}\cot^2\theta
[-2g_3(\partial_{\mu}B^A_{\nu}-\partial_{\nu}B^A_{\mu})\cot\theta
f^{ABC}B^{B,\mu}B^{C,\nu}+g_3^2\cot^2\theta
f^{ABC}B^{B,\mu}B^{C,\nu}f^{AB'C'}B^{B'}_{\mu}B^{C'}_{\nu}]\nonumber\\
&&-\frac{g_3^2}{4}\mathcal{K}\tan^2\theta[2g_3(\partial_{\mu}B^A_{\nu}-\partial_{\nu}B^A_{\mu})\tan\theta
f^{ABC}B^{B,\mu}B^{C,\nu}+g_3^2\tan^2\theta
f^{ABC}B^{B,\mu}B^{C,\nu}f^{AB'C'}B^{B'}_{\mu}B^{C'}_{\nu}]
\bigg]\;,\nonumber
\end{eqnarray}



\begin{thebibliography}{}

\bibitem{EWCL0}
T.Appelquist and C.Bernard, Phys. Rev. {\bf D22}, 200(1980);\\
A.Longhitano, Phys. Rev. {\bf D22}, 1166(1980); Nucl. Phys. {\bf
B188},118(1981)

\bibitem{EWCL}
 T.Appelquist and G-H. Wu, Phys. Rev. {\bf D48},
3235(1993); {\bf D51}, 240(1995)

\bibitem{EWCLmatter}
E.Bagan, D.Espriu and J.Manzano, Phys. Rev. {\bf D60}, 114035(1999)

\bibitem{Alam98}
S.Alam, S.Dawson, R.Szalapski, Phys. Rev. {\bf D57}, 1577(1998)

\bibitem{He97}
H.-J. He, Y.-P.Kuang, C.-P.Yuan, hep-ph/9704276

\bibitem{Dobado95}
A.Dobado and M.T.Urdiales, hep-ph/9502255

\bibitem{GaugeInv}
A.Nyffeler, A.Schenk, Phys. Rev. {\bf D62}, 113006(2000)

\bibitem{HeavyHiggs}
M.J.Herrero and E.R.Morales, hep-ph/9308276

\bibitem{STU}
M.E.Peskin and T.Takeuchi, Phys. Rev. Lett. {\bf 65}, 964(1990);
Phys. Rev. {\bf D46}, 381(1992)

\bibitem{PDG06}
W.-M.Yao, {\it et al.} (Particle Data Group), J. Phys. {\bf G33},
1(2006)

\bibitem{Wang00}
 Q.Wang, Y.-P.Kuang, M.Xiao and X.-L.Wang,
  Phys. Rev. {\bf D61}, 054011(2000)

\bibitem{Wang02}
 H.Yang, Q.Wang, Y.-P.Kuang and Q.Lu,
  Phys. Rev. {\bf D66}, 014019(2002)

\bibitem{Yang02}
 H.Yang, Q.Wang and Q.Lu,  Phys. Lett. {\bf B532}, 240(2002)

\bibitem{Ma03}
Y.-L.Ma, Q.Wang, Phys. Lett. {\bf B560}, 188(2003)

\bibitem{WQTechnicolor01}
Z.-M.Wang and Q.Wang,  Commun. Theor. Phys. 36 ,417(2001)\\
H.-H.Zhang, W.-B.Yan, J.K.Parry and X.-S.Li, arXiv:0704.1075
[hep-ph]

\bibitem{Weinberg7579}
S.Weinberg,  Phys. Rev. {\bf D13}, 974(1976); {\bf D19}, 1277(1979);

\bibitem{Hill02}
C.T.Hill and E.H.Simmons, Phys. Rept. {\bf 381}, 235(2003);
Erratum-ibid. {\bf 390}, 553(2004)

\bibitem{Susskind78}
L.Susskind,  Phys. Rev. {\bf D20}, 2619(1979);

\bibitem{Farhi80}
E.Farhi and L.Susskind,  Phys. Rept. {\bf 74}, 277(1981);

\bibitem{Gasser8384}
J.Gasser and H.Leutwyler,  Ann. Phys. {\bf 158}, 142(1984); ~ Nucl.
Phys. {\bf B250}, 465(1985);

\bibitem{Hill94}
C.T.Hill,  Phys. Lett.  {\bf B345}, 483(1995)

\bibitem{PDG}
W.-M.Yao, {\it et al.}, J. Phys. {\bf G33}, 1(2006)

\bibitem{Sparameter}
M.J.Dugan and L.Randall, Phys. Lett. {\bf B264}, 154(1991);\\
H.-H.Zhang, Y.Cao and Q.Wang, hep-ph/0610094

\bibitem{Tanabashi}
T.Inami, C.S.Lim, B.Takeuchi and M.Tanabashi, Phys. Lett. {\bf
B381}, 458(1996)
\end{thebibliography}
\end{document}